\documentclass[preprint,12pt]{elsarticle}

\usepackage{amssymb}
\usepackage{amsmath}
\usepackage{array}
\usepackage{epstopdf}
\usepackage[dvipsnames]{xcolor}

\journal{Physics Reports}

\begin{document}

\newcommand{\Teff}{T_{\rm eff}}
\newcommand{\teff}{T_{\rm eff}}
\newcommand{\Zbar}{\langle Z \rangle}
\newcommand{\sss}{\scriptscriptstyle}
\newcommand{\delad}{\nabla_{\rm ad}}
\newcommand{\kb}{ k_{\rm\sss B} }

\begin{frontmatter}



\title{Current Challenges in the Physics of White Dwarf Stars}


\author[lanl]{Didier Saumon}

\affiliation[lanl]{organization={Los Alamos National Laboratory},
            addressline={P.O. Box 1663}, 
            city={Los Alamos},
            postcode={87545}, 
            state={NM},
            country={USA}}

\author[lanl,uvic]{Simon Blouin}
\author[warwick]{Pier-Emmanuel Tremblay}

\affiliation[uvic]{organization={Department of Physics and Astronomy},
            addressline={University of Victoria}, 
            city={Victoria},
            postcode={V8W~2Y2}, 
            state={BC},
            country={Canada}}
                        
\affiliation[warwick]{organization={Department of Physics},
            addressline={University of Warwick}, 
            city={Coventry},
            postcode={CV4 7AL}, 
            country={UK}}

\begin{abstract}
White dwarfs are a class of stars with unique physical properties. They present many challenging problems whose solution requires the application of advanced theories of dense matter, state-of-the-art experimental techniques, and extensive computing efforts. New ground- and space-based observatories will soon provide an increasingly detailed view of white dwarf stars and reveal new phenomena that will challenge our models. This review is an introduction for researchers who are not in the field of white dwarf astrophysics with the intent to entice them to contribute their expertise to advance our knowledge of these exotic stars. We discuss a wide variety of currently unsolved or partially resolved problems that are broadly related to equations of state, transport processes and opacities.   
\end{abstract}




\end{frontmatter}


\section{Introduction}
\label{sec:intro}
At the end of their long lives, the vast majority of stars become white dwarf stars,
and such is the fate of our Sun.  These common denizens of our Galaxy have been
studied by astrophysicists for about a century. White dwarfs are the sites of exotic physical
conditions that are not usually encountered in normal stars.  From the tenuous surface
with a temperature of the order of 1 eV to the center where it reaches $\sim$~100\,eV and the
density rises to $\sim$~$10^6\,$g\,cm$^{-3}$, white dwarfs present a number of challenging physics
problems that need to be addressed to properly model their structure, evolution and
spectra.

Historically, theoretical work in dense plasma physics has found fertile applications
in white dwarf models.  A famous example is one of the first applications of quantum mechanics
to astrophysics---by R. H. Fowler in 1926---which revealed that the abnormally small radius of white
dwarfs follows from the equation of state of the degenerate electron gas. Further progress in
understanding white dwarfs required the development of models for equations of state, phase diagrams,
radiative opacities, electron conductivity and diffusion coefficients in complex physical
regimes. Some of these models remain approximate and sometimes inadequate to account
for the observations. On the other hand, the composition of white dwarf matter at a given depth in the star is typically mono-elemental or a binary mixture, which offers a degree of simplification compared to other types of stars.

Thousands of white dwarfs have been studied in detail and their ranks have grown
considerably to about a quarter million, most of them identified in the last few years.
The astrophysics of white dwarfs is a mature field and there is an abundance of exquisite data
that calls for new and refined physics models and also provides actual constraints on
theory. The spectra of white dwarfs reveal conditions at the surface, where the gas can
be dense enough to be interacting, with observable effects on the chemical/ionization equilibrium, the
equation of state, and the continuum and line opacities. The pattern of surface abundances reveals the
crucial importance of inter-species diffusion in the star's high gravity field. Some white dwarfs, like many other types of stars,
display periodic variations in their brightness that arise from global oscillations. Most white dwarfs go through at least one
pulsation phase during their evolution. In a manner entirely analogous to seismology for
the Earth, the study of these global oscillation modes---known as asteroseismology---probes the interior structure
of the white dwarfs in remarkable detail.

Beyond their intrinsic astrophysical interest, white dwarfs ought to command the attention
of physicists. They present exotic physical conditions
that are not seen on Earth and that challenge theories of dense matter and even not-so-dense matter.
While state-of-the-art experimental techniques are nibbling at the edges of the least understood
physical regimes, there is an abundance of astrophysical data that bears strongly on the underlying
physical theories and models. The current quantity and precision of the astrophysical data are
expected to rise dramatically in the next few years, thanks to new space-based and ground-based
observatories and large-scale surveys of the sky. The field is ripe with interesting problems
whose solution can be validated against data.

This review is aimed at scientists with an interest in modeling and measuring the properties of materials under
unusual conditions. Enough information is given about white dwarfs to provide context and motivation,
but the emphasis is on physical processes and unsolved problems of immediate interest. We introduce basic concepts of stellar structure, evolution and atmospheres as needed for clarity
and context but these are kept to a minimum. Several excellent monographs at the graduate study
level provide introductions to stellar astrophysics and we cite recent reviews of white dwarf stars
that expound on this rich field of research.

The review is structured in layers, with each main section delving deeper into the phenomenology and physics of white dwarfs.  We first describe the astrophysical context and significance of white dwarfs in broad terms (Section~\ref{sec:WD_stars}). This is a high-level introduction to white dwarfs as a class of stars. The principal physical processes responsible for their structure, evolution and spectral
appearance are introduced in Section~\ref{sec:phys_process}.  This is followed by a discussion of the  physical regimes encountered in white dwarfs that sets the context for the constitutive physics needed to
construct white dwarf models (Section~\ref{sec:constit_phys}). The stage is then set for introducing several contemporary problems in the
physics of white dwarfs (Section~\ref{sec:recent}), some whose current treatment needs refinement and some that remain unsolved. This section covers the more detailed physical theory of many topics including phase diagrams of dense plasmas, transport coefficients 
in partially ionized dense plasmas, collision-induced absorption opacity, 
line profile theory, and simulations of convection. Finally, we conclude with a 
list of problems that we hope will compel readers to contribute to 
the advancement of the astrophysics of white dwarf stars (Section~\ref{sec:conclusion}).

Many phenomena involving white dwarfs, particularly in interacting binary star systems, are quite complex and are
part of exciting branches of modern astrophysics. This review focuses on the somewhat simpler case of single white 
dwarfs as the potential for gains in understanding by the application of physical theory and models is, in our opinion, greatest and more immediate.


\section{White Dwarfs as Stars}
\label{sec:WD_stars}
White dwarfs represent the final evolutionary stage of stars less massive than $\approx 10 M_{\odot}$ \footnote{Convenient astronomical units are the mass, radius and luminosity of the Sun, $M_\odot=1.988 \times 10^{30}$\,kg, $R_\odot = 6.957 \times 10^8$\,m and $L_\odot = 3.828 \times 10^{26}$\,W, respectively, and the radius of the Earth, $R_\oplus = 6.378 \times 10^6$\,m.} ---more massive objects will collapse into neutron stars or black holes. This includes about 97\% of stars in the Milky Way. For most stars, this ultimate state is reached after fusing H into He (first on the main sequence and later on the red giant branch), then fusing He into C and O (first on the horizontal branch and later on the asymptotic giant branch, AGB), and finally losing their outer layers until only a dense C-O core remains. Without any nuclear energy source, except for residual H burning in some cases (Section~\ref{sec:physics_evolution}), these dense stellar embers simply cool down for the rest of time.

\subsection{Anatomy of a white dwarf}
\label{sec:WD_anatomy}
A typical white dwarf has a mass of $0.6\,M_{\odot}$ and a $0.013\,R_{\odot}$ ($=1.4\,R_{\oplus}$) radius, implying an extremely high density ($\rho \sim 10^6\,{\rm g\,cm}^{-3}$ throughout most of its structure). Under those conditions, matter is fully ionized and the electrons are completely degenerate throughout the bulk of the star. The pressure of this degenerate electron gas dominates the structure of the star and prevents it from collapsing under the influence of its strong gravity\footnote{In contrast, the lower-density main sequence stars are supported by thermal pressure, a negligible quantity in white dwarf interiors.}. The hydrostatic equilibrium between the inward pull of gravity and the outward push of the electron degeneracy pressure leads to the well-known mass--radius relation of white dwarfs. For a given chemical composition, the radius of a white dwarf is uniquely determined by its mass and temperature. Since the degeneracy pressure increases rapidly with density, the higher the mass of a white dwarf, the smaller its radius. For white dwarfs supported by the pressure of fully degenerate but non-relativistic electrons, the mass-radius relation is $M_\star \, \propto \, R_\star^{-3}$, where $M_\star$ and $R_\star$ are the mass and radius of the star, respectively.
There is however a limit to the mass that electron degeneracy pressure can support. At very high densities, the electron gas becomes relativistic and the equation of state softens (i.e., goes from $P \, \propto \, n_e^{5/3}$ in the non-relativistic case to $P \, \propto \, n_e^{4/3}$ in the strongly relativistic limit, where $P$ is the pressure and $n_e$ the electron density). This reduces the capacity of electron degeneracy pressure to support the white dwarf, leading to an upper limit on the mass of a stable white dwarf. This limit of $\approx 1.4\,M_{\odot}$ is known as the Chandrasekhar mass \citep{chandrasekhar1931}.

\begin{figure}
\begin{center}
\includegraphics[width=0.75\columnwidth]{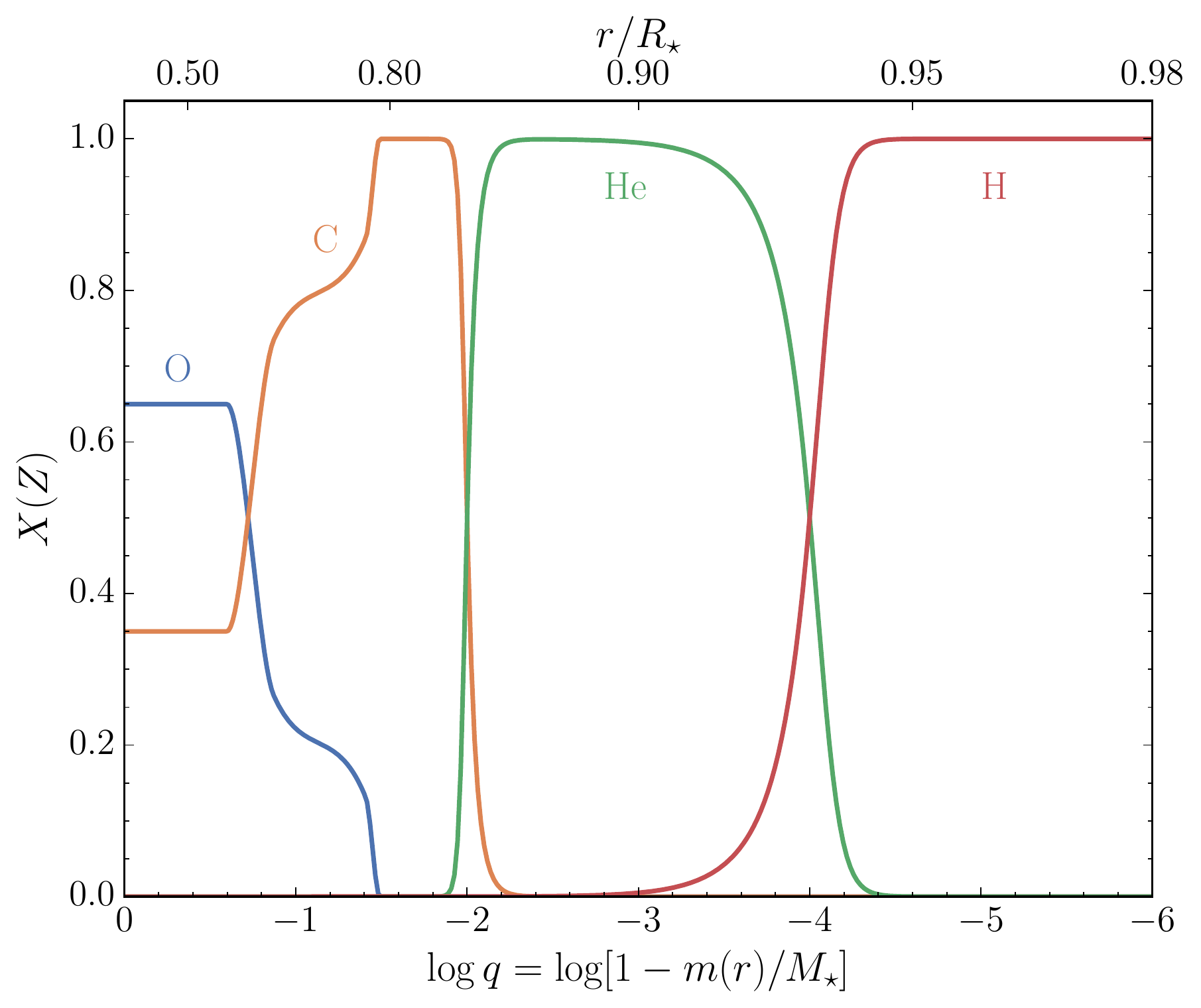}
\caption{Chemical layering in a representative model of a 0.6\,$M_{\odot}$ H-dominated atmosphere (DA) white dwarf with a surface temperature $T_{\rm eff} \simeq 12\,000\,{\rm K}$. The X-axis uses a mass coordinate where $m(r)$ is the mass within a sphere of radius $r$, hence $\log q=\log(1-m(r)/M_\star)$ is the log of the mass fraction outside of radius $r$. The center of the star is at the left, the surface is towards the right. The atmospheric layers, located at $\log q < -15$, are not shown here. \label{fig:Fontaine1}}
\end{center}
\end{figure}

For a typical white dwarf, the C and O leftover after the post-main sequence phases of the evolution form a core that represents 99\% of the total mass  (Figure~\ref{fig:Fontaine1}). Owing to the large thermal conductivity of its degenerate electrons, this core is nearly isothermal, with $T=10^2$ to $10^4\,{\rm eV}$ depending on the age of the white dwarf. The remaining outer 1\% is made of residual He and H that was not fused during the earlier evolutionary phases. An He envelope ($\sim$1\% of the total mass) surrounds the C-O core, which is itself surrounded by a thin H layer ($<0.01$\% of the total mass). The exact thickness of those H and He layers depends on details of nuclear burning during the late phases of stellar evolution. The uppermost layers of the H/He envelope form the atmosphere of the white dwarf, the only region we can directly observe. Its temperature ranges from $T=10$ to $0.1\,{\rm eV}$, and its density remains $\lesssim 1\,{\rm g\,cm}^{-3}$. While the H/He/C-O structure outlined above is representative of most white dwarfs, there are many exceptions. Notably, about 20--25\% of white dwarfs have little or no H present \citep{bedard2020} because they burned virtually all their H during previous evolutionary phases \citep{althaus2005}.


The highly stratified chemical structure of white dwarfs is due to their compact nature. Their high densities confer on them a very strong surface gravity ($g\sim 10^8\,{\rm cm\,s}^{-2}$, compared to $3\times 10^4\,{\rm cm\,s}^{-2}$ for the Sun), with the consequence that heavier elements quickly sink toward the core and lighter ones float to the surface \citep{schatzman1945}. While gravitational settling is the dominant force shaping the chemical structure of a white dwarf, it is often in competition with other  mechanisms such as convection \citep{dantona1979,pelletier1986,macdonald1991,rolland2018}, accretion \citep{dupuis1993,jura2014}, and radiative levitation \citep{chayer1995}. In any case, the distinct H/He/C-O layering of white dwarfs greatly simplifies their modeling compared to most other types of stars, where difficult-to-model convective mixing usually plays a large role.

\begin{figure}
\begin{center}
\includegraphics[width=0.75\columnwidth]{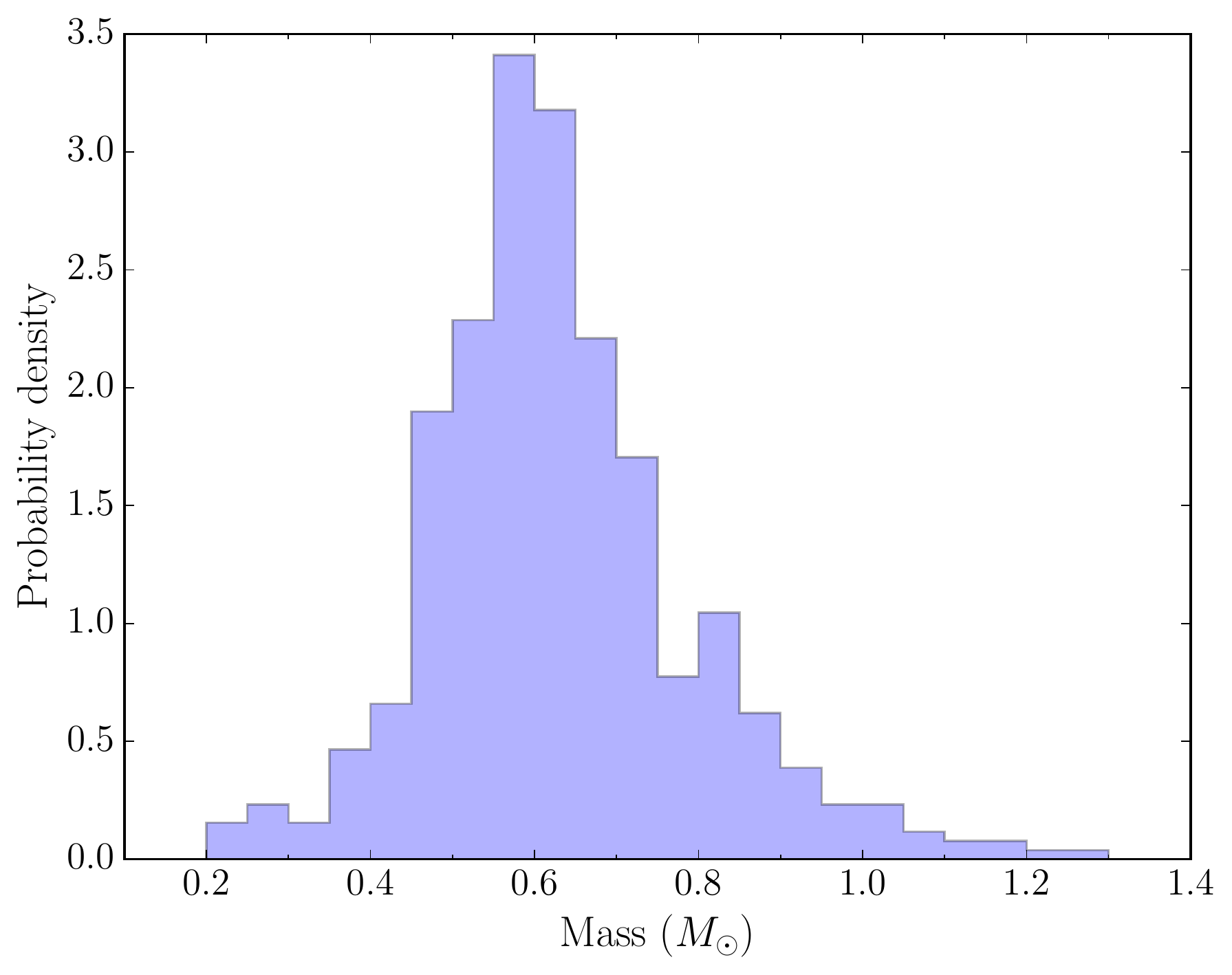}
\caption{Mass distribution of white dwarfs, based on the sample of Ref.~\cite{mccleery2020}. \label{fig:mass_distrib}}
\end{center}
\end{figure}

A C-O core composition is not a universal property of white dwarfs. If the progenitor star is close to the $10\,M_{\odot}$ limit, it may reach high enough temperatures to fuse C, thereby leading to an O-Ne white dwarf. The so-called ultra-massive white dwarfs ($M_{\star} > 1.1\,M_{\odot}$) are thought to have O-Ne cores \cite{siess2007} (although recent evidence points to the existence of ultra-massive C-O white dwarfs as well \cite{cheng2019,bauer2020,camisassa2020,althaus2021,blouin21}). At the other extreme, very low-mass progenitors ($ \lesssim 0.6 M_{\odot}$) will not reach the conditions required to fuse He into C and O, leading to a He-core white dwarf. In isolation, the low-mass progenitors of He-core white dwarfs are expected to remain in the H-fusing main sequence phase longer than the age of the Universe. Any He-core white dwarf that we observe today likely formed by mass loss in a close binary system \cite{iben1993,marsh1995}. This is notably the case of the extremely low-mass (ELM) white dwarfs ($M_{\star} < 0.3 \,M_{\odot}$). While white dwarfs with $M_{\star} \approx 0.2\,M_{\odot}$ all the way to $1.35\,M_{\odot}$ have been identified, the vast majority have masses remarkably close to $0.6\,M_{\odot}$. Indeed, $\approx$ 60\% of white dwarfs have a mass comprised between 0.5 and 0.7$\,M_{\odot}$ \cite{kilic2020,mccleery2020} (see Figure~\ref{fig:mass_distrib}), and $\approx$ 95\% are between 0.45 and 1.1$\,M_{\odot}$, where a C-O core is expected.

Around 10\% of known single white dwarfs are found to harbor a detectable magnetic field. The strength of this field spans many orders of magnitudes, from 1\,kG ($=0.1\,$T) for the weakest fields detected to 1000\,MG for the most strongly magnetic white dwarfs known (for comparison, sunspots can reach 3\,kG) \cite{ferrario2015}. The origin of those fields remains a fundamental unsolved problem of stellar astrophysics. A remnant fossil field inherited from their progenitors, the generation of a field during close binary evolution and stellar mergers, and a dynamo process in the white dwarf itself are all possible explanations.

\subsection{White dwarf cooling}
\label{sec:WD_cooling}
Once a star reaches the white dwarf state, it simply cools down for the rest of its evolution. Accurately modeling this cooling process has been one of the central goals of white dwarf research for over half a century \cite{mestel1952,lamb1975,hansen1999,fontaine2001,althaus2010}. While the evolution of white dwarfs is relatively simple compared to other astrophysical objects, it demands a precise description of a variety physical processes, which take place under extreme conditions seldom found elsewhere in the Universe. The rate at which a white dwarf cools down depends on the neutrino production rate in its core, the thermal conductivity of its core and envelope, the radiative opacity of its atmosphere, element transport in its interior, and the physics of phase transitions. The problem of white dwarf cooling is therefore far from trivial, but the unique potential of white dwarf cosmochronology (Section~\ref{sec:implications}) motivates theoretical and experimental efforts to improve our description of the physical processes at play.

\subsection{White dwarf atmospheres}
\label{sec:WD_atmos}
All the information we receive from white dwarfs comes from their atmospheres, making this very thin surface layer disproportionately important to our overall understanding. The atmosphere also plays the role of outer boundary condition for interior and evolution models, providing further motivation to achieve an accurate description of this region \citep{hansen1999,rohrmann2001,rohrmann2012}.

A first way to characterize the atmosphere of a white dwarf is through photometry which consists of measuring the spectral flux density $f_{\lambda}$ emitted by the star and received on Earth within a given bandpass. This emergent flux mostly comes from a region known as the photosphere, where 50\% of the photons escapes from the atmosphere without being absorbed or scattered. By measuring the flux within a series of different bandpasses, the spectral energy distribution of the star is constrained. This spectral energy distribution is used to determine the surface temperature (reported as the effective temperature $T_{\rm eff}$, which corresponds to the temperature of a blackbody that would emit the same total amount of radiation). In first approximation, the spectral energy distribution is given by the spectrum of a blackbody with temperature $T_{\rm eff}$. However, strong deviations from this approximation are common due to the various opacity sources in the atmosphere, and in practice detailed atmosphere models are used \citep{koester2010}. In some cases, the atmospheric composition (normally dominated by H or He) can be constrained from the spectral energy distribution alone \citep{bergeron2001,blouin2019,cunningham2020,lopez2022}.

If the distance $D$ of the white dwarf to the Earth is known (which is now the case for most known white dwarfs, thanks to the {\it Gaia} space observatory), the mass of the white dwarf can also be inferred. The radius $R_\star$ of the white dwarf is obtained using
\begin{equation}
    f_{\lambda}^{\rm observed} = \left(\frac{\pi R_\star^2}{D^2}\right) f_{\lambda}^{\rm intrinsic}~,
\end{equation}
where $f_{\lambda}^{\rm intrinsic}$ is the intrinsic flux given by a model atmosphere, and $f_{\lambda}^{\rm observed}$ is the flux measured at the Earth. From $R_\star$, the mass of a white dwarf (and surface gravity, $g=GM_\star/R_\star^2$) is then directly obtained from the theoretical mass--radius relation.

Another way through which the atmosphere of a white dwarf can be characterized is with spectroscopy. Spectroscopy generally enables a more precise characterization of the atmosphere as it unlocks the wealth of information contained in spectral lines and molecular bands. It gives access to the detailed chemical makeup of the atmosphere, to its temperature and pressure conditions (through the strengths and widths of the different spectral features), and to its magnetic field (via the Zeeman effect). The analysis of those spectral lines and molecular bands requires a detailed description of a variety of phenomena that affect the spectral line shapes (Stark effect, pressure broadening, etc.), as well as a good model for the vertical stratification of the atmosphere since the spectrum is formed from the contributions of many distinct layers that each have their own temperature and pressure conditions.

\begin{figure}
\begin{center}
\includegraphics[width=0.75\columnwidth]{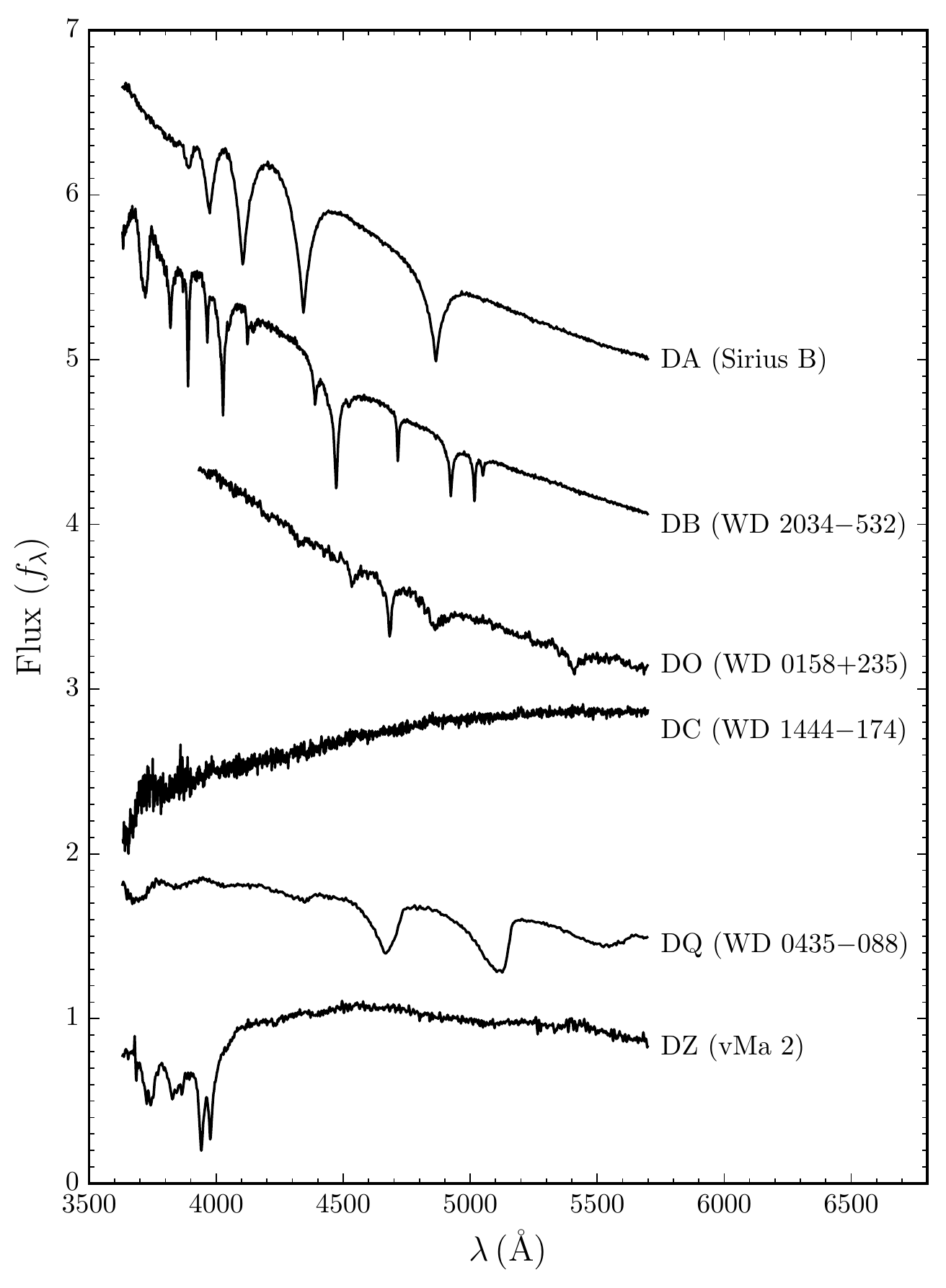}
\caption{Example spectra of DA, DB, DO, DC, DQ, and DZ white dwarfs. All spectra are normalized at 5000\,{\AA} and vertically shifted for clarity. Spectra taken from \cite{bergeron2011,gianninas2011,giammichele2012,limoges2015,bedard2020}. \label{fig:sptype}}
\end{center}
\end{figure}

The diversity of atmospheric compositions and surface temperatures gives rise to a variety of ``spectral types''. The most important ones are listed in Table~\ref{tab:sptype}, and an example of each one is shown in Figure~\ref{fig:sptype}. The DA spectral type is by far the most common and is characterized by H lines from the Balmer series in the visible portion of the electromagnetic spectrum. Then, there are the types DB and DO, which show He and He$^+$ spectral lines, respectively. DB and DO white dwarfs have very little or no residual H, which explains the visibility of He/He$^+$ lines. At low temperatures (below $T_{\rm eff} \approx 11\,000\,{\rm K}$ for He and 5000\,K for H), He and H spectral lines are no longer detectable in the visible portion of the spectrum, yielding a continuous, featureless spectrum. Those objects are known as type DC. There are also objects with C features in their spectra, known as DQ white dwarfs. Those C features can either be C/C$^+$ spectral lines (above $T_{\rm eff}=10\,000\,{\rm K}$) or C$_2$ molecular bands (at lower temperatures). How can C be found in white dwarf atmospheres if gravitational settling is so efficient? Two distinct evolutionary channels are thought to explain the DQ phenomenon. On the one hand, hot and massive white dwarfs with C-dominated atmospheres (known as hot DQs \cite{dufour2007}) are thought to be formed after the merger of two white dwarfs in a binary system \citep{dunlap2015,coutu2019}. On the other hand, the presence of C in cooler objects with C-polluted (but He-dominated) atmospheres is explained by the dredge-up of C from the deep interior by a convection zone in the He layer \citep{pelletier1986,bedard2022}. Finally, there are the DZ white dwarfs, characterized by spectral lines from elements heavier than C. In an exciting recent development, it is now accepted that the presence of low-Z elements (e.g., Ca, Mg, Fe) in their atmospheres is due to the recent accretion of rocky debris from their own planetary system \citep{jura2014,farihi2016,zuckerman2018}. This scenario is supported by the identification of debris disks around metal-polluted white dwarfs \citep{rocchetto2015,wilson2019,manser2020}, many of which have excess infrared emission due to warm dust \citep{farihi2016}, the detection of recurring transit events caused by planetary debris \citep{vanderburg2015,vanderbosch2020,vanderbosch2021,farihi2022} and the measurement of X-ray emission induced by the accretion event \cite{cunningham2022}.

\begin{center}
\begin{table}
\begin{tabular}{>{\raggedright\arraybackslash}p{0.18\linewidth}>{\raggedright\arraybackslash}p{0.38\linewidth}>{\raggedright\arraybackslash}p{0.42\linewidth}}
\hline
Spectral type & Defining characteristics & Origin\\
\hline
DA & H spectral lines & Standard single-star evolution\\  
DB & He spectral lines & H deficient\\
DO & He$^+$ spectral lines & H deficient\\
DC & No spectroscopic features & Cool atmosphere \\
DQ & C or C$^+$ spectral lines or C$_2$ molecular bands & C dredge-up or stellar merger\\
DZ & Spectral lines from elements heavier than C (e.g., Ca, Mg, Fe) & Accretion of rocky debris\\
\hline
\end{tabular}
\caption{The most common white dwarf spectral types.}
\label{tab:sptype}
\end{table}
\end{center}

The spectral type of a white dwarf is not a direct indication of its atmospheric composition. For example, a DA white dwarf can have a substantial amount of He in its atmosphere and still show only H lines, and most DQ and DZ white dwarfs have atmospheres dominated by He. The spectral type only denotes which species dominates the spectrum. Note also that the spectral type is not a fixed property of a white dwarf. Not only does the continuously decreasing temperature of the atmosphere determine which spectral features can be detected, but, more interestingly, the chemical composition of the atmosphere also changes. A He-dominated atmosphere can transform into a H-dominated atmosphere, and vice versa, due to transport mechanisms (chiefly, convection) that compete with the ever-present gravitational settling. This constitutes a unique window into processes that occur below the directly observable portion of the atmosphere, and it is the subject of intense observational and theoretical efforts \citep{rolland2018,blouin2019,coutu2019,genest2019,koester2019,ourique2019,bedard2020,cunningham2020,lopez2022}.

\subsection{Pulsating white dwarfs}
Little can be learned about the deep interiors of white dwarfs from the analysis of their atmospheres alone. Fortunately, the pulsations of white dwarfs can be used to infer the properties of their deeper layers. In white dwarfs, the pulsations are gravity modes where buoyancy is the restoring force, with periods of $\approx 100$--2500\,s. Each pulsating star vibrates in a unique set of normal modes, the frequencies of which are determined by its chemical stratification, rotation profile, mass, radius, temperature, and magnetic field. Asteroseismology is the study of those normal modes with the aim of extracting information on the inner structure of stars. Excellent reviews are devoted to white dwarf asteroseismology \citep{fontaine2008,winget2008,althaus2010,corsico2019}; here we only provide the context needed for the following sections.

Only about $\approx 2-3\%$ of white dwarfs pulsate, but most will pulsate at some point in their evolution. The cause of the onset of pulsations in white dwarfs is the partial ionization of the main constituents of their envelopes. Partial ionization leads to a formidable increase of the envelope opacity, which impedes the outgoing radiation flow and causes pulsational instabilities. The composition of the envelope determines when the partial ionization regime is reached. Hence, DB white dwarfs pulsate between $T_{\rm eff}=32\,000\,{\rm K}$ and $22\,000\,{\rm K}$ due to the partial ionization of He (forming the DBV pulsator class), and DA white dwarfs pulsate in the $T_{\rm eff}=12\,500-10\,500\,{\rm K}$ range when H is partially ionized (forming the DAV class). There are at least six different types of white dwarfs pulsators currently known, each falling within a narrow range of parameters---predominantly $\Teff$---that defines an instability strip. All of those pulsator classes correspond to different phases of the evolution, thereby offering unique peeks at the inner structure of white dwarfs through specific windows during their evolution. 

\subsection{White dwarfs and modern astrophysics}
\label{sec:implications}
Besides being interesting objects in themselves, white dwarfs are also useful to other branches of astrophysics. As they age, the observable properties of the H-burning main sequence stars remain almost unchanged, rendering age determinations very uncertain \cite{soderblom2010}. In contrast, because of their continuous cooling, the observable properties of white dwarfs change drastically as they age, making them uniquely precise cosmic clocks. They have been used to measure the age of a variety of stellar populations \citep{winget1987,oswalt1996,hansen2007,garciaberro2010,kalirai2012,kilic2017,kilic2019} and to reconstruct the history of our Galaxy \citep{rowell2013,tremblay2014,fantin2019,isern2019}. To determine the age of a white dwarf, one must first determine its mass, temperature, and atmospheric composition, which requires atmosphere models. Those atmospheric parameters are then compared to the predictions of cooling models to obtain a white dwarf cooling age (to which the appropriate pre-white dwarf lifetime can be added to obtain the total stellar age). Precision white dwarf cosmochronology therefore requires accurate atmosphere and cooling models. Considering the current limitations of those models, the uncertainty on the ages of very old white dwarfs ($\approx 10$\,Gyr\,\footnote{\ 1\,Gyr = $10^9$\, years}) likely exceeds 1\,Gyr. Improving this figure is one of the chief goals of white dwarf astrophysics. Achieving an accuracy of $\approx 4$\% would provide useful constraints to distinguish between competing cosmological models \cite{boylan2021}.

As a fossil record of previous stellar evolutionary phases, the interiors of white dwarfs contain precious information on the physical processes at play in other types of stars. As such, white dwarf asteroseismology is a promising tool to answer current questions in other subfields of stellar astrophysics. In particular, the C-O abundance profiles of white dwarfs contain critical information on the difficult-to-model convective mixing at the boundary of the He-fusing cores of red giant stars \citep{salaris2017,giammichele2018,giammichele2022}, as well as on the poorly constrained $^{12}{\rm C}(\alpha,\gamma)^{16}{\rm O}$ reaction rate, dubbed ``the holy grail of nuclear astrophysics'' \citep{horowitz2010,deboer2017, fang2017, shen2020,pepper2022,chidester2022}. Furthermore, the internal structure of high-mass white dwarfs that descend from double white dwarf mergers (including hot DQs) \citep{cheng2020} may provide valuable constraints for numerical simulations of such mergers. Double white dwarf mergers are one of the channels through which a type Ia supernova (the powerful explosion of a C-O white dwarf that exceeds the Chandrasekhar limit) can occur \citep{iben1984,webbink1984}. High-mass white dwarfs issued from mergers that failed to exceed the Chandrasekhar limit represent precious observational endpoints to calibrate the intricate numerical simulations of binary interaction.

Another field to which white dwarfs have a lot of exciting contributions to make is that of exoplanets. White dwarfs polluted by debris from their own planetary systems represent our only detailed window on the composition of rocky planetary material outside our solar system. Through a detailed modeling of the atmospheres of those polluted objects and with a good understanding of element transport mechanisms in the envelope, it is possible to precisely trace back the elemental composition of the infalling rocky bodies. So far, most objects have been found to have bulk compositions similar to that of rocky bodies in the solar system. Among the most interesting cases discovered to date are systems with water-bearing planetesimals \citep{farihi2011,farihi2013,raddi2015,gentilefusillo2017,hoskin2020}, differentiated bodies (i.e., bodies large enough to form a core, mantle, and crust) \citep{zuckerman2011,melis2011,gansicke2012,wilson2015,melis2017,harrison2018,hollands2018,harrison2021,buchan2021}, and objects that have intriguingly high Li \citep{kaiser2021,hollands2021,elms2022} and Be \citep{klein2021} abundances.

\section{Physical Processes in White Dwarfs}
\label{sec:phys_process}


Here we briefly describe the structure and cooling of a typical C-O core white dwarf. We refer to several reviews of white dwarf evolutionary models for a more detailed discussion \citep{tassoul1990,fontaine2001,hansen2004,althaus2010,bedard2020}. For the purpose of clarity, we define the nomenclature that we will use hereafter to refer to certain types of white dwarfs and models. Our choice is intentionally simplified from the actual practice but is adequate for our discussion. We divide the star in three regions: the {\it core}, which is made of C-O and constitutes 99\% of the mass; the {\it envelope}, which includes the thin layers (the remaining 1\%) of helium and hydrogen wrapped over the core; and the {\it atmosphere} which refers to the very thin outer layer that form the visible surface of the star. The  spectral type, e.g. DA, DB, etc., is used in the strict sense of its empirical definition (Table \ref{tab:sptype}) and does not imply a particular atmospheric composition. Because the spectral type is not a proxy for the composition of the atmosphere (although they correlate strongly), the latter  will be described as H- or He-atmosphere when it is dominated (including pure composition) by H and He, respectively. Finally, to distinguish white dwarfs that have only an outer He layer from those with both a He and H layers in their envelope, we will use H-rich and He-rich white dwarfs, respectively, even though 99\% of the star is made of C-O. Note that  the composition of the atmosphere is usually but not always the same as that of the outermost layer of the envelope. While these terms have strong overlap among white dwarfs, they are not interchangeable, hence the need for these distinctions.

\begin{figure}
\begin{center}
\includegraphics[width=\columnwidth]{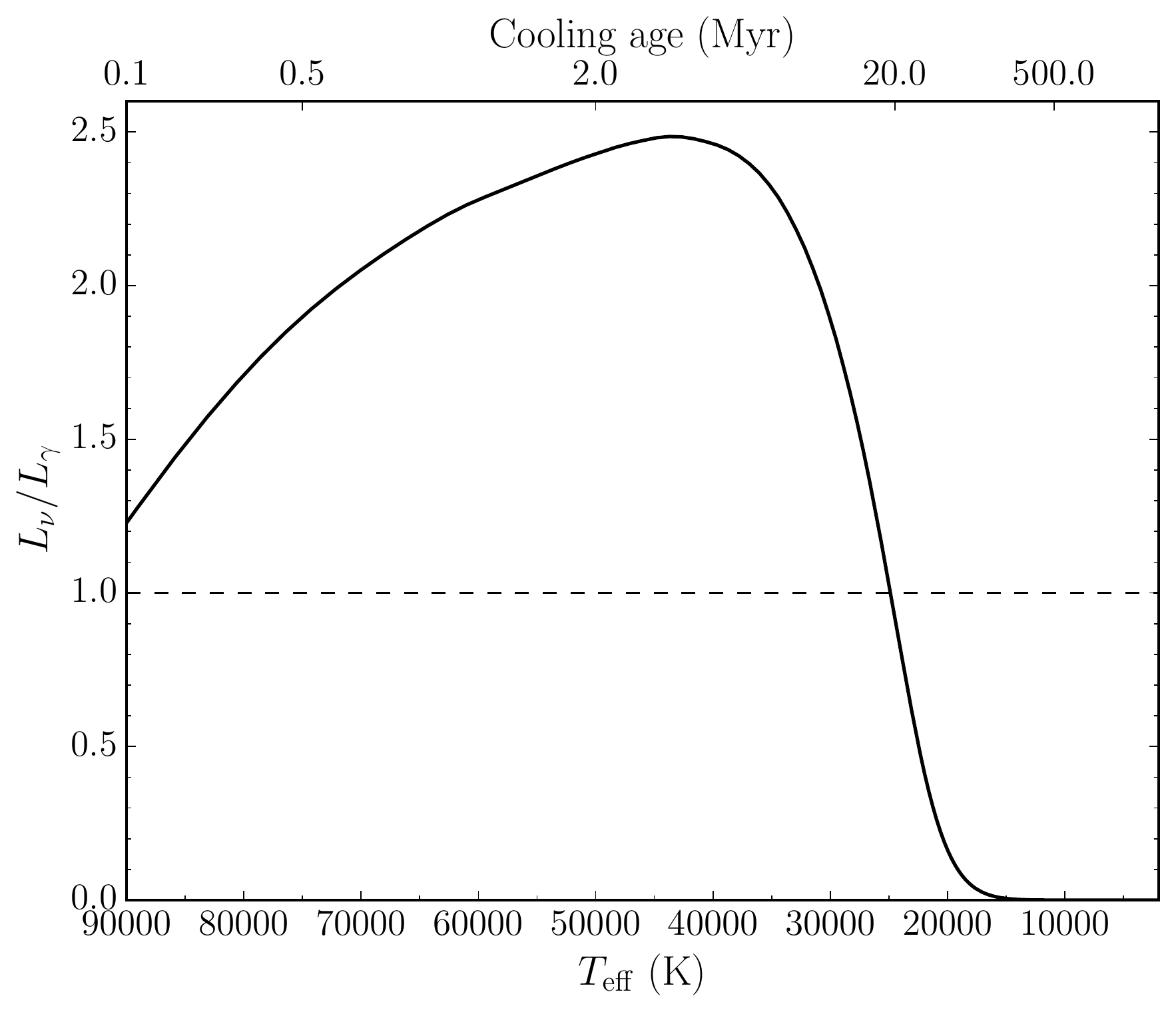}
\caption{Cooling track illustrating the behavior of the neutrino luminosity $L_\nu$ in relation to the photon luminosity $L_\gamma$ in
a representative DA model of 0.6 $M_{\odot}$. The star cools from left to right. For the first 20 Myr, the star loses more energy by neutrino emission than by radiation. This figure is based on the evolutionary tracks of Ref.~\cite{bedard2020}. \label{fig:Fontaine2}}
\end{center}

\end{figure}

\subsection{Evolution}
\label{sec:physics_evolution}

The evolution phases that follow the main sequence (red giant, horizontal branch, AGB, and post-AGB) represent a very small fraction of  the lifetime of a star \citep{hansenkawaler,kippenhahnweigert}, but they set most of the initial conditions for the subsequent white dwarf evolution, most importantly the mass fractions of H, He, C, and O and of traces of heavier elements, and the relation between the mass of the newborn star and its final mass as a white dwarf\footnote{This is the ``initial-to-final mass relation.'' The final mass is always smaller due to extensive mass loss during the life of the star, particularly during the post-main sequence phases.} \citep{weidemann1977,catalan2008,williams2009,cummings2018,el-badry2018,marigo2020}. These properties depend themselves on stellar rotation, the efficiency of convective mixing, nuclear reaction rates, and the metallicity\footnote{In astrophysical nomenclature, which can be arcane, ``metals'' refers to all the elements heavier than H and He. The metallicity of a star is its mass fraction of $Z>2$ elements, which is $\approx 0.013$ for the Sun.} at the time of star formation  \citep{werner2006,salaris2010,cummings2018,marigo2020}. 

While it takes only $\approx 3.5$\,Myr for a white dwarf to cool from $\Teff \sim  120\,000$\,K to 40\,000\,K \citep{bedard2020} (compared to a full lifetime of several Gyr) this early phase of the evolution has been studied extensively due to the ease of observing these bright, hot, young stellar remnants. The high luminosity and negligible nuclear energy production imply very fast cooling rates. Residual nuclear burning is negligible in hot white dwarfs formed from solar-metallicity progenitors, but is predicted to be a moderate source of energy at very low metallicities \citep{althaus2017}.

Element sedimentation works very fast in young white dwarfs, resulting in a chemical stratification of the envelope and nearly pure-H atmospheres in less than 1 Myr. In contrast, H-deficient stars have atmospheres showing a mixture dominated by He, C, and O in roughly comparable proportions for up to 3 Myr. This surface composition is determined by the balance between stellar winds and gravitational settling. The top He and H layers are initially non-degenerate and expanded by thermal pressure. As these layers cool from $\Teff = 10^5$ to 40\,000\,K,  the white dwarf shrinks by about 60\%  \citep{bedard2020}. This gravitational contraction is a minor source of energy, given the small mass of the outer layers.

Neutrino cooling is an important energy sink during the early stage of evolution \citep{althaus2010}, where the interior is hot enough for neutrinos to be created by  electroweak interactions. The vast majority of the neutrinos escape directly from the central regions to the stellar exterior without interacting with the stellar plasma, thus resulting in a net energy loss. Figure~\ref{fig:Fontaine2}  demonstrates that neutrino processes dominate the cooling time scale until the surface reaches a temperature of $\approx$ 25\,000\,K corresponding to a cooling age of 20 Myr. 

\begin{figure}
\begin{center}
\includegraphics[width=\columnwidth]{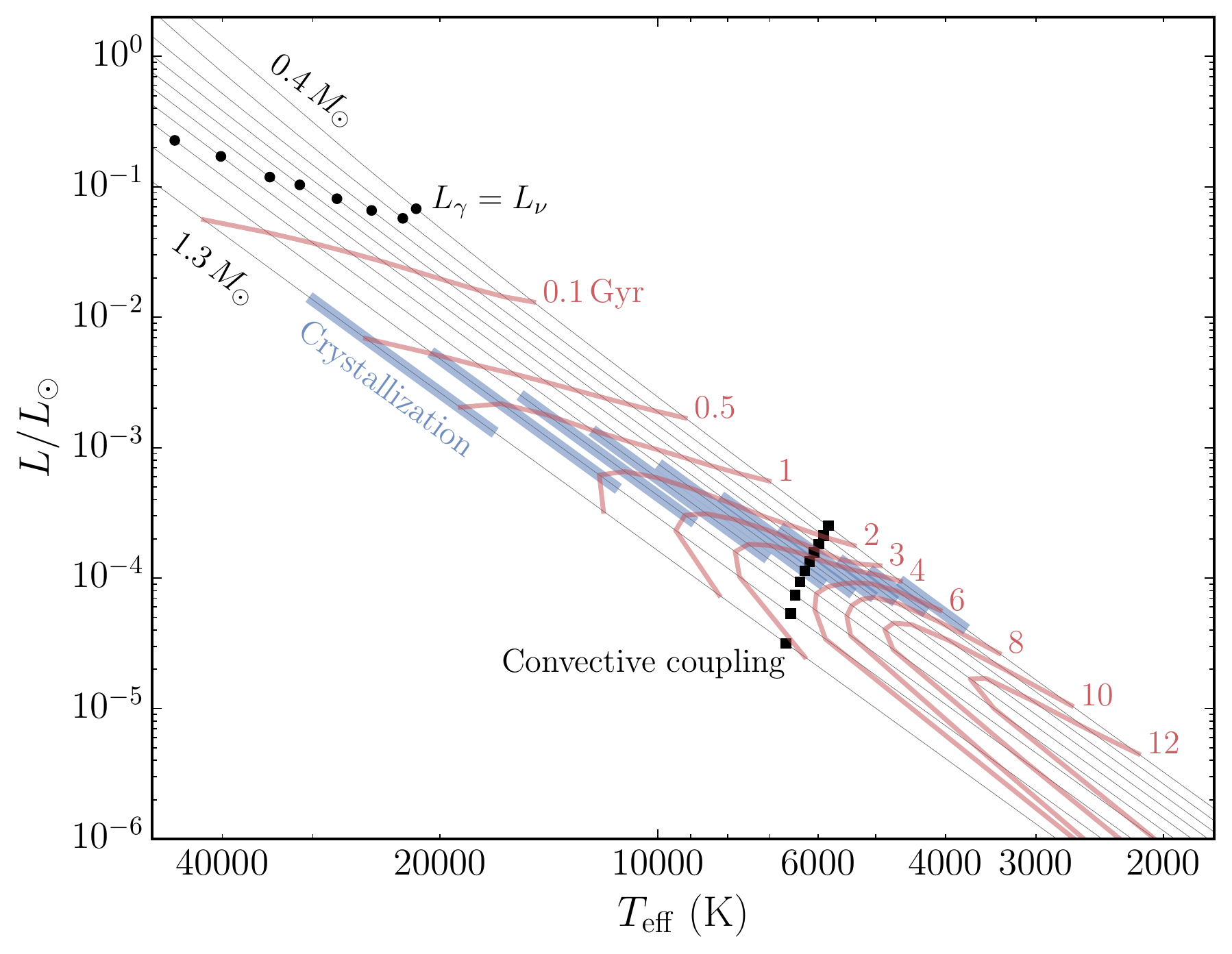}
\caption{Evolutionary tracks (thin grey lines) of C-O core DA white dwarf models ($M_{\star}=0.4$ to $1.3\,M_{\odot}$ in steps of $0.1\,M_{\odot}$). The red curves are isochrones, labelled with the cooling time in units of Gyr. The small filled circles indicate the transition between the neutrino cooling phase, which dominates at higher luminosities,  and the thermal
cooling phase. The blue section of each evolutionary track corresponds to the crystallization of the C-O core. The set of square symbols indicates the onset of convective coupling between the surface and the degenerate core. The stars evolve from left to right, hence the $\Teff$ axis is backwards. This figure is based on the evolutionary tracks of Ref.~\cite{bedard2020}. \label{fig:Fontaine3}}
\end{center}
\end{figure}

The next phase of evolution is thermal cooling, where the structure depends on the properties of the degenerate electrons and thermal ions. The slow leakage of thermal energy from the ions in the plasma largely sets the cooling rates. The interior initially acts as a moderately coupled plasma, but Coulomb interactions
grow as the star cools and the plasma eventually  crystallizes through a first-order phase transition (Section~\ref{sec3:crystallization}). 

The He and H envelope contains relatively little mass, but plays an important role in the cooling of a white dwarf because its large radiative opacity regulates  the flow of energy to outer space. The final phase of cooling is when superficial convection, initially restricted to the photosphere, grows deeper to reach the degenerate interior, an event named convective coupling (Section \ref{sec3:transport}).

Figure~\ref{fig:Fontaine3} illustrates the thermal phase of the cooling process for evolutionary models in the mass range 0.4 -- 1.3 $M_{\odot}$ and $T_{\rm eff}$ range 50\,000\,K -- 2000\,K. It highlights the three important events in the lifetime of a white dwarf: the end of the neutrino cooling phase and the beginning of thermal cooling (small filled circles), crystallization in the core (blue region), and convective coupling between the surface and the nearly isothermal core (small squares). 
The red isochrones show the strong dependence of the cooling rate on the mass and effective temperature of the white dwarf and how the mass determines the timing of the three main events in white dwarf cooling. A striking feature of the cooling tracks is that they are very nearly linear in this log-log figure and almost parallel. Since a star's luminosity $L_\star = 4\pi R_\star^2 \sigma \Teff^4$, where $\sigma$ is the Stefan-Boltzmann constant, this indicates that they cool at very nearly constant radius $R_\star$ and that the mass-radius relation is essentially fixed after a few tens of Myr of cooling, with only a few percent deviation from that of a fully degenerate star. Notably, the most massive white dwarfs cool fastest even though they have a higher thermal energy content and a smaller surface to radiate heat. This comes from the early release of energy due to crystallization at high luminosities and the subsequent Debye cooling (Section \ref{sec3:crystallization}).

In summary, the computation of cooling ages suffers from three main uncertainties. First, the constitutive physics (equations of state, radiative and conductive opacities, neutrino production rates, diffusion coefficients) impacts the three important events described above. The second uncertainty is the amount of ion thermal energy that is stored in the interior of the star, which depends on the total number of C and O ions. Their relative proportion follows from the nuclear fusion rates and convective mixing during the pre-white dwarf evolution. Finally, cooling rates also depend on how fast this energy is transferred from the interior to the surface across the thin but opaque envelope, which depends on the total residual masses of H and He, that are set by the post-AGB evolution.

\subsection{Crystallization}
\label{sec3:crystallization}

The onset of crystallization, the phase transition from liquid to solid of the C-O plasma at the center, is indicated by the upper end of the blue sections of the tracks in Figure~\ref{fig:Fontaine3}.  Because the internal energy is discontinuous between the liquid and solid phases, this predicted phase transition is of the first order. This corresponds closely to the first order liquid-solid phase transition of the one-component plasma and releases a substantial  $\approx \frac{3}{4}\,k_{\rm\sss B}T$ per ion of latent heat \citep{vanhorn1968,salaris2000,potekhin2013}.  We note that because massive white dwarfs have larger internal densities (for comparable temperatures), they
develop a crystallized core at higher effective temperatures and earlier in their evolution. It takes of the order of 1--3 Gyr for crystallization to grow and encompass $>$90\% of the degenerate core (Figure \ref{fig:Fontaine3}). The amount of energy released during this process, including an associated increase in gravitational binding energy (of comparable magnitude to the energy released by latent heat, see Section \ref{sec:phase_diag}), introduces a cooling delay of about 1--2 Gyr. 

It was demonstrated fifty years ago that the slow down of the cooling process due to the phase transition would result in an observable statistical excess of white dwarfs at certain temperatures and masses, corresponding to core conditions where crystallization occurs \citep{vanhorn1968}. The first evidence of such an accumulation of white dwarfs was found in the cooling sequence of the globular star cluster NGC 6397, which was interpreted as the release of latent heat from crystallization \citep{winget2009, campos2016}. 
More recently, the very precise data from the \textit{Gaia} space observatory revealed clear evidence of the early crystallization among white dwarfs with a mass higher than the average in the disk of the Milky Way (Figure \ref{fig:Fontaine3}) \citep{tremblay2019_crystal,cheng2019}. 

Theoretical studies of crystallization following the initial prediction 
characterised the crystallization process and associated physics. In particular, crystallization can induce phase separation, gravitational sedimentation and distillation in a mixture of C, O, Ne and traces of higher mass elements \citep{stevenson1980,isern1991,segretain1993,isern1997}.  The {\it Gaia} observations spurred new, detailed studies \citep{bauer2020,caplan2020,blouin20,blouin21,caplan2021}. The physical processes associated with crystallization are linked to the atomic weight of several key elements and isotopes crystallizing at different temperatures and inducing a change in chemical stratification that releases additional gravitational energy. These new developments are discussed in Section \ref{sec:phase_diag}.

Figure~\ref{fig:Fontaine3} shows that more massive models have a rapid final phase of cooling following core crystallization. This produces the extensive deformation of the later isochrones in the figure. This phase of ``Debye cooling'' occurs when the core temperature falls below the Debye temperature of the Coulomb solid $\theta_{\sss\rm D} = \hbar \omega_{\rm p}$, where $\omega_{\rm p}$ is the plasma frequency of the ions in the lattice. At such low temperatures, the discrete nature of the phonon spectrum of the lattice becomes apparent and the specific heat of the solid, initially given by the classical high-temperature  Dulong-Petit value of $C_{\sss\rm V} =3\kb$ per ion now decreases as $\sim (T/\theta_{\sss\rm D})^3$. In other words, the ions in the solid have lost most of their capacity to store thermal energy, which initiates this final and relatively rapid cooling phase \citep{vanhorn1968, lamb1975}. In white dwarfs with a typical mass of $0.6\,M_{\odot}$, Debye cooling starts very late in the evolution and the oldest such stars in our Galaxy have just recently entered the phase of Debye cooling.  

\subsection{Energy transport}
\label{sec3:transport}

There are three main mechanisms of energy transport in white dwarfs: radiation, conduction and convection. These methods of transport can occur in parallel. This is augmented by neutrino cooling in early stages, which can transport energy directly from the core to the outside. Conduction from degenerate electrons dominates energy transfer within the degenerate core throughout the cooling process, resulting in fairly isothermal cores. Convection and radiation dominate in the non-degenerate portion of the envelope, which we now discuss.

\begin{figure}
\begin{center}
\includegraphics[width=\columnwidth]{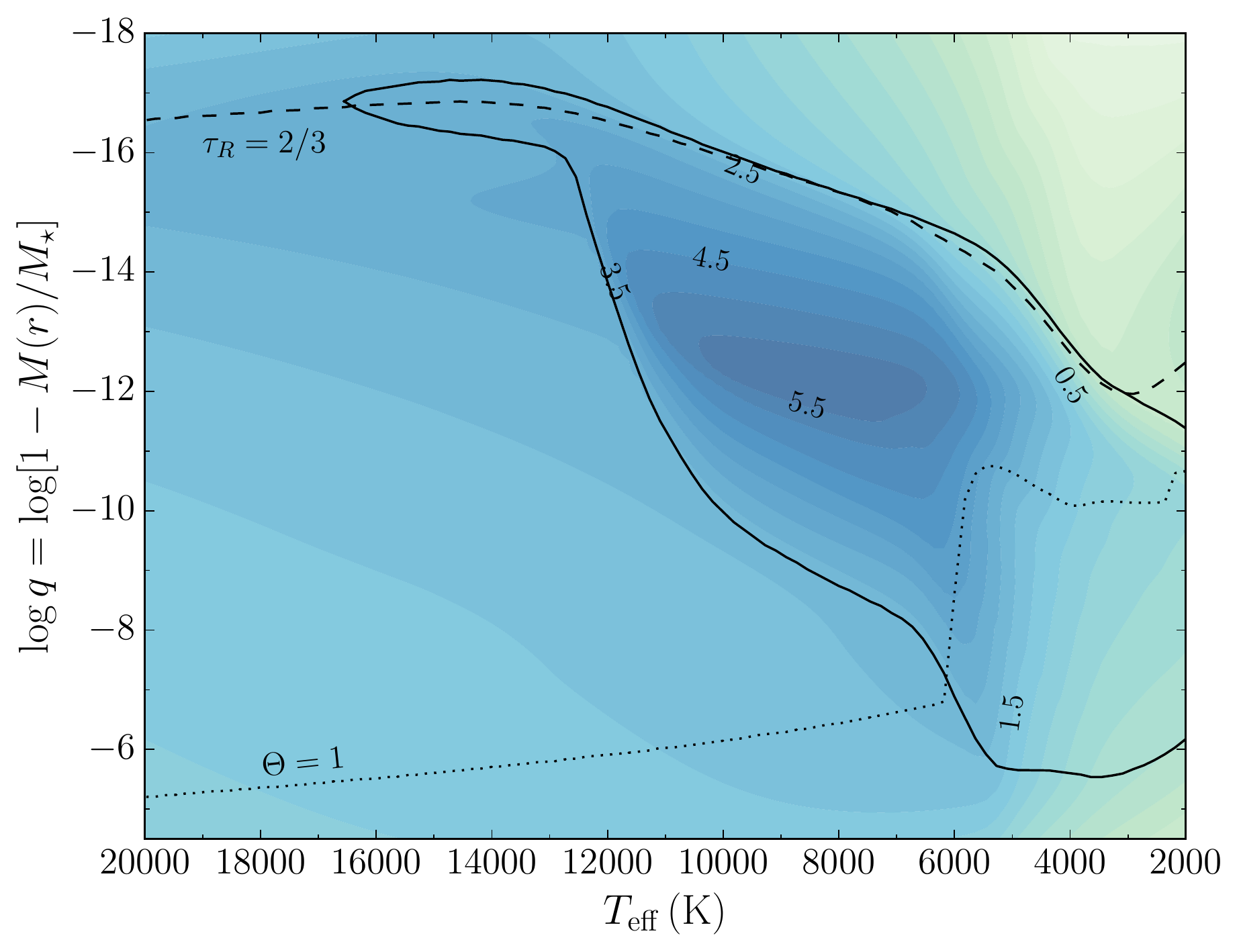}
\caption{Evolution of the convection zone and radiative opacity (blue contour surfaces, labeled with $\log \kappa_{\rm\sss R}$ in cm$^2$\,g$^{-1}$) in the outer envelope of a 0.6~$M_{\odot}$ H-rich white dwarf as a function of the fractional mass above a given layer in the star (a vertical coordinate) and $T_{\rm eff}$. The star cools from left to right.
The convection zone is bounded by the black solid contours and is largely determined by regions of high radiative opacity that occur when H is partially ionized. The depth of the photosphere (defined as the level where the Rosseland mean optical depth $\tau_{\rm\sss R}=2/3$; dashed line) and of the onset of electron
degeneracy ($\Theta = T/T_{\rm\sss F} =1$, where $T_{\rm\sss F}$ is the Fermi temperature; dotted line) are also shown. This figure was generated using outputs from the STELUM white dwarf evolution code \cite{bedard2021}.
\label{fig:pier2}}
\end{center}
\end{figure}

As they cool, models of H-rich white dwarfs with a thick H layer ($M_{\rm H}/M_{\star} \sim 10^{-4}$) start to develop convective instabilities at the photosphere due to the partial ionization of H at an effective temperature of around 18\,000\,K \citep{cunningham2019}, or a cooling age of 100 Myr.\footnote{In Figure \ref{fig:pier2}, which is based on a simple 1D model of convection, the onset of convection occurs at 16\,500\,K.}  Initially convection is inefficient and characterised by low P\'eclet numbers (see Figure 16 of Ref.~\citep{cunningham2019}). In Figure~\ref{fig:pier2}, the convection zone eventually grows deeper to sub-photospheric layers at around 12\,000\,K where it becomes efficient at transporting energy (P\'eclet number of order unity). At $T_{\rm eff} \lesssim$ 10\,000\,K, the convective photosphere becomes increasingly more adiabatic\footnote{Convection is said to be adiabatic when a rising fluid eddy does not exchange energy with its surroundings before dissolving higher up in the star. In this limit, the vertical temperature-pressure profile corresponds to an adiabat in the equation of state.} and less dependent on the assumed efficiency in the 1D model of convection \cite{tremblay2015conv}, while the convection zone continues to grow deeper where the dense flows are almost fully adiabatic. At the same time, the degeneracy front moves upward as the white dwarf cools (Figure~\ref{fig:pier2}). This creates the conditions for the so-called convective coupling event \cite{tassoul1990,fontaine2001} as the convection zone reaches the degenerate layers at a $\Teff$ of about 6000\,K, following a rapid change in envelope opacity. This is an important event in the evolution of a white dwarf since this allows, for the first time, a direct coupling between the energy reservoir in the degenerate interior with the radiating surface. 
The transport of energy from the degenerate core to the photosphere no longer depends on the opacity of an intermediate radiative layer. The efficient convection allows internal heat to escape rapidly, temporarily slowing down the decrease in luminosity.
Eventually, the main effect is however to accelerate cooling. After convective coupling, the convective zone stops growing inward (Figure~\ref{fig:pier2}). Convective layers never reach deeper than the superficial $M_{\rm H}/M_{\star} \sim 10^{-6}$ mass fraction. Compared to main sequence stars, these relatively shallow convection zones never reach densities large enough to produce observable magnetic dynamo effects \cite{fontaine1973}. Hence, and unlike the solar magnetic field, the population of magnetic white dwarfs cannot be explained by a dynamo induced by an outer convection zone (Section \ref{sec:conv-magnetic}). 

He-atmosphere white dwarfs follow a similar pattern of development of convective instabilities due to the partial ionization of He. Convection appears near the photosphere at $\approx 50\,000$\,K, the convection zone deepens rapidly  at $\approx$~30\,000\,K, and convective coupling occurs at $\approx$~12\,000\,K \citep{tassoul1990,rolland2018}. Compared to H-dominated atmospheres, these processes are shifted to much larger effective temperatures owing to the higher ionization potential of He. 

Even though convection involves complex three-dimensional flow patterns, most stellar models rely on the mixing-length theory of convection (MLT), a simple 1D phenomenological model of the convective flux \cite{bohm-vitense, hansenkawaler,coxgiuli, kippenhahnweigert}. The MLT considers a fluid element that, due to some local perturbation, is displaced vertically. As it rises and moves down the star's temperature and pressure gradients, it expands and cools, and suffer radiative losses to the surrounding fluid. If after this initial perturbation the fluid element is buoyant, it will continue to rise and the region is convectively unstable. The MLT provides estimates of the convective velocity, the convective energy flux and the resulting temperature gradient in the convective region, all functions of a single free parameter determining the vertical length scale of convective flows. This parameter is known as the mixing length, which is typically 1-2 times the local pressure scale height $H_{\sss\rm P}$ in the star, which is defined by
\begin{equation}
P(r) \propto \exp(-r/H_{\sss\rm P})~,
\end{equation}
where, for an ideal gas in hydrostatic equilibrium, $H_{\sss\rm p} = P/ (dP/dr) = \kb T/g \mu$, where $g$ is the gravitational acceleration and $\mu$ the average mass per particle. Except in regions of low optical depth ($\tau \lesssim 1$), the fluid element rises and expands nearly adiabatically. Such adiabatic convection is insensitive to the choice of mixing length and can carry very large fluxes. On the other hand, white dwarf model spectra are very sensitive to the efficiency of convection at the surface, which is largely unconstrained by the MLT \citep{bergeron1995,tremblay2013}. Nevertheless, the MLT provides a good description of the importance of convection in the evolution of a white dwarf. Modern 3D radiation-hydrodynamics simulations of convection in white dwarfs are discussed in Section 5.1.

Convection is thought to play a major role in spectral evolution, as vertical mixing can change the composition of the atmosphere (Section \ref{sec:WD_atmos}). Here we focus on major transitions from H- to He-dominated atmospheres. Convective dilution is a process in which the underlying He layer becomes convective in white dwarfs with very thin outer H layers ($M_{\rm H}/M_{\star} \sim 10^{-15}$). This results in a full mixing of the H layer into the larger underlying He layer at $\Teff$ of  30\,000--20\,000\,K, resulting in a transition from H-dominated to He-dominated atmospheres \citep{rolland2020}. Another important process is convective mixing, where models indicate that the top convective H layer, if $M_{\rm H}/M_{\star} < 10^{-6}$, reaches the underlying He layer (Figure \ref{fig:pier2}). This also results in the full mixing of the H layer with the much larger underlying He layer \cite{rolland2018,blouin2019,cunningham2020}. For thicker H layers, the superficial H convection zone never reaches the underlying He layer and convective mixing cannot happen at any point during the evolution. We note that the accretion of solid planetary debris can also induce more moderate changes of atmospheric composition, such as variations in the H/He ratio in He-dominated atmospheres from the accretion of water-rich asteroids as a source of H \citep{gentilefusillo2017}. 

The internal reservoir of thermal energy in a cooling white dwarf is insulated by the radiative H and He envelope. It is the vertically integrated opacity of this non-degenerate envelope that sets the rate of energy loss. Figure~\ref{fig:pier2} illustrates this opacity barrier for a DA white dwarf, highlighting the contribution of the insulating layers with large opacity (colored contours) above the degeneracy boundary which is defined as $\Theta = T/T_{\rm\sss F} =1$, where $T_{\rm\sss F}$  is the Fermi energy (dotted line). The opacity maximum is caused by the partial ionization of H. The upper boundary of the core's reservoir of ion thermal energy roughly corresponds to the degenerate boundary, where the running radial integral of the specific heat $\int C_{\rm\sss V} T\,dm$ reaches a plateau \citep{tremblay2015mag}. The strength of the insulating radiative layer is proportional to the optical depth $\tau_{\rm\sss R} = -\int \kappa_{\rm\sss R} \rho dr$ between the base of the convection zone and the degeneracy boundary $\Theta$ = 1. Because density increases with depth, the integrand is sharply peaked in the layers just above the degenerate boundary. This demonstrates how this region controls the rate of energy transfer from the degenerate core to the exterior. The low-density atmosphere ($\tau \approx 2/3$) provides a negligible contribution to the opacity barrier and does not influence the cooling process. This implies that the cooling rates, hence the relation between core and surface temperatures, are exactly the same for H-rich and He-rich white dwarfs until the convection zone ultimately reaches the degenerate boundary. This convective coupling results in the sudden disappearance of this insulating layer. From now on, the degenerate core is connected to the surface by a nearly adiabatic convection zone that can carry far more flux to the surface.  The opacity of the atmospheric layers now controls the cooling process, leading to age differences of $\approx$ 1 Gyr between H-rich and He-rich white dwarfs once they reach surface temperatures of 4000\,K \citep{bedard2020}.

Conductive opacities in the deep interior are relatively well established \citep{cassisi2007}. A better accounting of electron-electron scattering has resulted in an improved understanding of these opacities in the regime that corresponds to the core-envelope interface that largely governs cooling rates \citep{blouin2020con}.
Conduction works in parallel with radiative transfer at the base of the insulating layer. In optically thick layers, radiation transport is  described as a diffusive process and only depends on the Rosseland mean opacity, which is rather well known in this regime (see Section \ref{sec:constit_phys}).

The atmosphere does not impact cooling rates for most white dwarfs, but important work has also been performed to understand radiative transfer, its effect on line formation, its interaction with convection, as well as the radiative levitation of trace metals. In particular, non-local thermal equilibrium effects have a major impact on the structure of the upper atmosphere in young hot white dwarfs ($T_{\rm eff} \gtrsim 40\,000$\,K) \citep{wesemael1980, werner1996}. However, apart from the upper atmosphere of hot white dwarfs, local thermal equilibrium prevails in the entire star.

\subsection{Diffusion}
\label{sec3:diffusion}
Atomic diffusion is a key process in our understanding of white dwarf structure and evolution. It is responsible for the relaxation of the composition profile resulting from pre-white dwarf evolution into a chemically stratified young white dwarf with a C-O core surrounded by He and H layers (Figure \ref{fig:Fontaine1}).  In competition with stellar winds and radiative levitation, diffusion plays a central role in the subsequent spectral evolution \citep{bedard2020}. Even once the white dwarf has cooled below 30\,000\,K and the internal structure has largely settled, atomic diffusion couples to new mechanisms that appear later in the evolution, such as convection, accretion and crystallization. In DZ white dwarfs, it differentially sorts the infalling planetary material and alters the observed elemental composition at the photosphere. The upward tail of carbon at the core/He transition (Figure \ref{fig:Fontaine1}) can be dredged up by the He convection zone to ``pollute'' the surface of DQ white dwarfs with traces of carbon. During crystallization, the associated fractionation of trace elements between the solid and liquid phases couples the phase transition to the release of gravitational energy and diffusion in the liquid phase (Section \ref{sec:trace_crystal}).

Diffusion in stellar plasmas occur in the presence of a gravitational field and a radial temperature gradient. The inter-diffusion of ions is governed by the combined effects of gradients of concentration, pressure, and temperature, and an induced electric field. Composition transition zones are the regions of continuously varying composition that separate homogeneous layers. For instance, a white dwarf with a H layer has two composition transition zones, between the H and He layers, and between the He layer and the C-O core. Although this is not necessary, most evolution models assume that the abundance profiles in the transition zones are specified by the condition of diffusive equilibrium and neglect thermal diffusion. The result is interpenetrating diffusion tails at the transition zones, as can be seen in Figure~\ref{fig:Fontaine1}.

Other phenomena in white dwarfs arise from the diffusion of a trace element in a (usually) fully ionized background of H and He. 
In the presence of surface convection that maintains a homogeneously mixed composition, diffusion allows heavier elements to drop out at the bottom of the mixed region. The diffusion of trace elements usually requires time-dependent calculations \citep{pelletier1986, paquette86a, paquette86b, dupuis1993, bauer2019}. These calculations show that the $e$-folding diffusion time scale at the bottom of the convection zone ranges from days to $\approx 10^4$\, years \citep{koester2009, bauer2019} and that diffusive equilibrium is reached in a matter of at most millions of years in white dwarf envelopes but can take much longer in the degenerate interior \citep{camisassa2019,rolland2020}. The short diffusion time scale near the surface is an important result as it implies that metal-polluted DZ white dwarf atmospheres must have accreted these elements very recently in their evolution or are currently accreting, as their sinking times are short compared to evolutionary time scales. Otherwise, they would have long disappeared below the photosphere. This led to the current understanding where these elements result from recent or continuous accretion of planetary material---perhaps in the form of asteroids that are tidally disrupted---onto the white dwarf where they can be observed in the spectrum \citep{farihi2016}. Conversely, the upward diffusion of small amounts of H in the He layer can counteract convective dilution, where the He convective layer reaches up to the H/He transition zone and otherwise results in a mixed composition layer \citep{rolland2020}.

\subsection{Pulsations}
\label{sec3:pulsations}

The properties of the H-atmosphere (DAV) and He-atmosphere (DBV) pulsating white dwarfs are closely linked to the appearance of convection, partial ionization zones and large peaks in opacities. The opacity bumps due to the photoionization of H or He become so large that convection sets in and the pulsation modes are driven (linearly unstable) mostly through the convective driving mechanism. This comes from the fact that the integrand $dW/dm$ of the work integral\footnote{The kinetic energy of a pulsation mode is sustained by the work done by both the pressure and the gravitational forces averaged over a pulsation cycle.} \cite{cox1980,shiba}, which gives the contribution of each stellar layer to the net driving/damping of a mode averaged over one oscillation period, has most of its driving contribution near the base of the superficial convection zone \citep{fontaine2008,corsico2019}. Because the convective turnover time scale at the base is much shorter than the pulsation periods of interest, the convective flux responds
almost instantaneously to pulsations \citep{brickhill83}.

The compact nature of a white dwarf implies that the period of a pulsation mode with eigenfunction of radial order $k$ and degree index $\ell$ is much smaller than in other stars. The observed modes are all internal, non-radial gravity waves, with periods $\approx$ 100--2500 seconds. The eigenfunctions of these g modes have large amplitudes mostly restricted to the external envelope layers, the exact opposite of what is seen in non-degenerate stars like the Sun where g modes probe the deep interior. 

The condition for local g mode propagation is that the square of the frequency of a mode has to be less, at a given depth in a model, than the smaller of the square of the Brunt-V\"ais\"al\"a frequency
\begin{equation}
    N^2 = g \left( \frac{C_{\rm\sss V}}{C_{\rm\sss P}}\frac{d \ln P}{dr} - \frac{d \ln \rho}{dr} \right)~,
\end{equation}
 and the Lamb frequency
\begin{equation}
    S_l^2 = \frac{\ell(\ell+1)c_s^2}{r^2}~,
\end{equation}
where $c_s$ is the sound speed, and $C_{\rm\sss V}$ and $C_{\rm\sss P}$ are the specific heats at constant volume and pressure, respectively. The Brunt--V\"ais\"al\"a frequency is the oscillation frequency of a perturbed fluid element in a convectively stable region and the Lamb frequency is the inverse of the sound crossing time over the horizontal pulsation wavelength. The large values of these characteristic frequencies in white dwarfs, allowing for propagation of gravity modes in the envelope, is a direct consequence of electron degeneracy on the Brunt--V\"ais\"al\"a frequency \citep{fontaine2008}.

The fact that g modes show maximum amplitudes in the outer non-degenerate envelope implies that the observed pulsations are most sensitive to the parameters of these layers, such as the extent of the convection zone and of the H and He layers. The composition transition zones are of critical importance for asteroseismology. Their influence is revealed by peaks in the Brunt--V{\"a}is{\"a}l{\"a} frequency at composition transitions that ``pinch'' the eigenfunction of the pulsation modes, leading to spatially confined ``trapped'' modes and resonances.  Diffusion determines the sharpness of the transition zones and their potential for mode trapping. Because of this phenomenon, asteroseismology provides a powerful method to constrain the internal stratification of white dwarfs and to determine the thicknesses of their He and H layers \citep{fontaine2008}. In contrast, the degenerate interior of a white dwarf is difficult (but not impossible \cite{giammichele2018,giammichele2022}) to survey using pulsations since the eigenfunctions have very low amplitudes in those regions.

In recent years, progress has been made to understand the boundaries of the DBV and DAV instability strips in terms of $\Teff$ and surface gravity. A current problem in white dwarf asteroseismology is that the predicted $\Teff$ of the lower boundaries are lower than the observed values by a factor of about two  \cite{vangrootel2012,vangrootel2017}. It has recently been discovered that DAV pulsators near the observed low-$\Teff$ edge ($\approx 10\,500$\,K) exhibit energetic outbursts that increase the luminosity by $\sim 10$\% on a non-periodic time scale of days \cite{bell2015,hermes2015}. The current explanation for this behaviour involves a resonant transfer of energy between driven and damped modes \cite{luan2018}.

\subsection{Magnetic fields}
\label{sec3:magnetic}

The origin of magnetic fields in single white dwarfs is still largely unexplained and we refer the reader to several reviews on the topic \citep{ferrario2015,ferrario2020,bagnulo2021}. Most of the major scenarios involve external processes or pre-white dwarf evolution rather than an origin during white dwarf cooling. They include binarity and common-envelope evolution, post-main sequence or main sequence internal dynamos, flux conservation of fossil fields from the pre-main sequence, and double white dwarf mergers. One additional  proposed scenario, consisting of an internal convective dynamo generated during the crystallization process, could happen naturally during the cooling process \cite{isern2017,schreiber2021,ginzburg2022}. However it is still unclear if this scenario, whose efficiency depends on the rotation velocity and the diffusion of internal magnetic fields to the surface, is a significant channel for the creation of magnetic white dwarfs \citep{bagnulo2021,bagnulo2022}.

There is evidence that magnetic field incidence increases with mass \citep{ferrario2015,mccleery2020,bagnulo2021} ---a trend that can be explained by all scenarios for the origin of magnetic white dwarfs. Recent studies have shown that all magnetic field strengths in the range 0.1--500 MG are equally likely \citep{bagnulo2021}. There is also evidence that younger white dwarfs ($<$0.5 Gyr) have a lower incidence of magnetic fields and smaller field strengths \citep{bagnulo2021,bagnulo2022}. Some of these changes with cooling age could be caused by the outward diffusion to the surface of a magnetic field generated during the pre-white dwarf phases or from a crystallization-induced dynamo, which must overcome Ohmic decay, with estimated time scales of 0.5--4 Gyr depending on white dwarf mass \citep{fontaine1973}.

 Magnetic fields can affect the structure (e.g. the mass-radius relation) of white dwarfs if the magnetic pressure is larger than the central pressure,
 \begin{equation}
     \frac{B^2}{8\pi} > P_c \sim 10^{24}\,{\rm erg\,cm}^{-3}
 \end{equation}
 or $B \gtrsim 5 \times 10^{12}$\,G. While this simple estimate only accounts for the direct contribution of the magnetic field to the total pressure and ignores its effect on the equation of state (EOS), it is consistent with more detailed calculations that show that the mass-radius relation is not affected unless the interior field reaches $\sim 10^{13} - 10^{14}$\,G  \cite{bhattacharya2021}.  Such fields are orders of magnitude higher than the strongest fields observed at the surface and are considered unlikely in single white dwarfs (Section \ref{sec:mag_parameters}). 
 
 Similarly, observed magnetic fields are too weak to influence white dwarf cooling. This is because the bottleneck in cooling rates is at the interface between the degenerate core and the envelope above. Magnetic fields smaller than 1000 MG are not strong enough to impact energy transfer rates in these deep layers. The interaction between convection and magnetic fields is discussed in Section~\ref{sec:conv-magnetic}.

Magnetic fields in white dwarfs are detected and characterized by their effect on their spectra through Zeeman splitting and shifts of spectral lines or polarization of the continuum. These effects, which can be quite strong, are surface manifestations of magnetic fields where the density and temperatures are low compared to the interior. The problem of modelling spectral line profiles in the presence of comparable Stark and Zeeman effects has remained largely unsolved until very recently \citep{ferri2022}. This means that spectral lines of strongly magnetic white dwarfs cannot yet be used for reliable atmospheric parameter determinations or for studying magnetic effects on the atmospheric structure. Work is ongoing for calculating line splitting for highly magnetic white dwarfs with field strengths above the Paschen--Back regime, with the goal of better understanding magnetic field geometries and atmospheric composition \citep{hampe2020}. On the other hand, analysis with the photometric method using the accurate distances obtained with the {\it Gaia} space observatory is a reliable alternative to determine the stellar radius of the cooler magnetic white dwarfs \citep{mccleery2020} rather than fitting the gravity from the observed line profiles (Section \ref{sec:WD_atmos}).

\subsection{General relativity in white dwarfs}
\label{sec:GR}

The importance of general relativistic effects on the structure of white dwarfs can be estimated from the ratio
\begin{equation}
    \gamma = \frac{R_{\rm s}}{R_\star}  = \frac{2GM_\star}{c^2 R_\star}
\end{equation}
where $R_{\rm s}$ is the Schwarzschild radius. For a typical white dwarf of 0.6\,$M_\odot$ with a radius of 8900\,km, $\gamma = 2 \times 10^{-4}$ and the effect is commonly neglected in white dwarf models. However, a massive white dwarf near the Chandrasekhar mass limit of 1.4\,$M_\odot$ with a radius of $\approx 1000\,$km
has $\gamma \approx 0.004$. Thus, the most massive white dwarfs are compact enough for weak general relativistic effects to affect their structure. In particular, this reduces the Chandrasekhar mass by 2--3\% \citep{carvalho2018}.

\section{Physical regimes in white dwarfs}
\label{sec:constit_phys}
In this section, we examine the physical regimes encountered in white dwarfs as a reference for the development of models, theories and experiments to address the various challenges they pose. The physical conditions inside white dwarfs vary over many orders of magnitude from the very dense, fully ionized C-O core to the tenuous gas typical of the visible photosphere. These conditions also evolve as the star cools. It follows that a wide range of physical regimes are represented.

\subsection{Models with an outer hydrogen layer}
\label{sec:DA_physics}
\begin{figure}
\begin{center}
\includegraphics[width=\columnwidth]{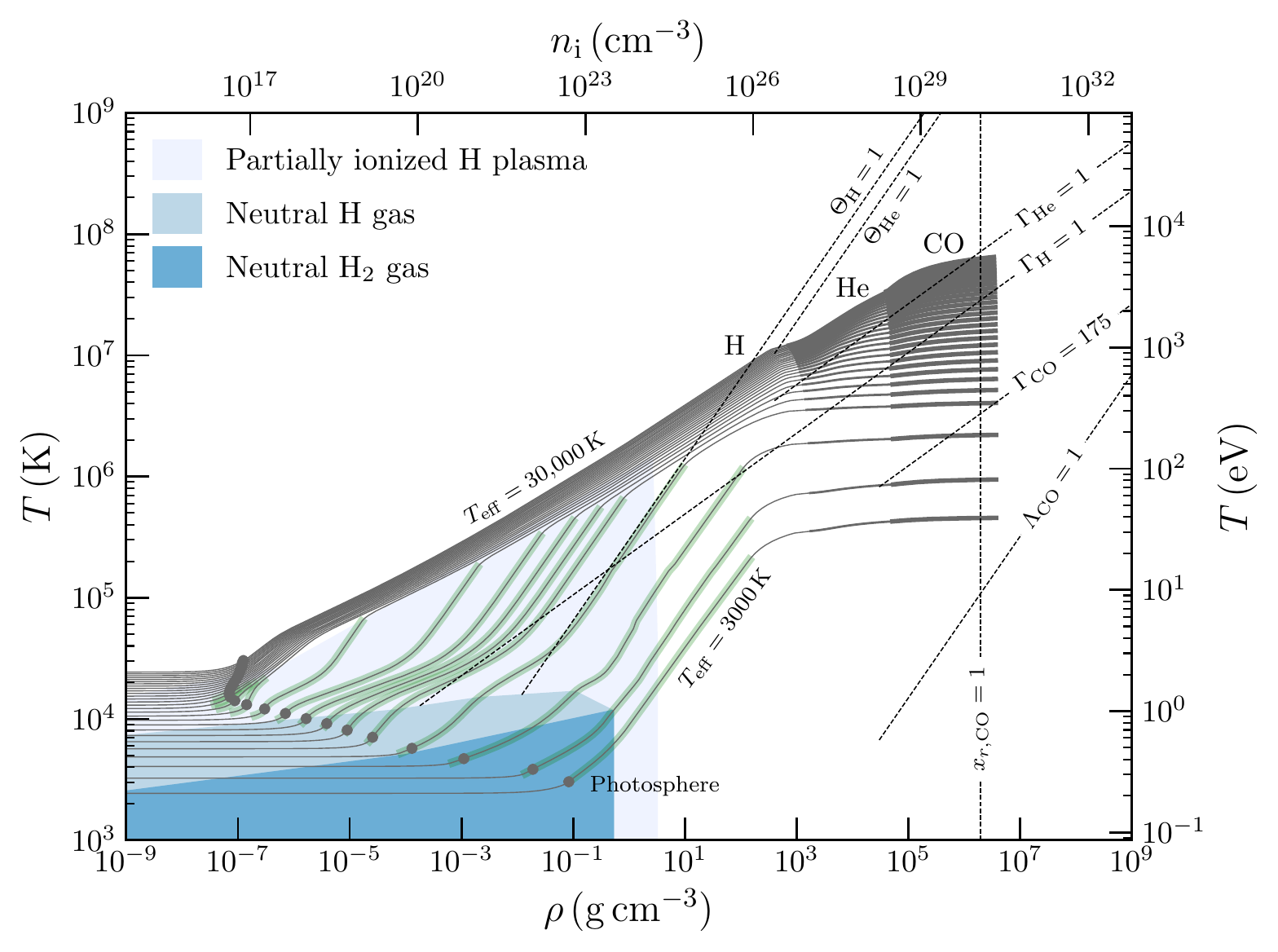}
\caption{Density--temperature profiles of a cooling H-rich white dwarf model. Profiles from $T_{\rm eff}=30\,000\,{\rm K}$ to 3000\,K in steps of 1000\,K are shown. The cooling sequence runs from top to bottom. The thickness of the line indicates the dominant atomic constituent, and regions that are convectively unstable are highlighted in green. The different shadings of blue indicate where the gas is mostly composed of H$_2$, atomic H gas, or a partially ionized H plasma. The white background indicates a fully ionized plasma of C, O, He or H. Finally, the circles on each profile mark the location of the photosphere, and the dotted lines indicate the values of some important plasma parameters (see text). Those profiles were obtained assuming a core with equal mass of C and O, an envelope with $\log (M_{\rm He}/M_{\star}) = -2$ and $\log (M_{\rm H}/M_{\star}) = -4$, and a total mass of $M_\star=0.6\,M_{\odot}$. In this log-log plot, the outer layers of the star are greatly expanded. The surface of the star is toward the left, the center is on the right.   
\label{fig:rhoT_DA}}
\end{center}
\end{figure}

Figure~\ref{fig:rhoT_DA} shows a series of density--temperature profiles of a representative model for a white dwarf with an outer H layer with total mass $M_\star=0.6 \, M_{\odot}$ (see Figure \ref{fig:mass_distrib}) as it cools from $T_{\rm eff}=30\,000$ to 3000\,K.\footnote{The models presented in Sections~\ref{sec:DA_physics} to \ref{sec:hDQ_physics} are computed with the white dwarf evolution code STELUM \citep{bedard2021}.} Such white dwarfs are of spectral types DA and turn into spectral type DC  below $\Teff \approx 5000$\,K (Section \ref{sec:WD_atmos}). The model assumes a core with equal mass fractions of C and O surrounded by a He layer of $10^{-2}\,M_\star$ and an outer H layer of $10^{-4}\,M_\star$. As is justified for the vast majority of single white dwarfs, rotation and magnetic fields are neglected. The 30\,000\,K model has a white dwarf cooling age of about 9\,Myr and the coolest and the oldest white dwarfs known have $T_{\rm eff}$ somewhat below 4000\,K and an age of about 10\,Gyr. This set of models spans all but the very earliest cooling history of H-rich white dwarfs.

The stratified structure of the star in C-O, He and H layers is revealed by the thickness of the grey lines in Figure \ref{fig:rhoT_DA}.  The C-O core is nearly isothermal due to the high thermal conductivity of the degenerate electrons. As the star cools, the central temperature decreases from 6\,keV to 40\,eV, but since it contracts very little during this part of the evolution, the density profile of the C-O core remains unaffected, ranging from $\log \rho\,$(g\,cm$^{-3}) \sim 4.6$ to 6.6. Under these conditions, both C and O are fully ionized. The He layer is also fully ionized and mostly isothermal. The temperature and density decrease by several orders of magnitude in the thin H layer, all the way to the photosphere (solid dot). Initially fully ionized in the hotter models, the H layer enters a regime of partial ionization around $T_{\rm eff} = 15\,000$\,K. At the photosphere, H atoms dominate below 10\,000\,K and H$_2$ molecules form in models cooler than 5000\,K. Pressure dissociation and ionization of H occurs in the coolest models at $\rho \sim 1\,$g\,cm$^{-3}$. Energy transport towards the surface is dominated by electron thermal conduction and radiation, but for $T_{\rm eff} \lesssim 18\,000\,$K a convectively unstable region develops and grows in the H layer (thick green line; compare to Figure \ref{fig:pier2}). 

The $(\rho,T)$ plane can be divided into regions of various plasma regimes. Electron degeneracy sets in when the degeneracy parameter, the ratio of the temperature to the Fermi temperature given by
\begin{equation}
\Theta = \frac{\kb T}{\epsilon_{\rm\sss F}} = \frac{2 \kb T m_e}{\hbar^2 (3 \pi^2 n_e)^{2/3}}~,
\label{eq:theta}
\end{equation}
is smaller than unity. Here, $\kb$ is Boltzmann's constant, $\epsilon_{\rm\sss F}$ the Fermi energy, $m_e$ the mass of the electron, $\hbar$ the Planck constant and $n_e$ is the electron number density. Two lines of $\Theta=1$ are shown in Figure~\ref{fig:rhoT_DA} corresponding to pure H and pure He compositions. The corresponding line for the C-O plasma is the same as that for He. Electrons are degenerate in the entire C-O core, the He layer and in the deeper regions of the H layer.  At the observable photosphere, electrons are always classical and recombine to form atoms and molecules in the cooler models. 

The ion plasma coupling parameter
\begin{equation}
\Gamma = \frac{Z^2e^2}{a k_{\rm\sss B} T}
\label{eq:Gamma}
\end{equation}
is the ratio of the electrostatic potential energy of two ions of charge $Ze$ at a separation of an ion sphere radius $a$ over their kinetic energy ($\sim k_{\scriptscriptstyle B} T$). Ions are strongly coupled and non-ideal when $\Gamma> 1$. Lines where $\Gamma=1$ are shown for pure H and pure He compositions. The corresponding line for the C-O plasma lies well above the $\Gamma_{\scriptscriptstyle \rm He}=1$ line and is omitted for clarity. The H layer is usually weakly coupled except in models cooler than 5000\,K. Similarly, the He layer is initially weakly coupled for models hotter than $\approx 20\,000\,$K but it eventually becomes strongly coupled. The plasma coupling is always strong throughout the C-O core. It is well known that a strongly coupled plasma, more specifically the one-component plasma model, freezes at $\Gamma \approx 175$ \citep{potekhin2000}. This freezing line for the C-O plasma intersects the core structures for models with $T_{\rm eff} \lesssim 6000\,$K. The entire C-O core has become solid once the star has cooled to $\approx 4000\,$K (Section \ref{sec3:crystallization}).

At the extreme densities found at the center of white dwarfs, the Fermi energy of the electrons becomes comparable to their rest mass energy. The relativity parameter is defined as
\begin{equation}
    x_r = \frac{p_{\sss\rm F}}{m_e c} = \frac{\hbar (3 \pi^2 n_e)^{1/3}}{m_e c}~,
    \label{eq:xr}
\end{equation}
where $p_{\sss\rm F}$ is the Fermi momentum and $c$ is the speed of light. The $x_{r,{\rm \scriptscriptstyle CO}}=1$ line indicates that electrons are moderately relativistic at the center of white dwarfs throughout their evolution.

Finally, ions can display quantum mechanical behavior at very high densities and low temperatures. The $^{12}$C and $^{16}$O nuclei are bosons of spin zero and are subject to Bose-Einstein statistics. Quantum mechanical effects  become manifest when the thermal de Broglie wavelength $\lambda_{\rm\sss th}$ of a particle is no longer negligible compared to the interparticle separation. In this case, ions no longer interact as classical point particles. This effect is measured by the quantum parameter
\begin{equation}
\Lambda = \frac{\lambda_{\rm th}}{a} =  \frac{1}{a} \Bigg[\frac{2\pi \hbar^2}{m k_{\rm\sss B} T} \Bigg] ^{1/2}~.
\label{eq:Lambda}
\end{equation}
The line of $\Lambda_{\rm\scriptscriptstyle CO}=1$ is below all the white dwarf structures shown in Figure~\ref{fig:rhoT_DA} but we can expect that weak quantum effects occur in the solid C-O core of the coolest white dwarfs.

\begin{figure}
\begin{center}
\includegraphics[width=\columnwidth]{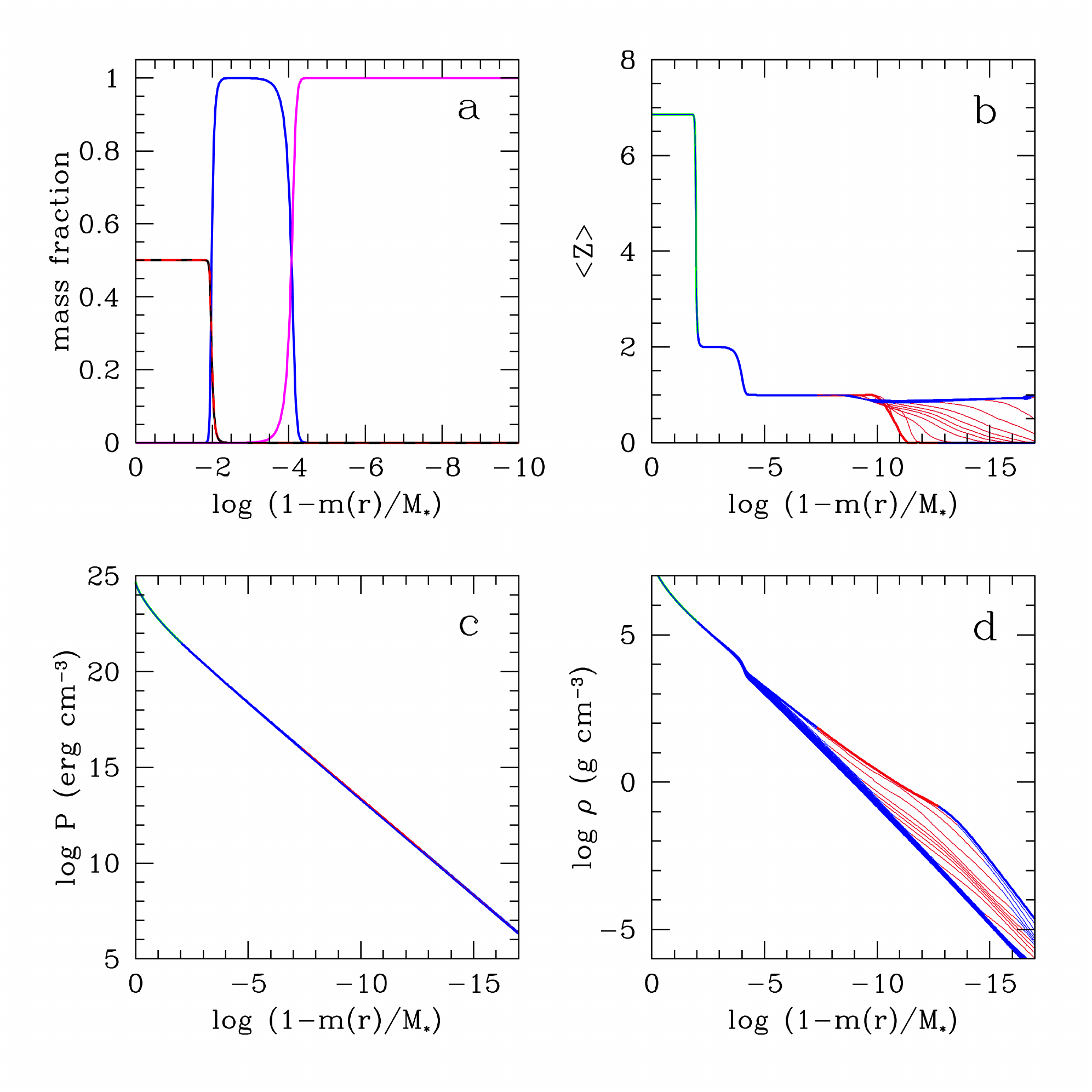}
\caption{Properties of a cooling  sequence of models of a H-rich white dwarf of $M_\star = 0.6\,M_\odot$ consisting primarily of a C-O core, surrounded by a $10^{-2}\,M_\star$ He layer and an outer H layer of $10^{-4}\,M_\star$. Models range from $\Teff = 30\,000\,$K to 3000\,K (heavier line) in steps of 1000\,K. The radial coordinate is a logarithmic mass coordinate (Equation \ref{eq:log_q}) with the center of the star located on the left at $\log q=0$ and the surface towards the right (note the ``backward'' X-axis). In panels b--d, conductive and radiative regions are in blue, convectively unstable regions in red and the crystallized core in green.
a) Composition profile showing the mass fractions of C (black), O (red), He (blue) and H (magenta), b) Average ion charge $\Zbar$, c) Pressure ($10^{12}\,$erg\,cm$^{-3}$ = $10^{12}\,$dyn\,cm$^{-2}$ = 1\,Mbar = 100\,GPa), d) Density profile. 
\label{fig:DA_panel1}}
\end{center}
\end{figure}

\begin{figure}
\begin{center}
\includegraphics[width=\columnwidth]{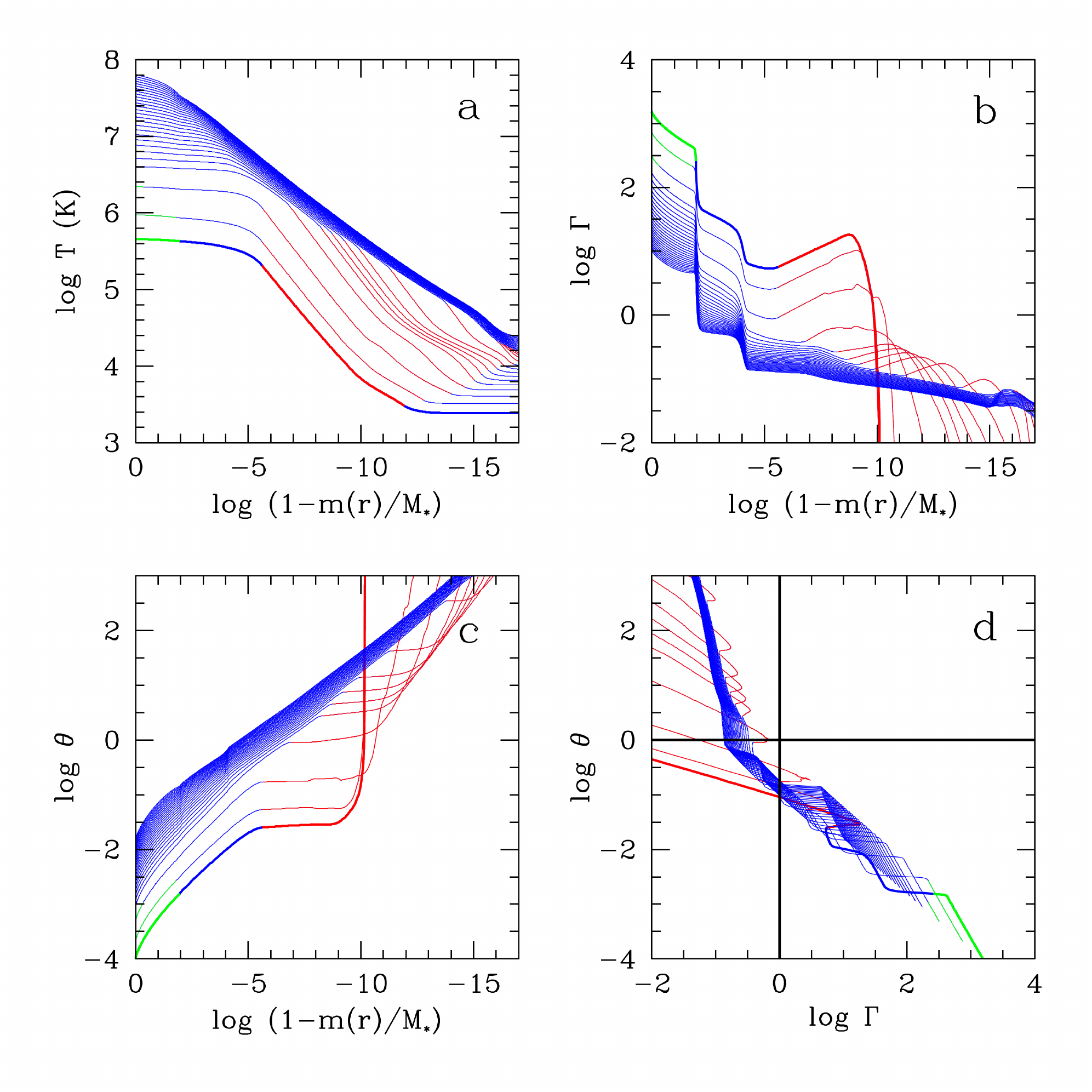}
\caption{Same as Figure \ref{fig:DA_panel1} but for the a) Temperature profile (1\,eV = 11605\,K), b) Ion coupling parameter $\Gamma$ (Equation \ref{eq:Gamma}), c) Electron degeneracy parameter $\Theta$ (Equation \ref{eq:theta}), and d) Ion coupling and degeneracy parameters.
\label{fig:DA_panel2}}
\end{center}
\end{figure}

\begin{figure}
\begin{center}
\includegraphics[width=\columnwidth]{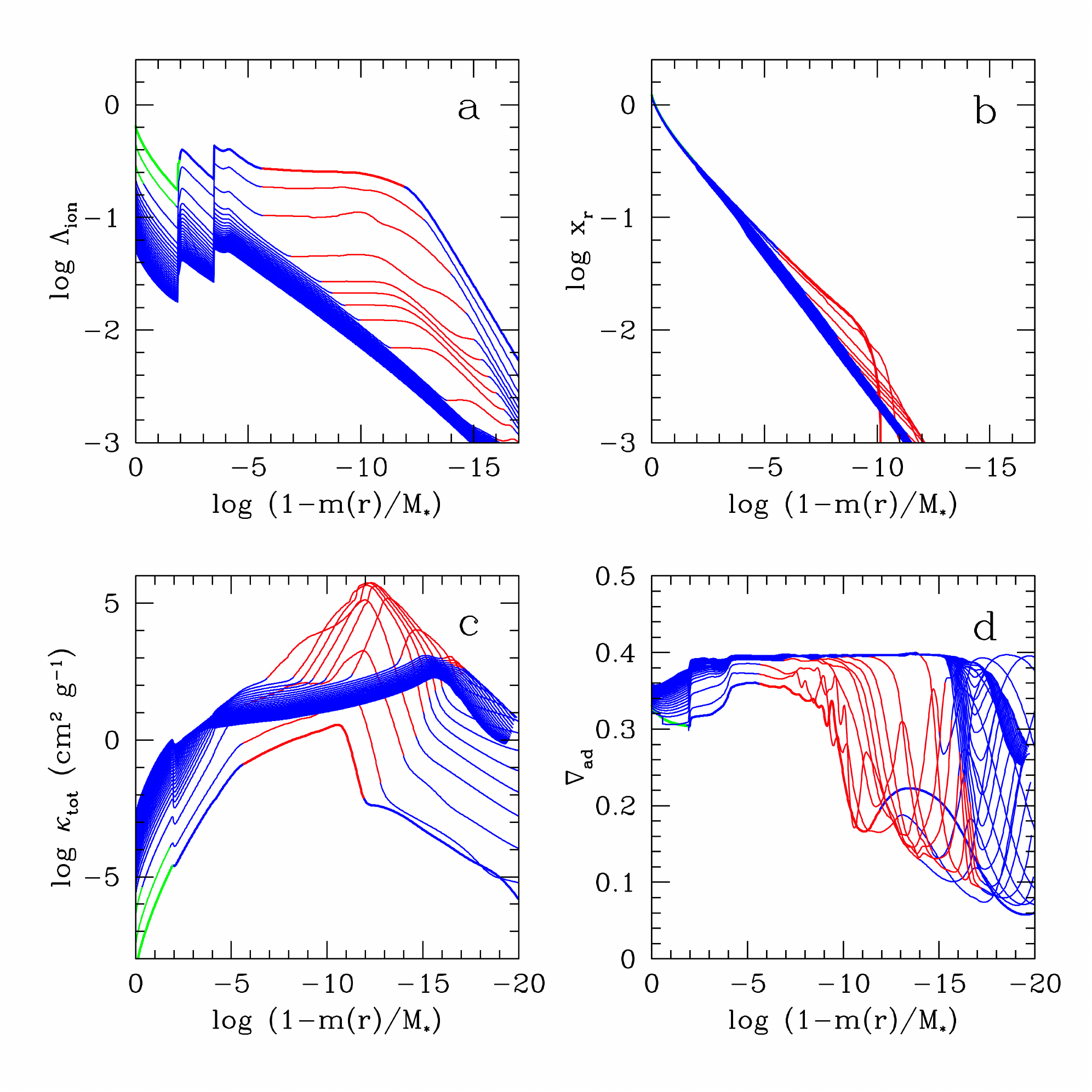}
\caption{Same as Figure \ref{fig:DA_panel1} but showing the a) Ion quantum parameter $\Lambda$ (Equation \ref{eq:Lambda}), b) Electron relativity parameter $x_r$ (Equation \ref{eq:xr}), c) Total opacity $\kappa_{\rm tot}$ (Equation \ref{eq:Ktot}), and d) Adiabatic temperature gradient $\nabla_{\rm ad}$ (Equation \ref{eq:delad}). 
\label{fig:DA_panel3}}
\end{center}
\end{figure}

Figures \ref{fig:DA_panel1} -- \ref{fig:DA_panel3} offer a closer look at key plasma parameters and properties throughout the models. 
As in Figure \ref{fig:Fontaine1}, they use a logarithmic mass coordinate $\log q$ for the radial position in the star
\begin{equation}
    \log q = \log \Bigg(1-\frac{m(r)}{M_\star} \Bigg)~,
    \label{eq:log_q}
\end{equation}
where $m(r)$ is the mass enclosed in a volume of radius $r$. This mass coordinate greatly expands the outer part of the star where much interesting physics occurs. As a point of reference, the photosphere (``surface'') of the star is located around $\log q \sim -17$ for the hotter models and can go as deep as $\log q = -12$ at $\Teff = 3000\,$K (Figure \ref{fig:pier2}). 

The mass fractions of C (black), O (red), He (blue) and H (magenta) are shown in
Figure \ref{fig:DA_panel1}a. As a result of the processes that occur in pre-white dwarf evolution and subsequent gravitational settling (Section \ref{sec:WD_anatomy}), the structure of the star is stratified in well-defined layers according to composition. The composition boundaries are not sharp however, which reflects the diffusive equilibrium between the species (Section \ref{sec3:diffusion}). In the subsequent panels of Figure \ref{fig:DA_panel1}, the regions where conductive or radiative transport dominate are shown in blue, the convection zone is in red and the central crystallized region in green. A quantity that is useful in interpreting subsequent figures is the average ion charge throughout the star (Figure \ref{fig:DA_panel1}b). In the fully ionized C-O core, the average charge $\Zbar=6.86$. It drops to $\Zbar =2$ in the He layer and then to $\Zbar =1$ in the outer H layer. Throughout most of the star ($\log q > -8$), $\Zbar$ is unaffected by the star's cooling.
In the outer reaches of the cooler models, H recombines and the gas becomes neutral ($\Zbar \rightarrow 0$). Convective energy transport (in red) is clearly associated with the partial ionization of H.

The pressure profiles of all the models in this sequence are nearly identical (Figure \ref{fig:DA_panel1}c). This is because the EOS is dominated by the degenerate electron pressure with $\Theta \ll 1$ and is insensitive to the temperature in the cooling star. Figure \ref{fig:DA_panel1}d shows a drop in density at the He/H boundary ($\log q=-4$) but there is no corresponding feature at the boundary between the C-O core and the He layer ($\log q=-2$). The requirement that the pressure be continuous in a star in hydrostatic equilibrium implies that $P_e \ (\approx P)$ and thus $n_e$ all be continuous in a highly degenerate plasma.  This feature in the mass density profile appears when $Z/A$ changes from 0.5 to 1 at the He/H boundary. Figure \ref{fig:DA_panel2}a shows the same temperature profiles as Figure \ref{fig:rhoT_DA} but in terms of a different radial coordinate.

The ion plasma coupling parameter (Equation \ref{eq:Gamma}) is quite large in the C-O core, with values of $\Gamma = 50$ in the hotter models to above 1000 in the crystallized core of the 3000\,K model (Figure \ref{fig:DA_panel2}b). The drop in $\Zbar$ from 6.86 in the C-O core to 2 in the He layer reduces $\Gamma$ by roughly a factor of 10, but the He plasma remains strongly coupled except in the hotter models. The strength of the coupling decreases again upon entering the H layer ($\log q < -4$). The coupling is weak but not negligible in the H layer of most models, with $\Gamma \gtrsim 0.05$. 
The base of the H layer of the cooler models, including in the convection zone, is strongly coupled.  The ion coupling drops rapidly once the plasma recombines to form H atoms.

The most well-known characteristic of white dwarf stars, that they are degenerate stars, is illustrated in Figure \ref{fig:DA_panel2}c where the electron degeneracy parameter $\Theta$ (Equation \ref{eq:theta}) is less than unity in every model through nearly the entire star and can be as small as $\sim 10^{-4}$ at the center. The degenerate region encompasses the entire C-O core ($\log q > -2$), the entire He layer ($-4 \le \log q \le -2$) and most of the mass of the H layer. A notable feature is that $\Theta$ is nearly constant in the convection zone (in red), owing to the identical $T(\rho)$ dependence of the constant specific entropy profile in the adiabatic convection zone and that of a line of constant $\Theta$. Degeneracy is lifted (and $\Theta$ rises rapidly) when the H plasma recombines near the surface.

From the point of view of plasma physics, it is interesting to look at white dwarf structures in terms of $\Gamma$ and $\Theta$ simultaneously (Figure \ref{fig:DA_panel2}d). The models fall along a broad diagonal band that spans from strongly degenerate and strongly coupled plasmas ($\Gamma,\ \Theta >1$, lower right quadrant) to the weakly degenerate, weakly coupled plasmas
($\Gamma < 1,\ \Theta <1$, upper left quadrant). Plateaus of nearly constant $\Theta$ along each curve reflect the rapid change in $\Zbar$ at the composition boundaries (Figure \ref{fig:DA_panel1}a). In these models typical of the most common white dwarfs, the plasma does not enter a regime of strong coupling and weak degeneracy (upper right quadrant) but it crosses the regime of warm dense matter ($\Gamma, \ \Theta \sim 1$) around the He/H transition region.

Carbon and oxygen ions display weak quantum mechanical behavior in the core with a quantum parameter $\Lambda \gtrsim 0.02$  that later rises steadily to moderate levels in the crystallized center of the coolest white dwarfs where $\Lambda \sim 0.6$ (Figure \ref{fig:DA_panel3}a). Jumps in $\Lambda$ occur at the composition transitions between layers where the ion mass decreases and the thermal wavelength increase. Remarkably, ionic quantum effects are non-negligible ($\Lambda > 0.01$) in all models cooler than $\Teff=10\,000\,$K everywhere except in the very outer regions. This includes all of the He layer, nearly all the mass in the superficial H layer, and most of the convective zone of these cool models. As a rule of thumb, a perturbative treatment of the ion quantum effects in the EOS, such as the Wigner--Kirkwood expansion to order $\hbar^2$, is usually adequate for $\Lambda \lesssim 0.1$.

Figure \ref{fig:DA_panel3}b shows the relativity parameter for the electrons (Equation~\ref{eq:xr}). Relativistic effects are important at the center of 0.6\,$M_\odot$ white dwarfs as $x_{\rm r} \sim 1$ but decrease steadily outward. Relativistic corrections are not negligible ($x_{\rm r} > 0.01$) throughout the entire C-O core, the full He layer and most of the H layer ($\log q > -7$). In more massive white dwarfs, the increased central density causes $x_{\rm r}$ to rise significantly (Section \ref{sec:hDQ_physics}).

Energy transport in stars can proceed through radiation, conduction or convection. Except near the photosphere, radiation transport can be described very accurately as a diffusive process of the radiation field in local thermodynamic equilibrium, i.e. a black body distribution at the local temperature $T(r)$. The flux of radiation through a layer of radius $r$ in the star can be written as \citep{hansenkawaler} 
\begin{equation}
    F_{\rm rad} (r) = - D_{\rm rad} \frac{dT}{dr} = - \frac{16 \sigma T^3}{3\kappa_{\rm\sss R} \rho} \frac{dT}{dr}~,
    \label{eq:Frad}
\end{equation}
where $D_{\rm rad}$ is the radiative diffusivity. This defines the frequency averaged Rosseland mean opacity $\kappa_{\rm\sss R}$ which is effectively an inverse diffusivity. Similarly, the conductive flux of is given by
\begin{equation}
  F_{\rm cond} = -D_{\rm cond} \frac{dT}{dr}~,
  \label{eq:Fcond}
\end{equation}
where $D_{\rm cond}$ is the thermal conductivity. By analogy with Equation \ref{eq:Frad}, we define a conductive opacity
\begin{equation}
    \kappa_{\rm cond}= \frac{16 \sigma T^3}{3 D_{\rm cond} \rho}~.
    \label{eq:Kcond}
\end{equation}
As radiative and conductive transport can occur in parallel, the total opacity $\kappa_{\rm tot}$ is
\begin{equation}
    \frac{1}{\kappa_{\rm tot}} = \frac{1}{\kappa_{\rm\sss R}} + \frac{1}{\kappa_{\rm cond}}~,
    \label{eq:Ktot}
\end{equation}
and the heat flux through a non-convective layer in the star becomes
\begin{equation}
    F(r) =  - \frac{16 \sigma T^3}{3\kappa_{\rm tot} \rho} \frac{dT}{dr}~.
    \label{eq:Ftot}
\end{equation}
The high degree of electron degeneracy in the white dwarf core favors conductive energy transport which eventually is superseded by radiative transport closer to the surface. The conductive opacity $\kappa_{\rm cond}$ is very small (very high thermal conductivity) in the C-O core and in the He layer for the coolest models (Figure \ref{fig:DA_panel3}c) but it increases rapidly towards the surface. The Rosseland mean opacity ($\kappa_{\rm\sss R}$) is many orders of magnitude larger than the conductive opacity in the C-O core and is roughly constant outward until the H plasma starts to recombine and $\kappa_{\rm\sss R}$ has a broad peak followed by a rapid decline near the surface where H is (nearly) all in atomic or molecular form (see Figure \ref{fig:DA_panel1}b). The net result is that the total opacity  $\kappa_{\rm tot}$ (Figure \ref{fig:DA_panel3}c) is very small in the C-O core and He envelope, has a broad peak in the H layer for models with $\Teff \lesssim 12\,000\,$K, and drops rapidly near the surface.
A striking feature of Figure \ref{fig:DA_panel3}c is that the region of convective energy transport (in red) coincides with the broad peak in $\kappa_{\rm\sss tot}$. This is a consequence of the Schwarzschild criterion for convective instability in stars \citep{kippenhahnweigert,hansenkawaler}
\begin{equation}
    \delad < \nabla~,
    \label{eq:schwar}
\end{equation}
where 
\begin{equation}
   \delad = \frac{\partial \ln T}{\partial \ln P}\Bigg|_{\sss S}
   \label{eq:delad}
\end{equation}
is the adiabatic temperature gradient,\footnote{The derivative is taken at  constant specific entropy $S$.} a thermodynamic quantity obtained from the EOS, and
\begin{equation}
    \nabla =  \frac{\partial \ln T}{\partial \ln P}
    \label{eq:del}
\end{equation}
is the dimensionless temperature gradient in the star. With some simple approximations that are justified in white dwarfs, the convective instability criterion can be rewritten as
\begin{equation}
    \nabla_{\rm ad} < \frac{3}{64\pi \sigma G} \frac{L_\star}{M_\star} \frac{P\kappa_{\rm tot}}{T^4}\  \propto \ \frac{P}{T^4} \kappa_{\rm tot}~,
    \label{eq:schwar2}
\end{equation}
where $G$ is the constant of gravitation and $L_\star$ the luminosity of the star. Thus, aside from the factor $P/T^4$, convective instability is favored when $\delad$ is small or $\kappa_{\rm tot}$ is large. 

Figure \ref{fig:DA_panel3}d shows that the convective region generally occurs when $\delad$ becomes small. Generally, in $(\rho,T)$ regimes where partial ionization occurs, $\delad$ decreases (see Figures 10 and 17 of Ref.~\cite{scvh1995} for example), $\kappa_{\rm cond}$ is very large because of the reduced electron degeneracy (hence, $\kappa_{\rm tot} \sim \kappa_{\rm\sss R}$), and $\kappa_{\rm\sss R}$ increases due to the contribution of bound-free absorption to the opacity.

\subsection{Models with an outer helium layer}
\label{sec:DB_physics}

\begin{figure}
\begin{center}
\includegraphics[width=\columnwidth]{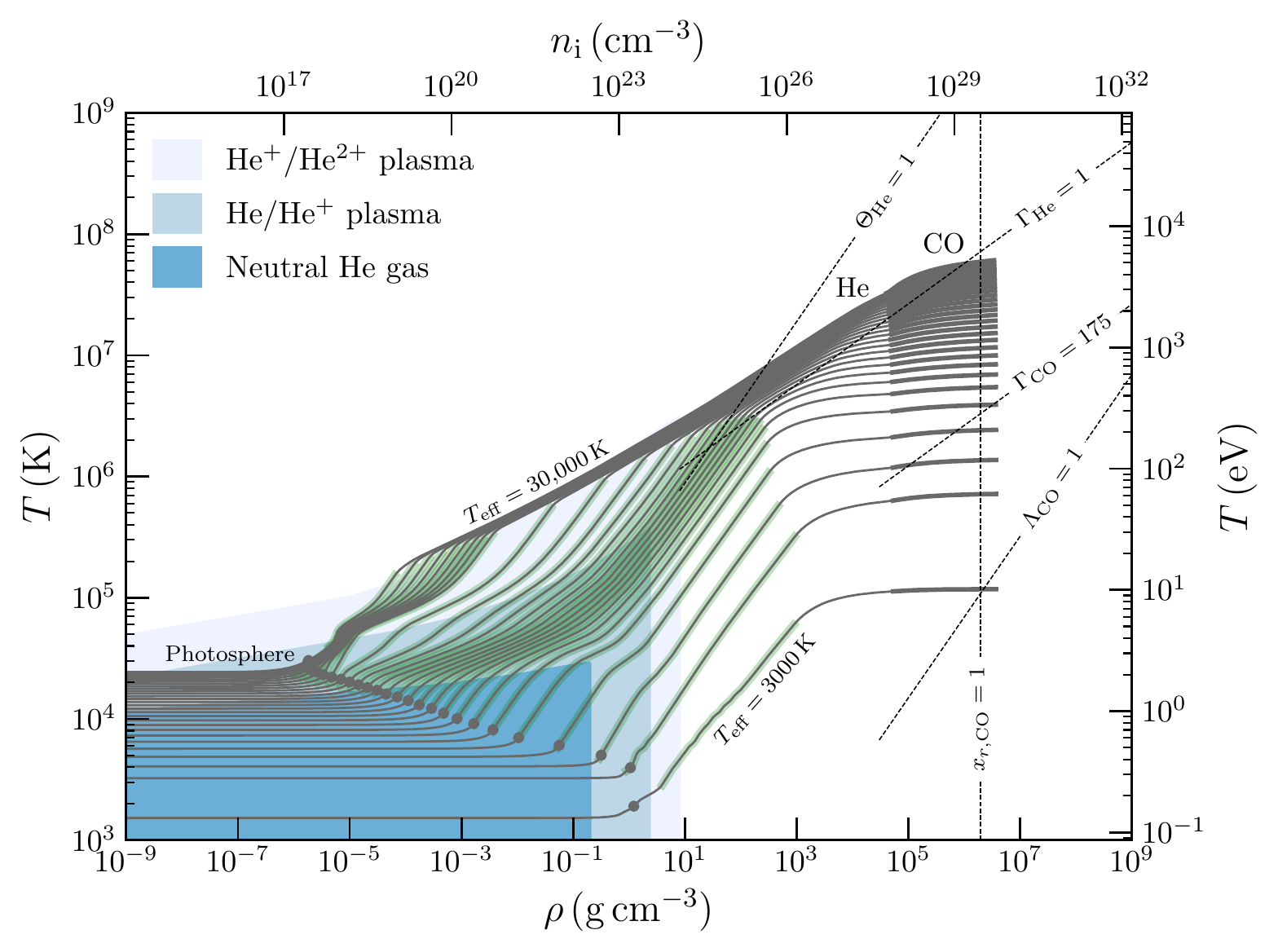}
\caption{Same as Figure~\ref{fig:rhoT_DA}, but for He-rich white dwarfs. The blue shadings now indicate the ionization of He, see the legend. An envelope with $\log (M_{\rm He}/M_{\star}) = -2$ 
is assumed, and the white dwarf mass is $0.6\,M_{\odot}$. The surface of the star is toward the left, the center is on the right.
\label{fig:rhoT_DB}}
\end{center}
\end{figure}

Internal structures of white dwarf models with only an outer He layer are shown in Figure \ref{fig:rhoT_DB}. These models are representative of about 20\% of all white dwarfs. The models shown have otherwise the same parameters as those in Figure \ref{fig:rhoT_DA}.  The physical properties of these He-atmosphere white dwarfs are very similar to those with a H atmosphere except near the surface where He has replaced H. A thin convection zone appears near the photosphere at $\Teff \approx 50\,000\,$K and becomes progressively deeper as the star cools. Pressure ionization of He (lighter blue shadings) occurs at densities of a few g\,cm$^{-3}$ and $T \lesssim 20\,$eV, conditions found in the convection zone of the cooler models shown and around the photosphere of models of 5000\,K or less. Generally, electron degeneracy and ion Coulomb coupling become significant at higher $T_{\rm eff}$ (earlier in the evolution) than in H-rich models. The photosphere of the coolest models is found at densities approaching 1 g\,cm$^{-3}$ where we expect strong non-ideal effects to leave an imprint on the spectra of those stars (Sections \ref{sec:atmosphere_models}, \ref{sec:He_atm} and \ref{sec:line_prof_cool}).

\begin{figure}
\begin{center}
\includegraphics[width=\columnwidth]{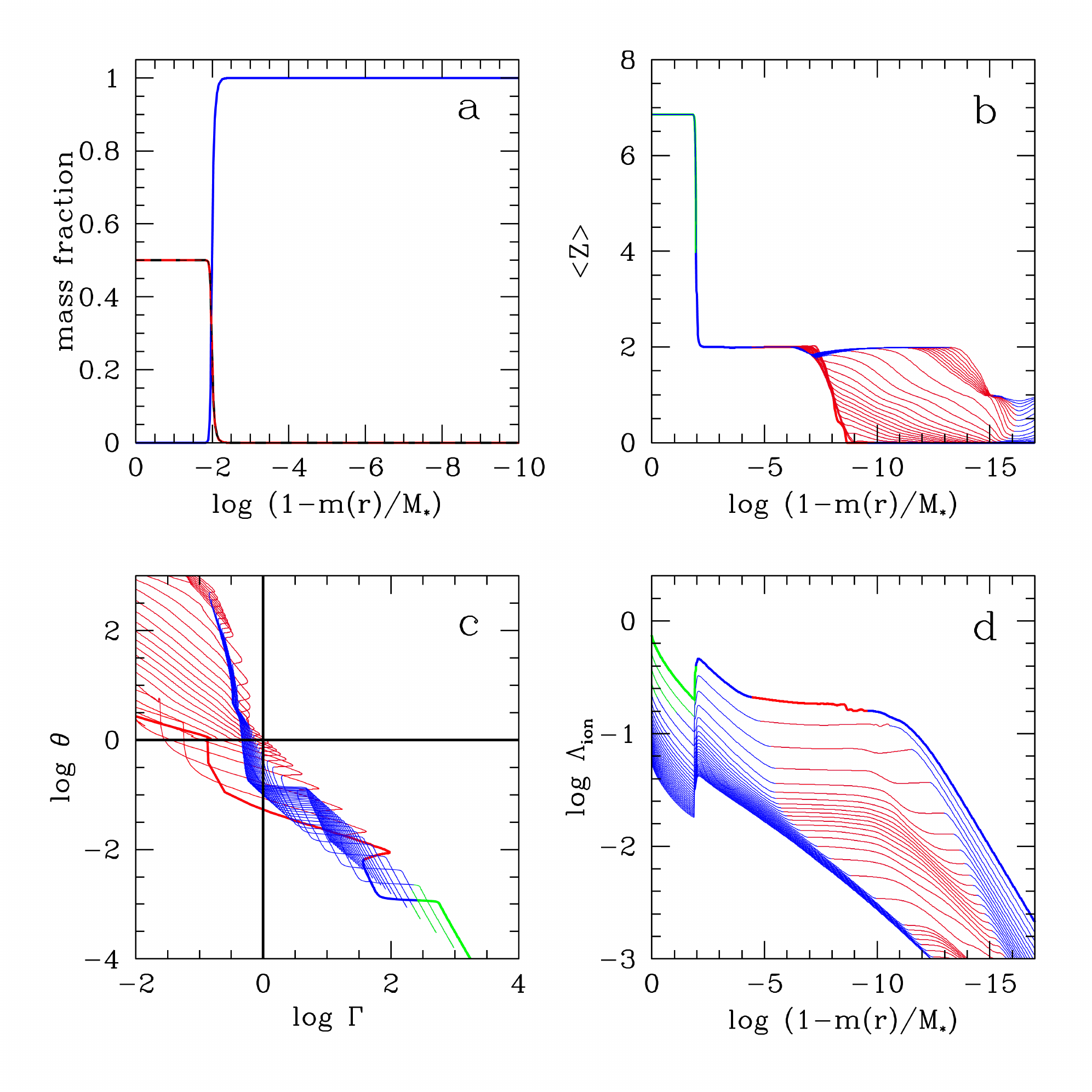}
\caption{Properties of a cooling  sequence of models of a He-rich white dwarf of $M_\star = 0.6\,M_\odot$ consisting of a C-O core, surrounded by a $10^{-2}\,M_\star$ He layer. Models range from $\Teff = 30\,000\,$K to 3000\,K (heavier line) in steps of 1000\,K. The radial coordinate is a logarithmic mass coordinate (Equation \ref{eq:log_q}) with the center of the star located on the left at $\log q=0$ and the surface towards the right (note the ``backward'' X-axis). In panels b--d, conductive and radiative regions are in blue, convectively unstable regions in red and the crystallized core in green.  Compare with the same sequence of models for H-rich white dwarfs (Figures \ref{fig:DA_panel1} -- \ref{fig:DA_panel3}).
a) Composition profile showing the mass fractions of C (black), O (red), and He (blue), b) Average ion charge $\Zbar$, c) Ion coupling parameter $\Gamma$ (Equation \ref{eq:Gamma}) and electron degeneracy parameter $\Theta$ (Equation \ref{eq:theta}), d) Ion quantum parameter $\Lambda$ (Equation \ref{eq:Lambda}).
\label{fig:DB_panel1}}
\end{center}
\end{figure}

The simple composition profile of these models is shown in Figure \ref{fig:DB_panel1}a. The behavior of $\Zbar$ within the C-O core and the base of the He layer is the same as in the DA models (Figure \ref{fig:DB_panel1}b). The He layer remains fully ionized from $\log q=-2$ to $-13$ in the hotter models, where it then drops to a plateau where He remains singly ionized ($\Zbar \approx 1$). In the cooler models, recombination of He starts steadily deeper in the model and decreases smoothly to neutral He. 

The pressure, density and temperature profiles are nearly identical as those of H-atmosphere models shown in Figures \ref{fig:DA_panel1}c,d and \ref{fig:DA_panel2}a. However, He has a lower opacity than H at low temperatures, resulting in a different atmospheric structure (Figure \ref{fig:atmprof}). The latter acts as the surface boundary condition for models of the interior and the relative transparency of the He atmosphere leads to faster cooling. The main effect is a slightly lower central temperature (less than a factor of 2) after convective coupling has occurred ($\Teff \lesssim 12\,000\,$K) and a larger crystallized region at a given $\Teff$. 

The higher charge of fully ionized He leads to stronger ion plasma coupling and a higher degree of degeneracy in the upper reaches of the models than in the H layer of the H-rich models. These two plasma parameters are shown in Figure \ref{fig:DB_panel1}c. In addition to the shift to higher $\Gamma$ in the upper reaches of the He layer ($\Theta \gtrsim 1$), a striking difference is that the convection zone (in red) of models with $\Teff \lesssim 12\,000$\,K extends well into the strongly coupled, strongly degenerate regime ($\Gamma > 1$, $\Theta < 1$, lower right quadrant). In all the models shown, the He layer crosses the warm dense matter regime.
Other physical parameters, such as the relativity parameter $x_r$, the ion quantum parameter $\Lambda$, the total opacity $\kappa_{\rm tot}$ and the adiabatic temperature gradient $\delad$ are very similar to those discussed for the H-atmosphere white dwarf models, both in their values and trends, with small differences due to the change in mass and charge when He is substituted for H. The ion quantum parameter is non-negligible ($> 0.01$) throughout most of the mass of the He layer (only the outer $10^{-10}$ of the mass is in the classical limit) and ionic quantum effects are apparent in the He convection zone below $\Teff =17\,000$\,K (Figure \ref{fig:DB_panel1}d). The convection zone is associated with the partial ionization of He, which corresponds to a maximum of the total opacity and a minimum of the adiabatic temperature gradient, just as we observe with the partial ionization of H in models of H-rich white dwarfs.

\begin{figure}
\begin{center}
\includegraphics[width=\columnwidth]{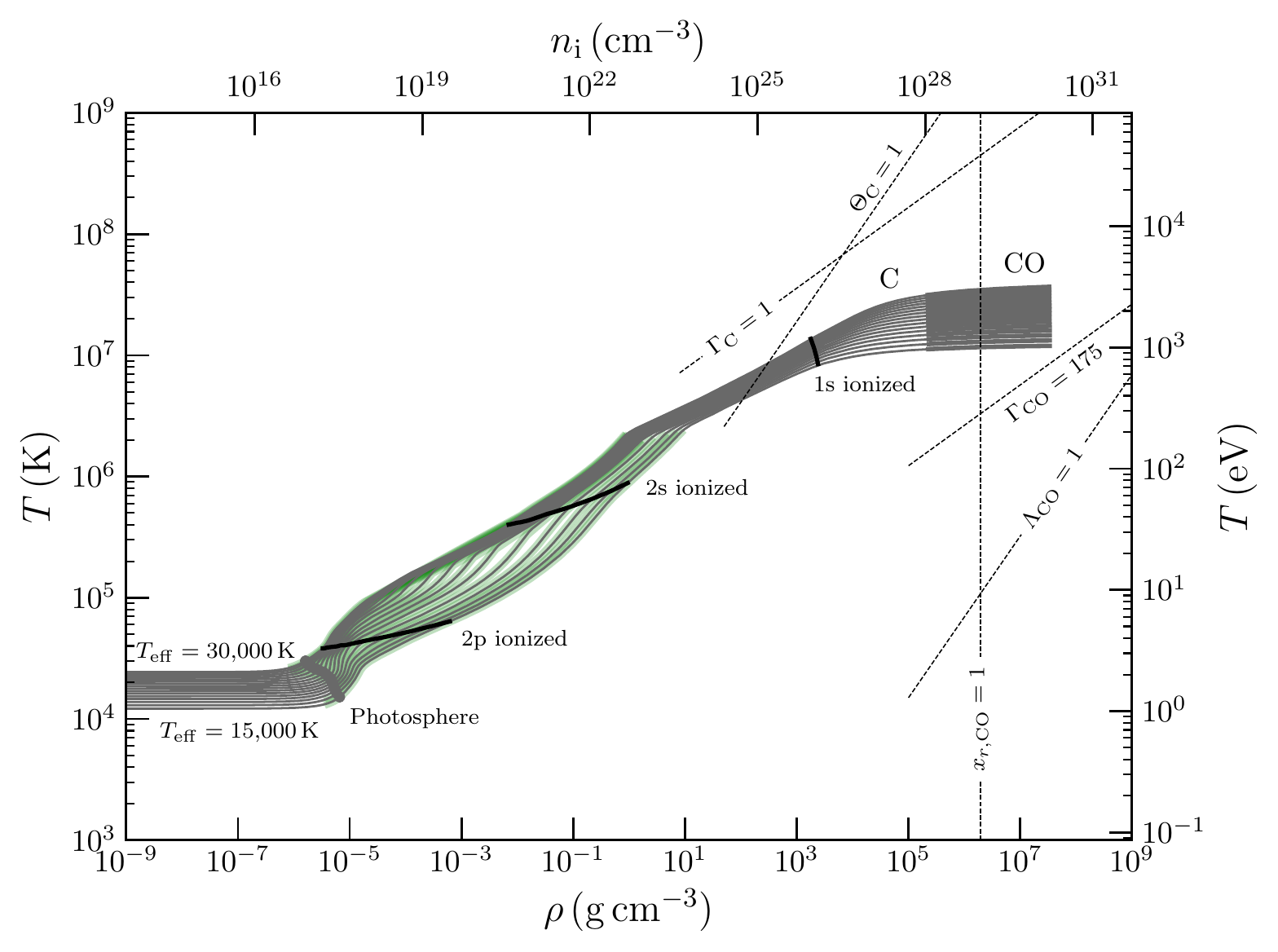}
\caption{Same as Figure~\ref{fig:rhoT_DA}, but for hot DQ white dwarfs ranging from 15\,000 to 30\,000\,K. The thick black lines indicate where the different electronic shells of C are ionized. The envelope is assumed to be made of pure C, and the white dwarf mass is $1\,M_{\odot}$. The surface of the star is toward the left, the center is on the right.
\label{fig:rhoT_HDQ}}
\end{center}
\end{figure}

\subsection{Models of hot DQ white dwarfs}
\label{sec:hDQ_physics}
\begin{figure}
\begin{center}
\includegraphics[width=\columnwidth]{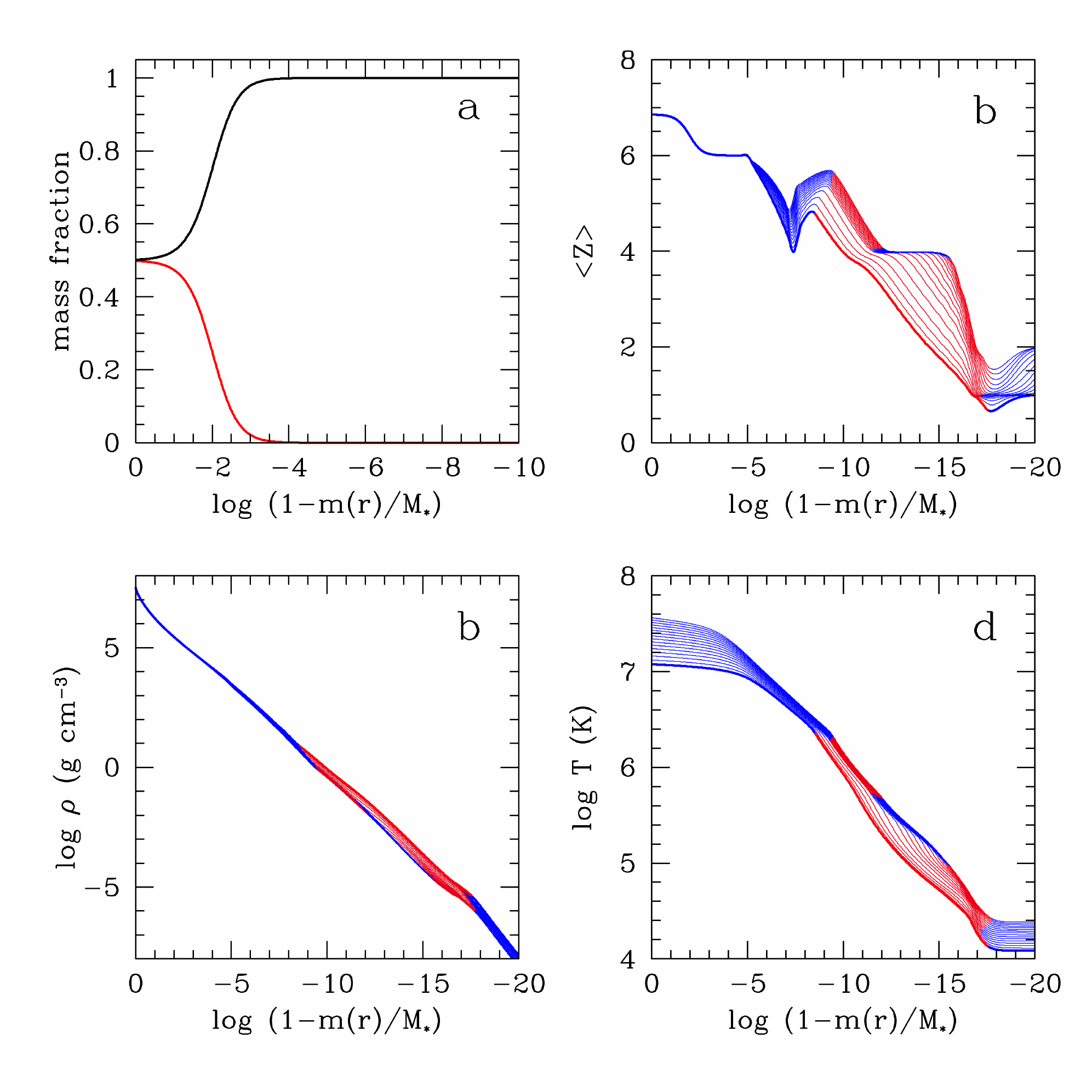}
\caption{Properties of a cooling  sequence of models of a hot DQ white dwarf of $M_\star = 1\,M_\odot$ consisting entirely of a C-O plasma. In the absence of a He layer, there is no clearly defined core. Models range from $\Teff = 30\,000\,$K to 15\,000\,K (heavier line) in steps of 1000\,K. The radial coordinate is a logarithmic mass coordinate (Equation \ref{eq:log_q}) with the center of the star located on the left at $\log q=0$ and the surface towards the right (note the ``backward'' X-axis). In panels b--d, conductive and radiative regions are in blue and convectively unstable regions in red.  a) Composition profile showing the mass fractions of C (black) and O (red), b) Average ion charge, c) Density profile, and d) Temperature profile. 
\label{fig:hDQ_panel1}}
\end{center}
\end{figure}

\begin{figure}
\begin{center}
\includegraphics[width=\columnwidth]{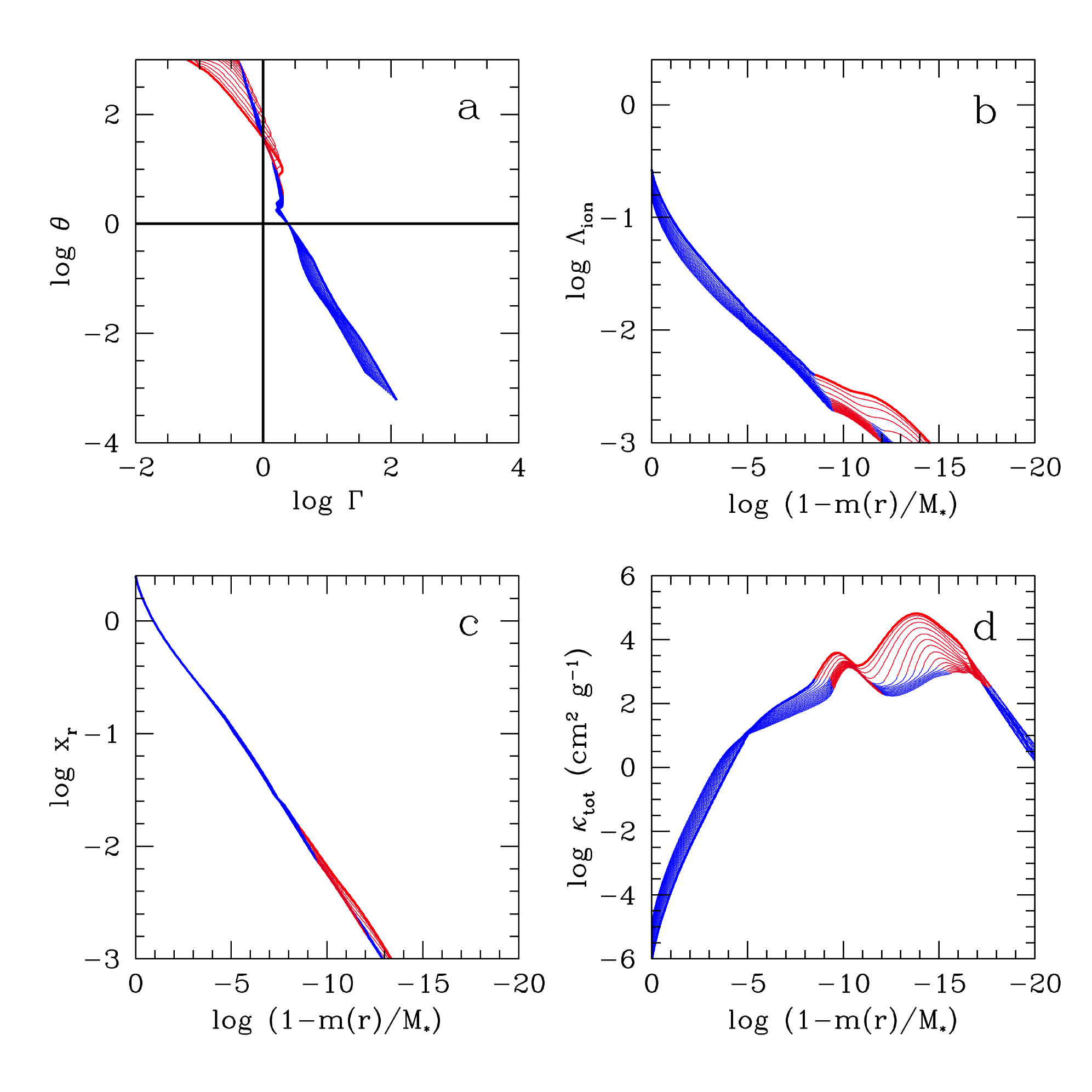}
\caption{Same as Figure \ref{fig:hDQ_panel1} but showing profiles of the a) Ion coupling ($\Gamma$) and the electron degeneracy ($\Theta$) parameters, b) Ion quantum parameter $\Lambda$, c) Electron relativity parameter $x_r$, and d) Total opacity $\kappa_{\rm tot}$.
\label{fig:hDQ_panel2}}
\end{center}
\end{figure}

Figure \ref{fig:rhoT_HDQ} shows the structure of models of hot DQ stars, composed of a C-O core overlain with a C envelope (no H or He). Hot DQ stars are believed to be the result of the merger of two white dwarfs in a binary system and are more massive than typical H-rich and He-rich white dwarfs. The models shown are for a mass of 1\,$M_\odot$. The details of the origin of hot DQs and their later evolution are rather uncertain and only models generously bracketing their observed temperature range ($\Teff$ from 18\,000 to 24\,000\,K) are shown. Little is known about the interior composition of hot DQ white dwarfs. In Figures \ref{fig:hDQ_panel1} and \ref{fig:hDQ_panel2}, we adopt a slightly different composition where, by construction, the models have equal mass fractions of C and O at the center and a decreasing amount of O going outward (Figure \ref{fig:hDQ_panel1}a). Given the low abundance of O seen in the spectra of hot DQ stars, its mass fraction is forced to vanish near the surface.  Our choice of C-O abundance profile is somewhat arbitrary but serves to illustrate significant differences with the more common and generally less massive white dwarfs with He and H layers. 
In particular, the higher mass leads to central densities that exceed $3\times 10^7\,$g\,cm$^{-3}$ where electrons are relativistic. Crystallization has not started in these relatively warm models and ions are nearly classical. 
All hot DQ models shown have an outer convection zone associated with the $L$ shell ionization and the onset of $K$ shell ionization of C, where the ions are strongly coupled but the electrons are non-degenerate. 

The behavior of the average ion charge is more complex than in the H-rich and He-rich  models (Figure \ref{fig:hDQ_panel1}b). It starts at $\Zbar = 6.86$ at the center and decreases to $\Zbar = 6$ where the O abundance becomes negligible. The ion charge drops to $\Zbar = 4$ as the $K$ shell electrons of C recombine. The sharp dip centered at $\log q=-7$ is likely an artifact of the specific C EOS table used in these models and hints that a better EOS model for partial ionization of C in the WDM regime would be valuable. Finally, most of the $L$ shell electrons recombine in the outer region of the models with $\Zbar \approx 1$ -- 2 at the surface, depending on the value of $\Teff$. The hotter models have two distinct convection zones, corresponding to the partial ionization of the $K$ and $L$ shell electrons. They are separated by a plateau at $\Zbar=4$. The two convection zones merge when the partial ionization of the two electronic shells are no longer distinct in the C EOS, in the $(T,\rho$) domain where temperature and pressure ionization become comparable. The details of the convection zone in hot DQs are closely tied to the modeling of the pressure ionization of C in the warm dense matter regime, which can vary considerably between EOS tables and has not been benchmarked by experiments.

The density profiles in the hot DQ models are nearly independent of $\Teff$ and are about one order of magnitude above those of H-rich and He-rich white dwarfs (Figure \ref{fig:hDQ_panel1}c), a direct consequence of the higher stellar mass. The temperature profile is similar to that of H-rich and He-rich models of the same $\Teff$ range. The merging of the two convection zones at $\Teff \approx 18\,000$\,K is quite apparent in Figure \ref{fig:hDQ_panel1}d. 

The higher density of the hot DQ models combined with a higher overall ion charge (except in the inner core inside $\log q = -2$) make for an overall shift to larger ion coupling (Figure \ref{fig:hDQ_panel2}a). At the center, $\Gamma$ is about 3 times larger than in He-rich models for the same $\Teff$ and $\Gamma \ge 1$ through almost the entire star ($\log q > -11$). While the bottom of the convection zone is always moderately coupled, it is not degenerate, unlike in the models of H-rich and He-rich white dwarfs. The larger density increases the importance of ionic quantum effects by about a factor of four (for a given $\Teff$) and quantum effects are non-negligible ($\Lambda > 0.01$) everywhere except for the outer $10^{-5}$ mass fraction of the star (Figure \ref{fig:hDQ_panel2}b). Similarly, relativistic effects on the electrons are now dominant at the center with $x_r \approx 2.4$ (Figure \ref{fig:hDQ_panel2}c) and remain significant up to the onset of convection. Finally, the profiles of the total opacity show the same trends as that of the H-rich and He-rich models, except that the two electronic shells of the C atom each produce a local maximum in $\kappa_{\rm tot}$ (Figure \ref{fig:hDQ_panel2}d). The strong correlation between opacity maxima and convective instability is also seen in hot DQ models.

The outer convection zone in white dwarfs plays an important role in the late phases of cooling after convective coupling, in driving the pulsation instability, and in mixing (Sections \ref{sec3:transport}, \ref{sec3:pulsations}). The astrophysics of white dwarfs thus depends on a proper modeling of their convective
regions, which in turn depends, among other factors, on the EOS and opacity of the partially ionized plasma.\footnote{While the location of the convection zone boundaries depend on both $\nabla_{\rm ad}$ and $\kappa_{\rm\sss R}$ through the Schwarzschild criterion (Equations \ref{eq:schwar}--\ref{eq:schwar2}), the structure and properties of the convection zone are largely insensitive to the opacity.} Unless the density is very low,  partial ionization is challenging to model. In H-rich, He-rich and hot DQ white dwarfs, the convective region crosses the regime of warm dense matter (Figure \ref{fig:rhoT_DA}) whose quantitative understanding is actively pursued experimentally and theoretically.

\subsection{Magnetic fields and white dwarf plasmas}
\label{sec:mag_parameters}
 Since about 10\% of magnetic white dwarfs have very strong surface fields  of $\sim 10^2$ -- $10^3$\,MG \citep{ferrario2015}, it is interesting to see what effect, if any, such  a field has on the interior plasma which is not directly accessible to observation. 

A key quantity in magnetized plasmas is the gyrofrequency of particles of mass $m$ and charge $Z$
\begin{equation}
    \omega_c = \frac{|Z|eB}{mc}~.
    \label{eq:gyro_freq}
\end{equation}
The corresponding energy $\hbar\omega_c$ is the spacing between quantum mechanical eigenstates of a free particle in a magnetic field (the Landau levels). Due to their much lower mass, electrons have a higher $\omega_c$ and are more affected by magnetic fields than ions.

The effect of magnetization on the EOS of dense plasmas has been studied extensively in the context of neutron stars whose typical surface fields are very large ($\sim 10^{12}$\,G) compared to those found in white dwarfs (e.g. Ref.~\cite{potekhin2013} and references therein). The kinetic degrees of freedom of particles in a magnetized plasma can be treated classically if $\zeta = \hbar \omega_c/\kb T \ll 1$. Assuming $B \lesssim 10^9\,$G, this is always true for ions in white dwarfs throughout the entire star with $\zeta \lesssim 10^{-6}$ in the C/O core and reaching $\sim 10^{-3}$ in the very outer part of the H layer. The criterion $\zeta \gtrsim 1$ is met for electrons only in the very outer regions of the star where $T \lesssim 13\,000$\,K, which falls within the atmosphere where the density is very low, the plasma coupling is weak and electrons are nearly classical. The corrections to the plasma EOS in the $\zeta \ll 1$ regime are small but not entirely negligible \cite{potekhin1999, potekhin2013}. Compared to the scale of atomic binding energies, a 100\,MG field corresponds to an energy of $\hbar \omega_c \sim 0.043\,$Ha. Such a field has a small effect on the more tightly bound energy levels of H and He, but excited states are moderately affected \citep{rosner1984, thirumalai2009, rueda2020}. The EOS of a partially ionized, magnetized H plasma encountered in the atmosphere has also been studied \citep{potekhin1999, rueda2020}. Thus, single white dwarfs with large fields fall in a weakly magnetic regime of the EOS where corrections are mostly at a perturbative level in the very outer region of the star and are negligible throughout the interior. To our knowledge, however, detailed calculations of the EOS of H and He in the relatively weak fields encountered in white dwarfs have not been applied to models of magnetic white dwarfs. 

On the other hand, highly magnetized white dwarfs with interior fields up to $10^{14}$\,G have been proposed as possible progenitors of rare, over-luminous supernovae of Type Ia \citep{das2012} and as an alternative model to magnetars to explain soft gamma-ray repeaters and anomalous X-ray pulsars \citep{malheiro2012}. A simple application of the scalar virial theorem indicates that the global field can be as large as $10^{12}$\,G \citep{lai1991} in white dwarfs. Even larger central fields are thus physically plausible. Such large fields affect the electron contribution to the EOS, which become quantized in terms of the Landau magnetic levels \citep{lai1991, potekhin2013, das2012, peterson2021}. Interestingly, the magnetic field stabilizes the solid phase  with respect to the liquid and the crystallization temperature increases, with the  plasma coupling parameter of freezing decreasing from $\Gamma=168$ ($B=0$) to 155 ($B=10^{13}\,$G) for a carbon plasma \citep{potekhin2013}.

With a field of $10^{14}$\,G, the magnetic pressure contributes significantly to the total pressure and affects the structure of the star, notably raising the maximum mass of a stable white dwarfs well above the canonical Chandrasekhar mass of 1.40\,$M_\odot$ \citep{rueda2013, carvalho2018} for a non-magnetic, non-rotating carbon white dwarf \citep{das2012}. While the Chandrasekhar mass is usually defined in terms of the gravitational stability of the star given its EOS (e.g. \citep{hamada1961, rueda2013}), many factors affect the models near the limit of stability, such as general relativity (Section \ref{sec:GR}), the inclusion of the electromagnetic energy in the Einstein--Maxwell stress-tensor, the core composition, the equation of state, the geometry of the magnetic field, the oblateness of the star and the surface boundary condition. At the very high central densities found in stars near the maximum mass, instability can arise from two nuclear processes: electron capture \citep{hamada1961} and pycnonuclear reactions \citep{yakovlev2006}. In models of highly magnetic white dwarfs, one of these two nuclear processes usually limits the maximum mass. Published estimates of the maximum mass of magnetic white dwarfs vary considerably depending on the model assumptions, ranging from 1.75 to 2.58\,$M_\odot$, with the more detailed and physically consistent models giving $\approx 2.0\,M_\odot$ \citep{das2012, das2013, bera2014, coelho2014, franzon2015, das2015, chatterjee2017, otoniel2017, otoniel2019, gupta2020}, with a corresponding maximum field of $\approx 4 \times 10^{14}$\,G \citep{otoniel2019}. So far, there is no direct evidence for white dwarfs with such high internal fields and there is no known white dwarf with a mass above 1.4\,$M_\odot$ \citep{kilic2021}.

Transport properties (diffusion coefficients, conductivity, opacity) are typically more sensitive to the details of the microphysics than the EOS. An analysis of the physical regimes of transport in magnetized plasmas \citep{baalrud2017} implies that it is unaffected by the magnetic field (unmagnetized regime) throughout nearly the entire star. In H-rich white dwarfs, H$^+$ ions become weakly magnetized only in the outer $\sim 10^{-11}$ of the mass of the star for fields of 100\,MG, and only in stars with $\Teff \gtrsim 20\,000$\,K. In the interior, the effects of the field are negligible mainly because of the high plasma density that results in $\omega_p \gg \omega_c$, where $\omega_p = \sqrt{4\pi Z^2 e^2 n /m}$ is the ion plasma frequency. Thus, ion inter-diffusion remains unaffected (and isotropic) even under such strong fields. The field has a stronger influence on electrons, which are weakly magnetized over a greater depth in the star, corresponding to the outer $\sim 10^{-9}\,M_\star$. Besides the obvious effect on absorption lines at the photosphere, the bound-free and free-free opacities are modified at these larger depths \citep{wickramasinghe1995, merani1995}. We can expect that the screening of the ions by the free electrons will become anisotropic and dependent on the field orientation. Such a distortion of the electron density surrounding an ion would have an indirect effect on the ion inter-diffusion through the effective ion-ion potential (Section \ref{sec:diffusion_env}). While the electron thermal conductivity would also be modified slightly by the presence of the field, it no longer plays a role in the outer layers of the star.

\subsection{Atmosphere models}
\label{sec:atmosphere_models}
To conclude this section, we examine in more detail the physical regimes encountered in the atmospheres of white dwarfs. Despite representing a very small fraction of the mass and radius of the white dwarf\footnote{The pressure scale height in white dwarf atmospheres is typically of the order of ten to a few hundred meters.}, this thin layer is of great importance since it is directly probed by astronomical observations via photometry and spectroscopy. Virtually all of the properties of white dwarfs are inferred from the analysis of this region. The effects of moderate non-ideal physics on the EOS, ionization and dissociation equilibria, atomic level populations and radiative absorption cross-sections are manifest in the spectra of white dwarfs, which offer a strong test of microphysics models.

\begin{figure}
\begin{center}
\includegraphics[width=\columnwidth]{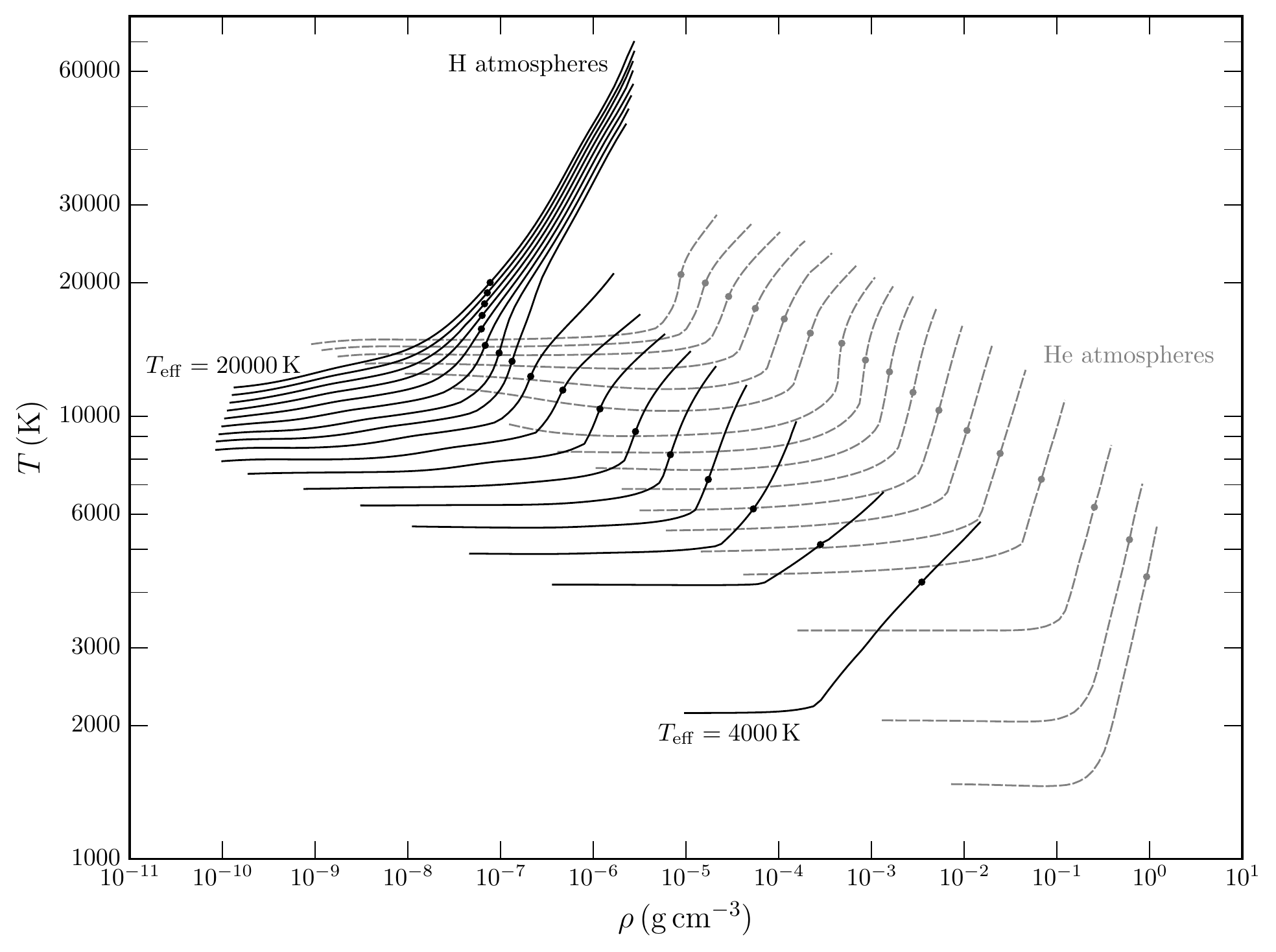}
\caption{Density--temperature profiles of white dwarf atmospheres. The left set of profiles (solid lines) corresponds to H-atmospheres from $T_{\rm eff}=20\,000\,{\rm K}$ to 4000\,K in steps of 1000\,K, and the right set (dashed lines) are He-atmospheres with the same $\Teff$. A surface gravity of $10^8\,{\rm cm\,s}^{-2}$ is assumed, which corresponds to a typical $\simeq 0.6\,M_{\odot}$ white dwarf. 
The level of the photosphere is indicated by a solid dot. The interior of the star is toward the upper right. Note that the profiles shown here differ somewhat from the uppermost layers of the profiles displayed in Figures~\ref{fig:rhoT_DA} and~\ref{fig:rhoT_DB}. These profiles were generated using a detailed atmosphere code \cite{blouin2018a,blouin2018b} that includes more accurate constitutive physics for the specific regime relevant to atmospheres than what is implemented in the more general structure code used for Figures~\ref{fig:rhoT_DA} and~\ref{fig:rhoT_DB}. \label{fig:atmprof}}
\end{center}
\end{figure}

Figure~\ref{fig:atmprof} shows density--temperature profiles for H- and He-atmosphere white dwarfs of different surface temperatures. For each profile, a circle indicates the location of the photosphere. The photosphere is located at a Rosseland optical depth of $\tau_{\rm\sss R}=2/3$, meaning that $e^{-2/3}\simeq 50\%$ of the photons emitted at the photosphere are absorbed or scattered before reaching the top of the atmosphere and the other half escapes directly to space. In other words, the photosphere is the region from which a typical photon detected on Earth originates, and it is therefore representative of the conditions that we are most interested in modeling correctly. Regions above the photosphere also contribute to the emergent flux, particularly at wavelengths where the (wavelength-dependent) opacity is high compared to the Rosseland mean opacity (e.g., in spectral lines). The wavelength dependence of the opacity implies that the spectrum is formed at different depths in the atmosphere. This property can be used to probe the vertical stratification of the atmosphere \cite{klein2020}.

\begin{figure}
\begin{center}
\includegraphics[width=0.7\columnwidth]{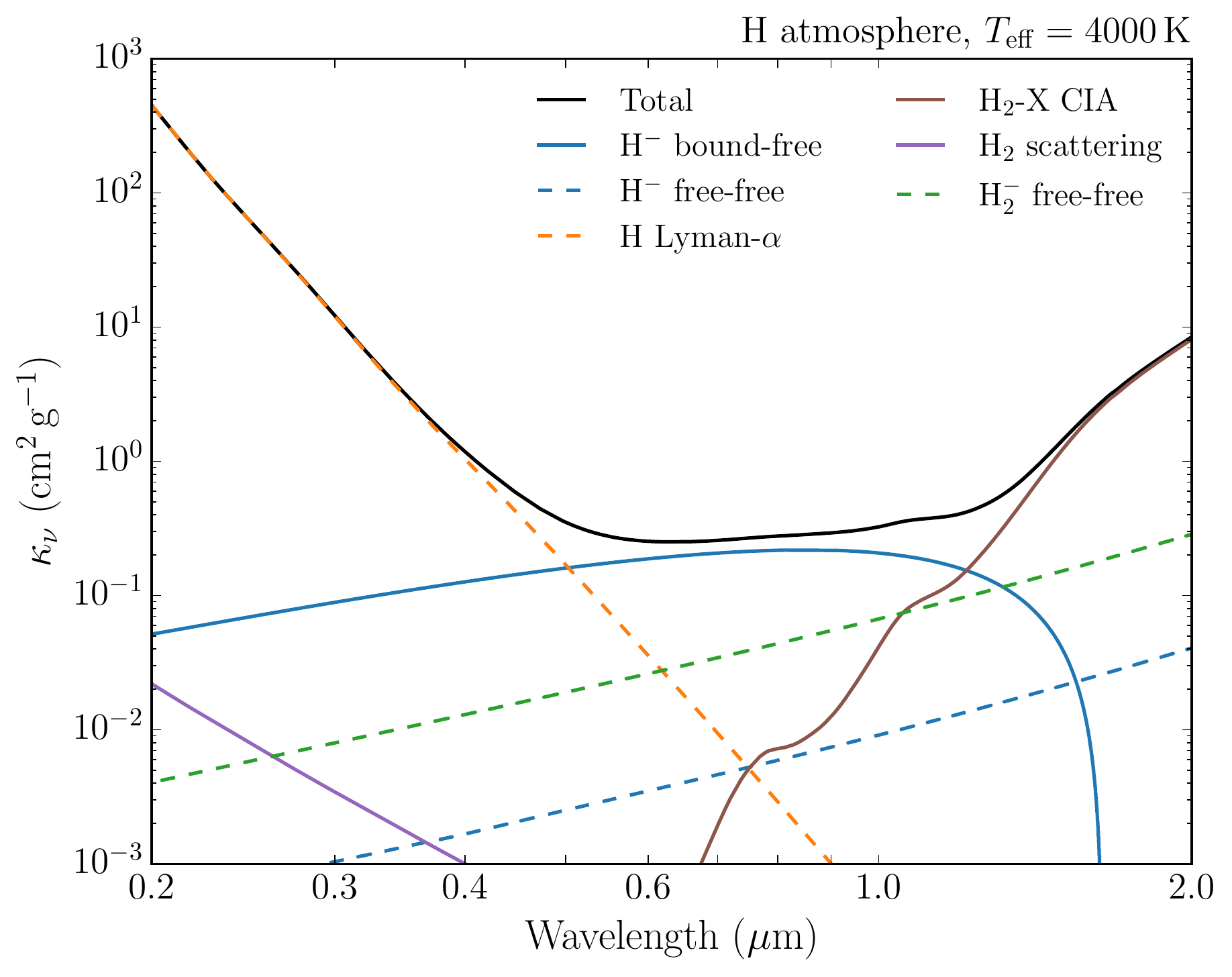}
\includegraphics[width=0.7\columnwidth]{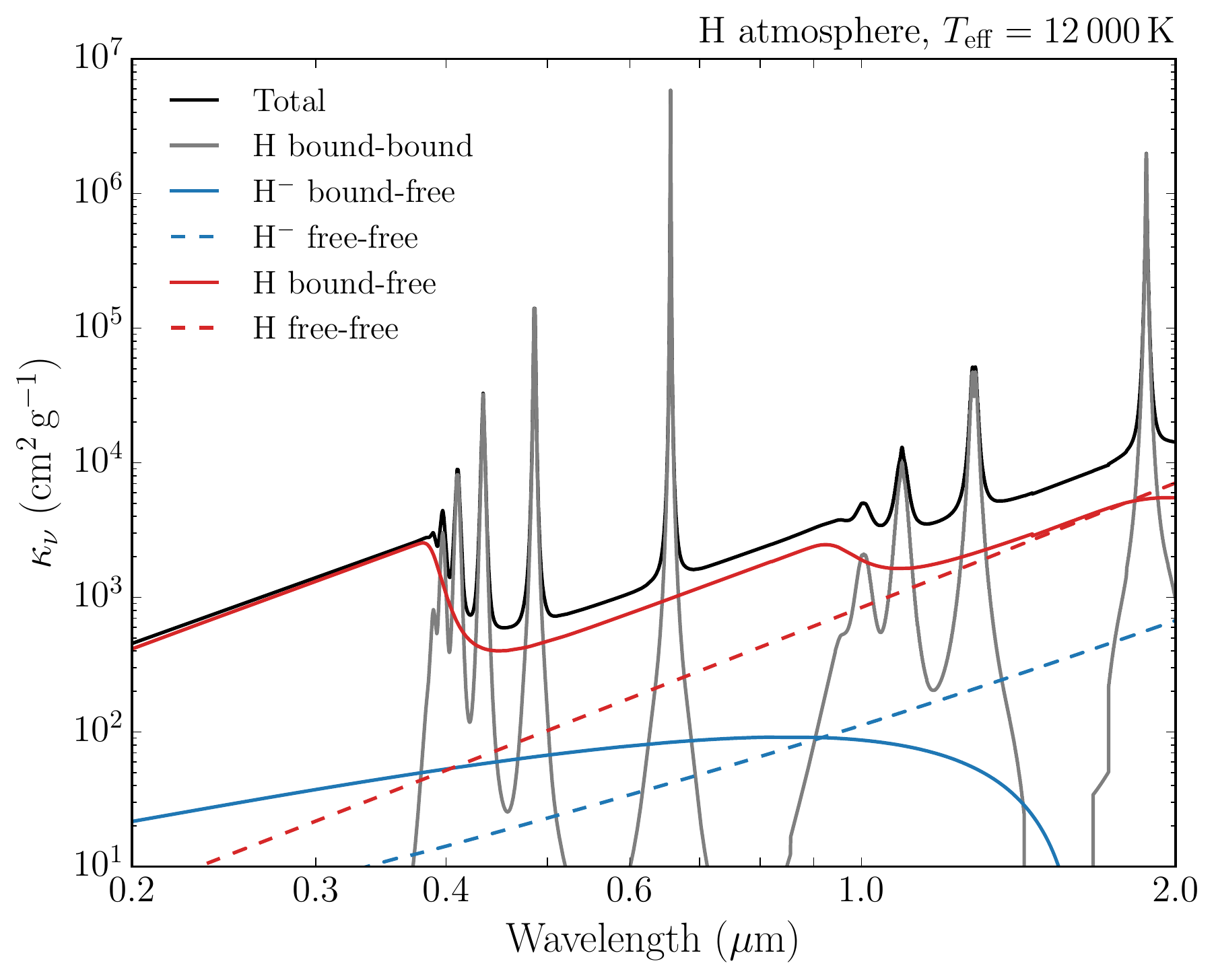}
\caption{Dominant contributions to the radiative opacity of H-atmosphere white dwarfs. The top panel shows the wavelength-dependent opacity of a cool ($T_{\rm eff}=4000\,$K) white dwarf, which would be classified as DC due to the absence of spectral lines, and the bottom panel is for a warmer object ($T_{\rm eff}=12\,000\,$K), which would be classified as DA due to the strong Balmer lines. In both cases, a surface gravity of $10^8\,{\rm cm\,s}^{-2}$ is assumed and the opacities are evaluated for the conditions at the photosphere. \label{fig:kappa_atm_H}}
\end{center}
\end{figure}

\begin{figure}
\begin{center}
\includegraphics[width=0.7\columnwidth]{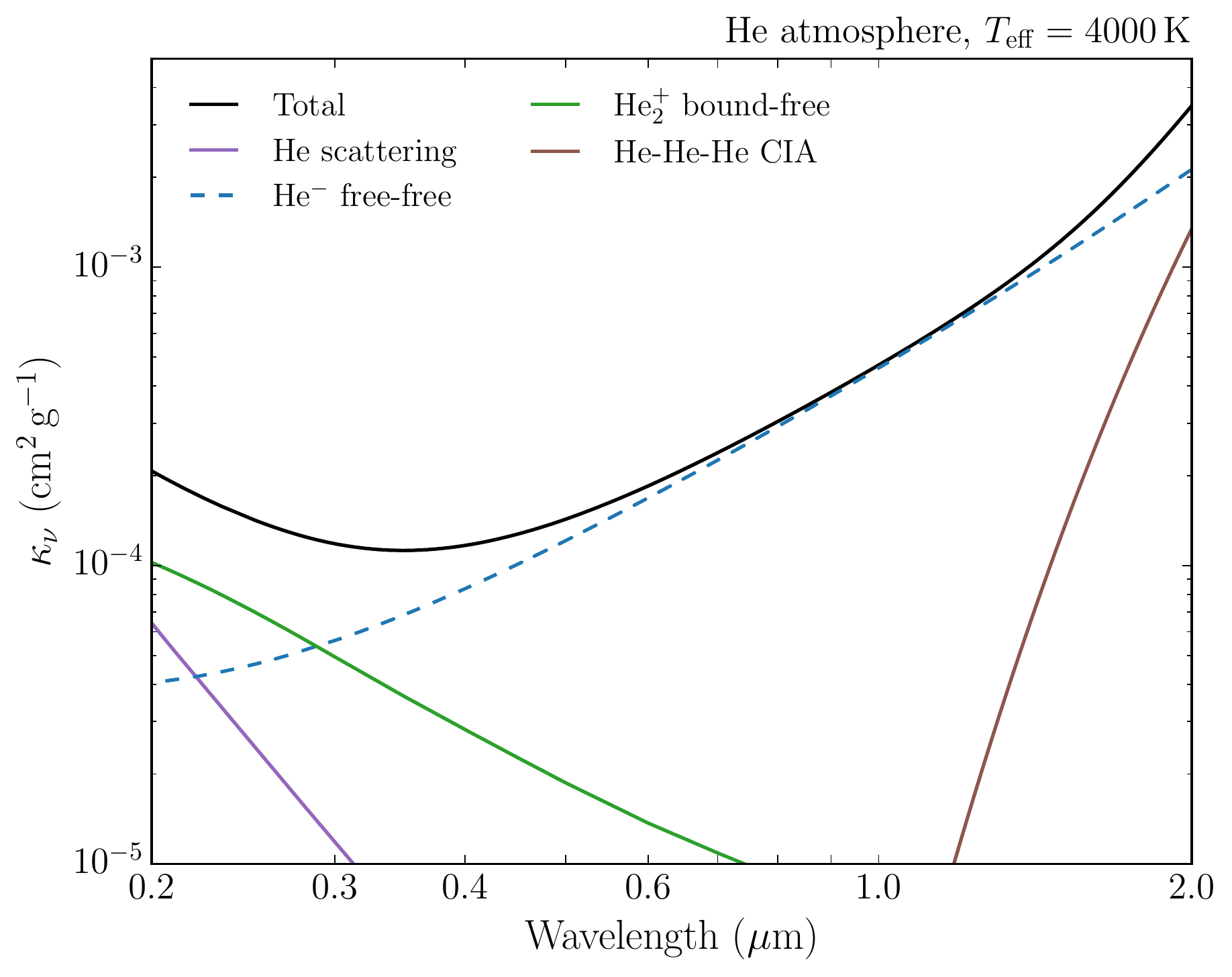}
\includegraphics[width=0.7\columnwidth]{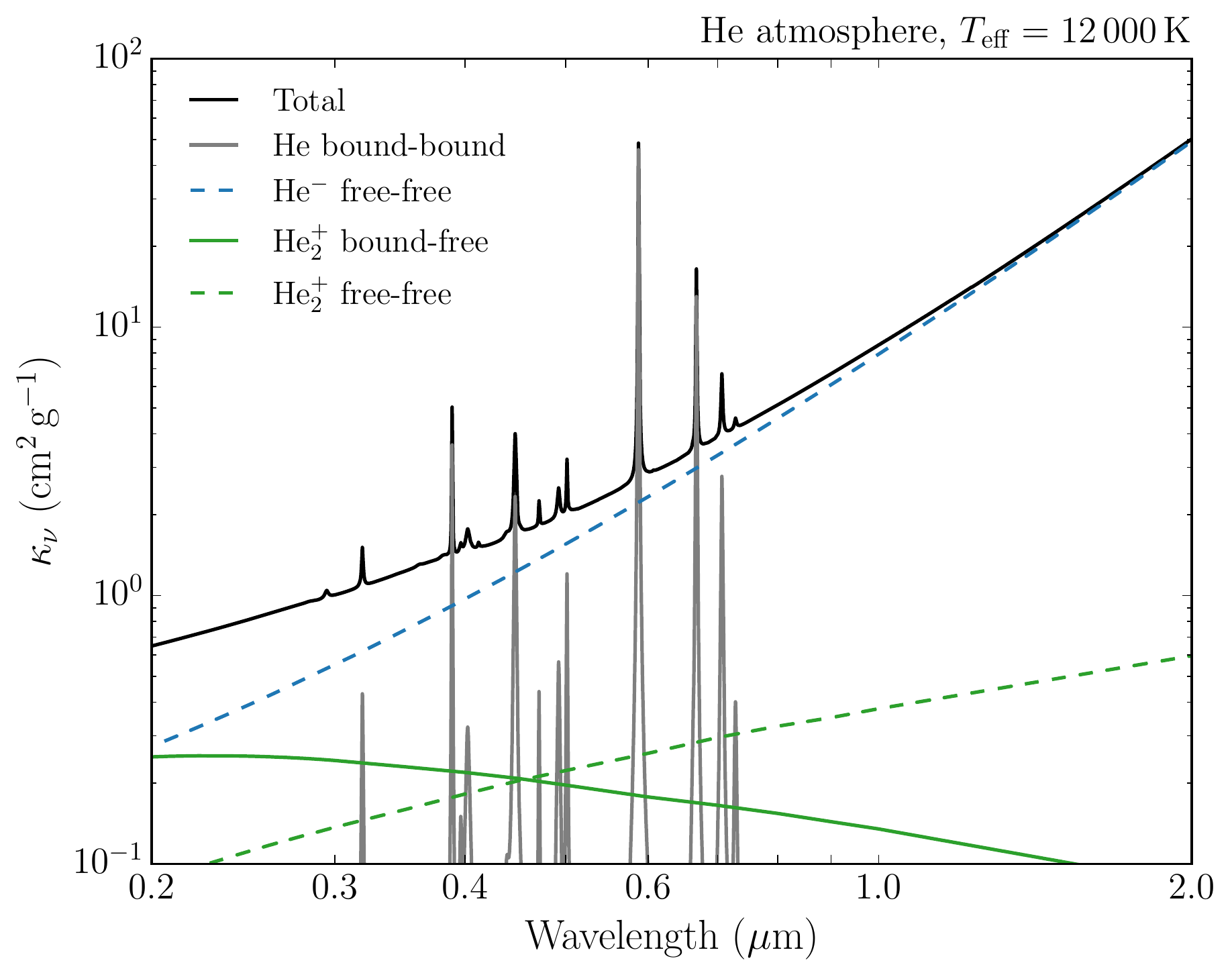}
\caption{Similar to Figure~\ref{fig:kappa_atm_H} but for He atmospheres. The $T_{\rm eff}=4000\,$K model corresponds to a DC white dwarf, while the hotter $T_{\rm eff}=12\,000\,$K model corresponds to a DB (He lines in the visible). \label{fig:kappa_atm_He}}
\end{center}
\end{figure}

The profiles shown in Figure~\ref{fig:atmprof} reveal two important trends: (1) the cooler the atmosphere, the denser it is and (2) He atmospheres are denser than H atmospheres. To understand why this is the case, we consider the definition of the optical depth $\tau$ in terms of the radiative opacity $\kappa$,
$d\tau = - \kappa \rho dr$ (the optical depth is measured inward from the surface). The pressure in a fluid in hydrostatic equilibrium is determined by 
\begin{equation}
    \frac{dP}{dr} = -\rho g
    \label{eq:hse}
\end{equation}
or
\begin{equation}
    dP = \frac{g}{\kappa} \,d\tau~.
\end{equation}
With the photosphere defined as $\tau=2/3$, we have, approximately, $P = 2g/3\kappa$. For an ideal gas, the density at the photosphere is of the order of
\begin{equation}
    \rho = \frac{2g \mu}{3\kappa \kb \Teff}~.
    \label{eq:rho_phot}
\end{equation}
Thus, the photospheric density increases with the surface gravity $g$ and average mass per particle $\mu$, and decreases as the photospheric 
temperature ($\teff$) and opacity increase. The $\teff$ and $\mu$ account for part of the trends we see in Figure~\ref{fig:atmprof} but the opacity plays a major role. Figures~\ref{fig:kappa_atm_H} and~\ref{fig:kappa_atm_He} show, for two different temperatures, the total wavelength-dependent opacity (black lines) of H- and He-atmosphere white dwarfs, respectively. Those figures show that (1) for a given atmospheric composition, the atmosphere becomes more transparent at low temperatures and (2) for a given temperature, a He atmosphere is more transparent than a H atmosphere (note the different scales for the opacity in Figures~\ref{fig:kappa_atm_H} and~\ref{fig:kappa_atm_He}). This is primarily due to the higher ionization potential of He, which results in a lower electron density and lower opacity. This explains why cool and H-poor atmospheres are more dense than warm and H-rich atmospheres.

Let us now examine the different contributions to the total opacity. In the case of a cool H atmosphere (top panel of Figure~\ref{fig:kappa_atm_H}), we see that three processes dominate. In the visible, the opacity is dominated by bound-free absorption from the H$^-$ ion. In the blue/UV, it is the far red wing of the Lyman-$\alpha$ line that contributes the most to the total opacity. Even if it is centered in the UV at $\lambda=0.1215\,\mu$m, the Lyman-$\alpha$ line strongly affects the emergent spectra of cool H-atmosphere white dwarfs in the visible because collisions with H atoms and H$_2$ molecules significantly broaden the absorption line profile at those densities. This opacity source has been shown to be crucial to reproduce the spectral energy distribution of cool white dwarfs \cite{kowalski2006lyman}. Finally, in the infrared, it is collision-induced absorption (CIA) caused by H$_2$-H$_2$ and H$_2$-H interactions that dominates. At sufficiently low temperatures, CIA becomes so strong that the peak of the spectral energy distribution shifts to shorter wavelengths as they cool --- the opposite of the behavior of a black body spectrum \cite{hansen1998,kilic2020}. The physics of CIA is discussed further in Section~\ref{sec:CIA}.

The dominant opacity sources completely change at higher temperature. The far red wing of Lyman-$\alpha$ and CIA are both absent from the bottom panel of Figure~\ref{fig:kappa_atm_H}. The photospheric density decreases by four orders of magnitude between $T_{\rm eff}=4000\,$ and $12\,000\,$K (Figure~\ref{fig:atmprof}), which implies fewer collisions in the plasma and in turn a less pronounced broadening of the Lyman-$\alpha$ line. Similarly, molecular hydrogen is fully dissociated at  the lower density and higher temperature and H$_2$ CIA goes away. Figure~\ref{fig:kappa_atm_H} shows that the opacity of an H atmosphere at $12\,000\,$K is dominated instead by bound-free and bound-bound absorption from H. These opacity sources were negligible at $4000\,$K because there was too little thermal energy to excite bound-free and bound-bound transitions. Thanks to these new opacity sources, the total opacity in the visible is now a few orders of magnitude higher than in the $4000\,$K case and the density is lower (Equation \ref{eq:rho_phot}). Finally, note that the opacity of the H spectral lines (grey line) can be a few orders of magnitude higher than the continuum opacity set by bound-free transitions (red solid line), implying that those lines are formed above the photosphere.

We now turn to the case of He atmospheres. The most striking aspect of Figure~\ref{fig:kappa_atm_He} is how small the total opacity is for cool He atmospheres. This leads to high, liquid-like photospheric densities (Figure~\ref{fig:atmprof}), where the ideal gas physics that is well justified for warmer, less dense atmospheres breaks down in the coolest atmospheres with $\teff \lesssim 7000\,$K. This complicates the modeling of those objects as discussed further in Sections~\ref{sec:He_atm}, \ref{sec:line_prof_cool}, and \ref{sec:mol_opac}. For both temperatures shown in Figure~\ref{fig:kappa_atm_He}, the dominant opacity source is He$^-$ free-free. This absorption mechanism has a linear dependence on the electronic density. This explains why it produces very little opacity at $4000\,$K when He is almost completely neutral, but much more at $12\,000\,$K when the Saha chemical equilibrium predicts a much higher ionization fraction. Other important opacity sources for He atmospheres are Rayleigh scattering from He atoms \cite{kowalski2014} (significant at low temperatures only), CIA from triple-He collisions (only at low temperatures, where the density is high enough for those three-body collisions to occur), and bound-free and free-free absorption from molecular He$_2^+$.

\section{Recent advances and challenges in the constitutive physics of white dwarfs}
\label{sec:recent}
In this section, we describe in greater detail several interesting problems where recent progress has been achieved in understanding the physical processes in white dwarfs with computer simulations, better models of the microphysics of plasmas, and experimental work that recreates the conditions in those stars. The topics cover computational fluid dynamics of convection in 3D, equations of state, liquid-solid phase transitions and their phase diagrams, atomic and molecular opacities, line broadening, inter-diffusion, and thermal conductivity. 

\subsection{Radiation-hydrodynamics simulations of surface convection}
\label{sec:rhd}
The physics of convective energy transport in stars, including the Sun, is relatively well understood \citep{nordlund2009}, although the complexities arising from the non-local and turbulent nature of convection has delayed the development of precise, quantitative models. Despite presenting a very simple description of convection, the mixing-length theory (MLT) has been reasonably successful for white dwarf modelling over the last 50 years. In this picture, a layer is either convective or non-convective according to the Schwarzschild criterion (Equation \ref{eq:schwar}) or, in the presence of a composition gradient, the Ledoux criterion \citep{ledoux1947,kippenhahnweigert}. In reality, the convective flows do not stop abruptly at well-defined boundaries at the top and bottom of the convection zone. Furthermore, in the non-adiabatic convective layers that define the atmosphere of most stars and white dwarfs, the convective efficiency predicted by the MLT is very sensitive to the underlying parameters describing the geometrical shape, radiative energy losses, and lifetime of convective cells \citep{ludwig1999,tremblay2015conv}. These quantities are not constrained by the MLT and must be calibrated from observations.

The emergence of powerful computing resources has enabled the generation of large grids of 3D radiation-hydrodynamics (RHD) simulations that have changed the paradigm for convection in the outer layers of stars \citep{nordlund1982,freytag1996,asplund2009}. Large grids of 3D simulations of convection are now covering almost all branches of stellar evolution including main sequence stars, brown dwarfs, giants, supergiants and white dwarfs \citep{freytag2012,magic2013,tremblay2013,cukanovaite2019}, enabling studies of large samples of stellar spectra with 3D atmosphere models. Two independent state-of-the-art codes have produced the bulk of the 3D RHD models for white dwarfs: CO5BOLD \citep{freytag2012} and ANTARES \citep{kupka2018}. 

These RHD simulations solve the time-dependent fluid equations of conservation of mass, momentum, and energy in three spatial dimensions for a compressible plasma in a constant gravitational field. This is solved together with non-local, frequency-dependent radiation transport and tensor viscosity using a realistic EOS. The radiative opacities can be either gray (i.e. frequency averaged) or chosen to preserve some level of frequency-dependent via an opacity-binning scheme. For both main sequence stars and white dwarfs, the {\it box-in-a-star} simulation geometry is used where the box represents a small portion of the surface of the star. A Cartesian grid is setup with periodic lateral boundary conditions and appropriate physical boundary conditions at the top and bottom (e.g. constant radiative or entropy flux). These simulations have been compared to observations of the surface of the Sun where resolved convective eddies are apparent as granulated brightness fluctuations at optical wavelength. They have been shown to agree within a few percent in brightness contrast \citep{wedemeyer2009}. 

\begin{figure}
\begin{center}
\includegraphics[bb= 150 15 400 200,width=0.55\columnwidth]{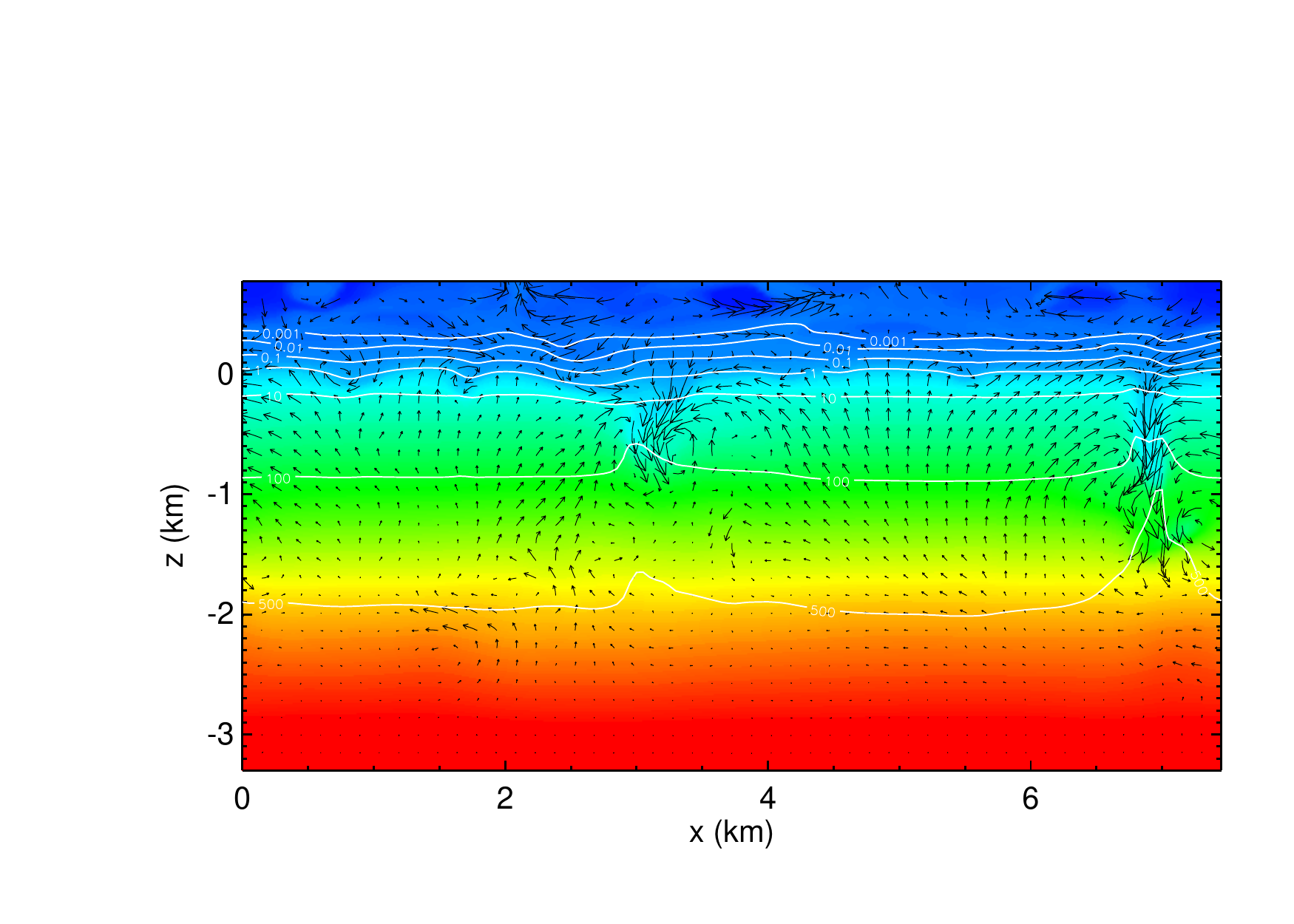}\\
\includegraphics[bb= 150 15 300 330,width=0.30\columnwidth]{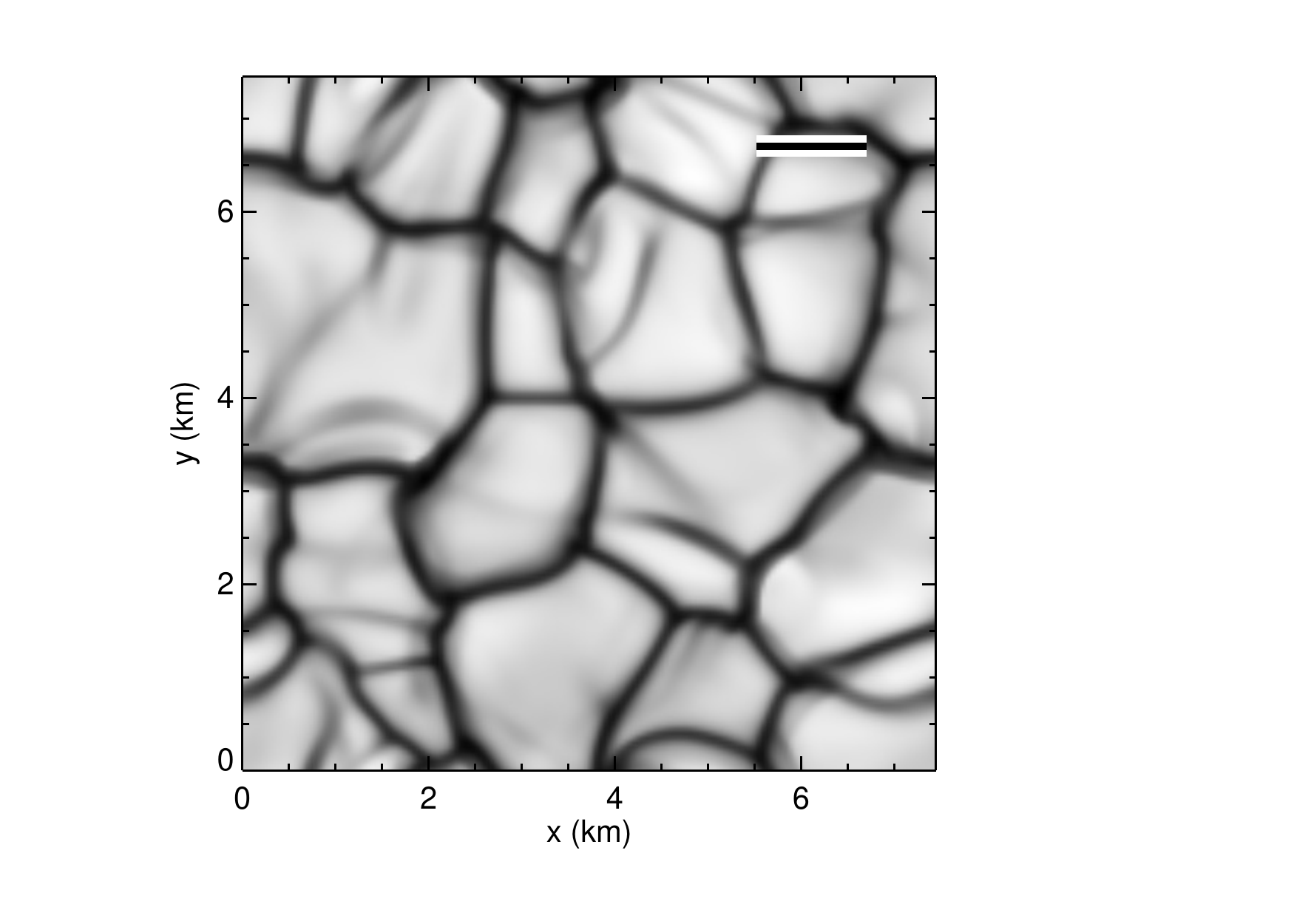}
\caption{Snapshot of a 3D radiation-hydrodynamics simulation of convection in a white dwarf at $T_{\rm eff}$ $\approx$ 12\,000\,K and $\log g = 8$. Top: Temperature structure for a slice in the horizontal-vertical $x-z$ plane through a rectangular box with coordinates $x, y, z$ (in km), with $z$ being in the vertical direction. The temperature is colour coded from 60\,000 (red) to 7000\,K (blue). The arrows represent convective velocities, while white lines correspond to contours of constant Rosseland optical depth $\tau_{\rm\sss R}$, with values given in the figure. Bottom: Frequency integrated emergent intensity at the top of the horizontal $x-y$ plane. This pattern of bright cells with dark borders, called granulation, is observed at the photosphere of the Sun. Hot rising convective flows form the bright areas, where the fluid cools by radiation and sinks along the darker edges. The rms intensity contrast with respect to the mean intensity is 18.8\%. The length of the bar in the top right is 10 times the pressure scale height at the photosphere ($\tau_{\rm R}$ = 2/3). Reproduced from Ref.~\citep{tremblay2013a} with permission \copyright ESO.
\label{fig:pier5}}
\end{center}
\end{figure}

Surface convection in white dwarfs (Figure~\ref{fig:pier5}) is quantitatively similar to that seen in main sequence stars including the Sun since the Mach (0.01--1.0) and P\'eclet (10$^{-2}$--10$^{2}$) numbers as well as the granulation intensity contrasts (0.1--25\%) are much alike \citep{tremblay2013b}, although with some notable differences for He atmospheres \citep{cukanovaite2019}. The major difference between white dwarfs and main sequence stars is in the extent of the internal convective envelope. The radial extent and total mass enclosed in white dwarf convection zones are orders of magnitude smaller \citep{tremblay2015conv}. Furthermore, the higher pressure at the surface of a white dwarf, combined with the usually pure-H or mixed H/He atmospheric compositions, results in different 3D effects on the spectral line shapes \citep{tremblay2013}.  

\begin{figure}
\begin{center}
\includegraphics[bb=88 118 522 650,width=0.58\columnwidth,angle=270]{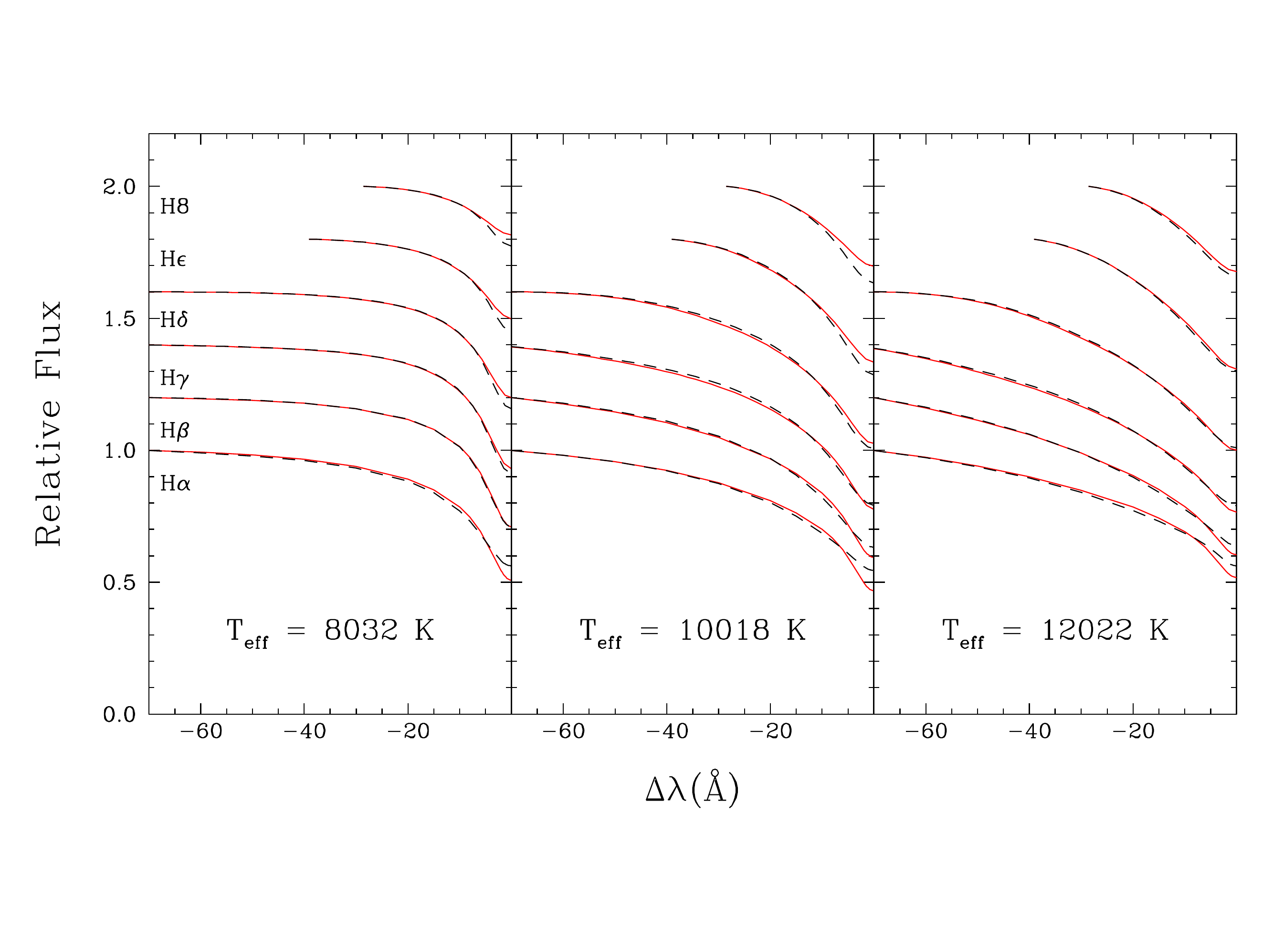}
\caption{Comparison of the blue wing of six Balmer line profiles (H$\alpha$ to H8) calculated from spatially and time-averaged 3D atmosphere structures (red, solid) and standard 1D structures (black, dashed). The values of $T_{\rm eff}$ are shown on the different panels and $\log g=8$. All line profiles were normalized and shifted vertically for clarity. The overall trend is that the 3D RHD simulations produce lower Balmer lines that are narrower, and the opposite for the higher lines in the series. Reproduced from Ref.~\citep{tremblay2013a} with permission \copyright ESO.
\label{fig:pier6}}
\end{center}
\end{figure}

Overall, these 3D simulations have demonstrated that it is possible to determine largely parameter-free, time- and spatially-averaged temperature and pressure stratifications for convection zones, in contrast to the 1D MLT theory where free parameters must be fitted to the observations. Typically, no single 1D MLT parametrization can reproduce the full surface-averaged emergent spectrum of a 3D RHD model of a white dwarf at all wavelengths (Figure~\ref{fig:pier6}). The reasons for this behaviour are complex and depend on how 3D fluctuations (e.g. cool downflows and warm updrafts) average over time, the stellar disk, and optical depth.

Current grids of 3D model atmospheres cover essentially the full range of convective white dwarfs, from 3500 to 18\,000\,K and $\log g$ from 5.0 to 9.0 for pure-H atmospheres, and from 10\,000\,K to 34\,000\,K and $\log g$ from 7.5 to 9.0 for mixed H/He atmospheres \citep{tremblay2015ELM,cukanovaite2019}. Only the warmest pure-H ($\gtrsim$11\,000\,K) and mixed atmospheres ($\gtrsim$24\,000\,K) currently have the full vertical extent of the convection zone simulated, allowing the study of physical processes at the bottom boundary layers, such as convective overshoot \citep{kupka2018,cunningham2019} and pulsation excitation mechanisms. With current computing capabilities, simulations of the full extent of the convection zone of cooler white dwarfs remain out of reach.

\begin{figure}
\begin{center}
\includegraphics[bb= 170 220 450 700,width=0.50\columnwidth]{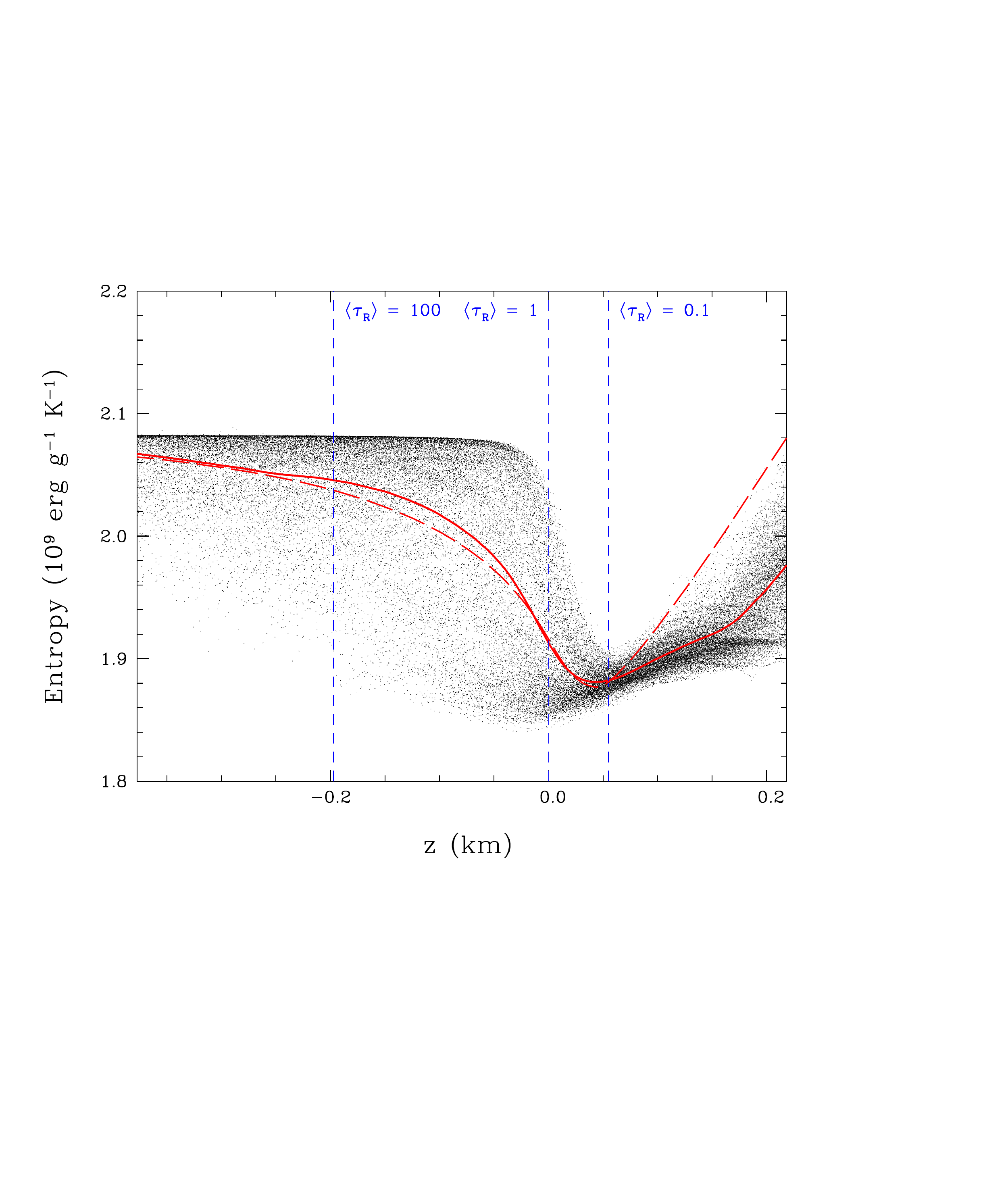}
\caption{Local 3D entropy values (black dots) as a function of geometrical depth for a subset of a simulation at $T_{\rm eff} = 10\,025$\,K and $\log g = 8$. The entropy profile of the 3D simulation, averaged over constant geometrical depth $z$ and time, is shown with a red solid line. The entropy of an equivalent 1D model with the best MLT parameterization is shown as a red dashed line. Average optical depth values of $\tau_{\rm\sss R} = 100$, 1, and 0.1 are highlighted with dashed blue lines. The surface of the star is to the right. Adapted from Ref. \citep{tremblay2015conv}, with permission \copyright AAS. 
\label{fig:pier3}}
\end{center}
\end{figure}

Accounting for 3D convection in calculations of stellar structures is desirable for many applications, such as stellar evolution and asteroseismology. The calibration of 1D envelope structures with 3D simulations as surface boundary condition for interior models has now been performed for white dwarfs \citep{tremblay2015conv,cukanovaite2019} using methods previously employed for main sequence stars \citep{ludwig1999,tramp2014}. 
For warmer convective white dwarfs, the full vertical extent of the convection zone can be simulated in 3D, which can then be matched directly with the appropriate 1D structure below the convection zone. For cooler white dwarfs, the 3D simulations of the upper part of the convection zone can be matched to 1D MLT models of convection to cover its full extent.  Deep convection becomes essentially adiabatic, i.e. constant specific entropy with depth, a limit whose thermodynamic stratification can be well-represented with the MLT. The constant value of the entropy in the regime of adiabatic convection can be extracted at the bottom of a 3D RHD simulation to calibrate the 1D MLT that is used to extend the convection zone along an adiabat all the way to the bottom. 
Figure~\ref{fig:pier3} demonstrates that significant entropy fluctuations are apparent at all geometrical depths in a 3D simulation, although there is a constant asymptotic upper limit below the observable surface. According to Ref. \citep{stein1989}, the fluid in central regions of broad ascending flows is still thermally isolated from its surroundings as it rises adiabatically until it reaches layers immediately below the photosphere. In other words, convective upflows keep an imprint of the physical conditions at the bottom of the convection zone. The asymptotic entropy value of the 3D simulations has therefore been employed to calibrate 1D structures in established white dwarf evolution codes.

\subsubsection{Convective overshoot}
\label{sec:overshoot}

The size of a 1D convection zone is based on the location of the Schwarzschild instability boundaries. It is well-known that this approach cannot properly characterize the actual hydrodynamic flows that exist at convective boundaries, instead producing a significant and unphysical discontinuity in the flow of kinetic energy \cite{spiegel1963,zahn1991,freytag1996}. 

\begin{figure}
\begin{center}
\includegraphics[width=1.00\columnwidth]{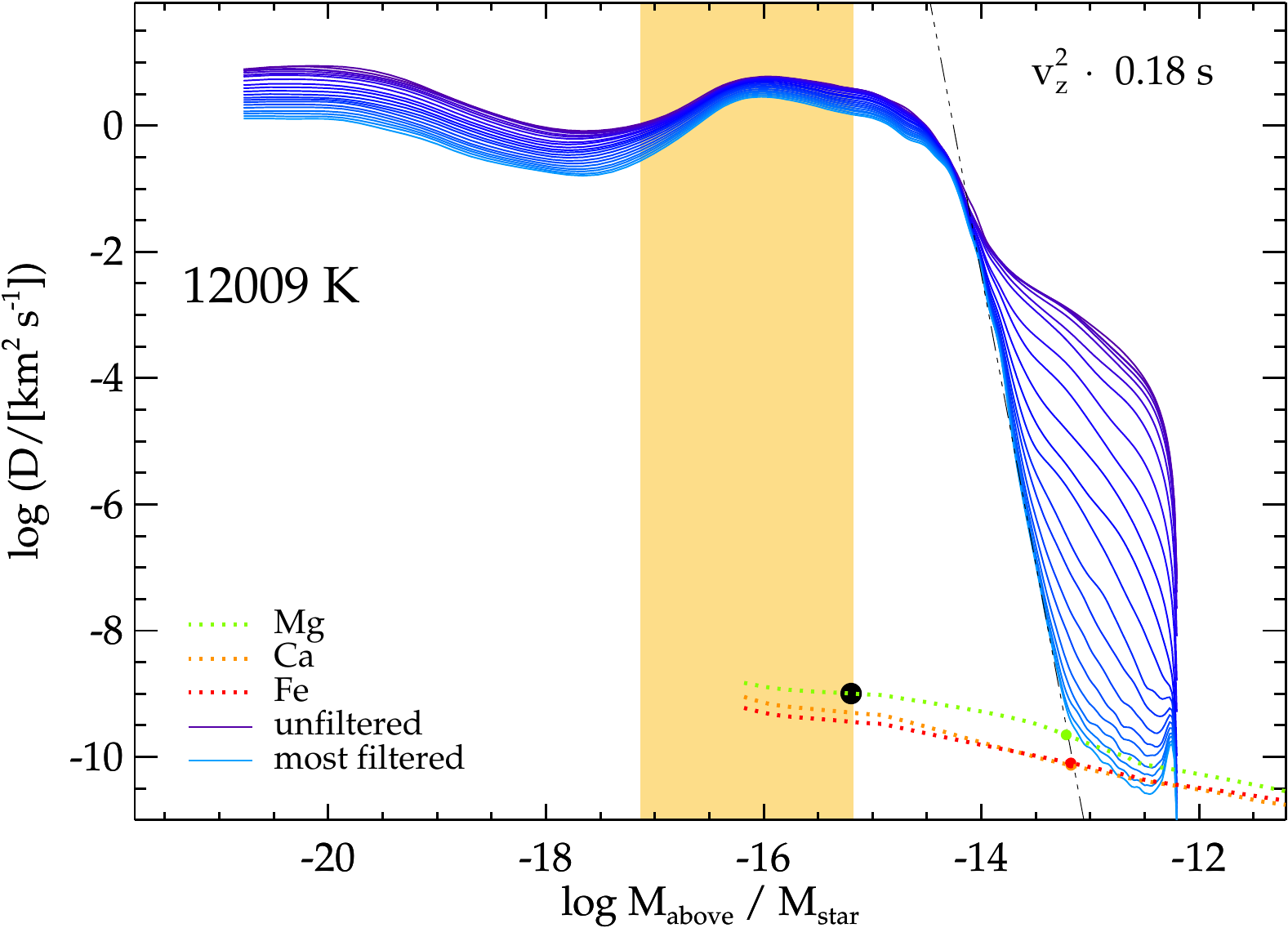}
\caption{Diffusivity as a function of depth  $\log q$ (Equation \ref{eq:log_q}) in a pure-H white dwarf envelope with $T_{\rm eff}$ = 12\,009\,K and $\log g=8$. The macroscopic (convective) diffusivity inferred from 3D convective velocities \cite{cunningham2019} is shown with solid lines. From top to bottom, the diffusivity is shown with increasing spatial filtering of the velocities to remove small scale features. This is done to reduce the effect of artificial standing waves.
The best fit to macroscopic diffusion coefficients is shown as a black dot–dashed line. The convectively unstable region defined by the Schwarzschild criterion is indicated by the orange shaded region, which has a much smaller extent than the high-diffusivity profile extracted from the simulation that includes convective overshoot. Also
shown are the microscopic diffusion coefficients for Mg, Ca, and Fe (dotted lines), taken as typical trace elements in white dwarf atmospheres. The base of the mixed region is located where the macroscopic and microscopic diffusion coefficients are equal (filled color points in 3D and black point in 1D). The surface of the star is to the left. Reproduced from Ref. \citep{cunningham2019} with permission. 
\label{fig:pier4}}
\end{center}
\end{figure}

In reality, plumes are accelerated near the edge of the unstable layers and are able to travel outside of the 1D convection zone into the convectively stable layers in a phenomenon known as \textit{convective overshoot}, which happens both at the top of the convection zone, usually in the photosphere, and at its base. A robust characterization of convective overshoot  requires hydrodynamics simulations.
The strength or characteristic length of convective overshoot is usually different when looking at thermal (convective flux) or mechanical quantities (convective velocities). One useful definition of the extent of overshoot corresponds to the region where the macroscopic overshoot diffusivity is larger than the microscopic diffusion coefficient. These regions are considered to be fully mixed in terms of chemical abundances, from which follows the definition of the mixed convective mass.
Recent independent studies \cite{kupka2018,cunningham2019} of warm H-rich white dwarfs with small convective zones have demonstrated that bottom convective overshoot has a dramatic impact as it  increases  the inferred mixed mass by up to 2--3 orders of magnitude (see Figure~\ref{fig:pier4}). Bottom convective overshoot has strong implications for spectral evolution (convective mixing) and accretion of planetary debris (inferred accretion rates and mass of accreted planetesimals). The impact of convective overshoot on the convective coupling to the degenerate core at lower $\Teff$ has not been explored but is likely small compared to existing uncertainties in the constitutive atomic and equation of state physics. 

Convective overshoot studies have so far been limited to white dwarfs with $T_{\rm eff} = 11\,000-18\,000$\,K and shallow convection zones. There is nevertheless evidence that overshoot becomes less important in terms of the increase in mixed mass as the convection zone reaches deeper layers \citep{cunningham2019} but more work is needed. A physically motivated analytical parameterization of convective overshoot \citep{bauer2019} is of major importance for white dwarfs and stellar evolution in general. 

In cases where material is accreted onto a white dwarf from rocky planetesimals or a stellar companion, a gradient in chemical composition is induced between the photosphere (or mixed layers) and the envelope just below. This gradient can lead to thermohaline instabilities \citep{deal2013,bauer2018,bauer2019}. This mechanism is distinct from convective overshoot, and can happen in warm white dwarfs that have no convective instabilities. While no 3D simulations have yet been performed for thermohaline mixing in white dwarfs, recent studies have shown that it is likely to be a significant effect adding up to convective overshoot \citep{bauer2019}. 

\subsubsection{Magnetic white dwarfs}
\label{sec:conv-magnetic}

\begin{figure}
\begin{center}
\includegraphics[bb= 70 15 390 330,width=0.48\columnwidth]{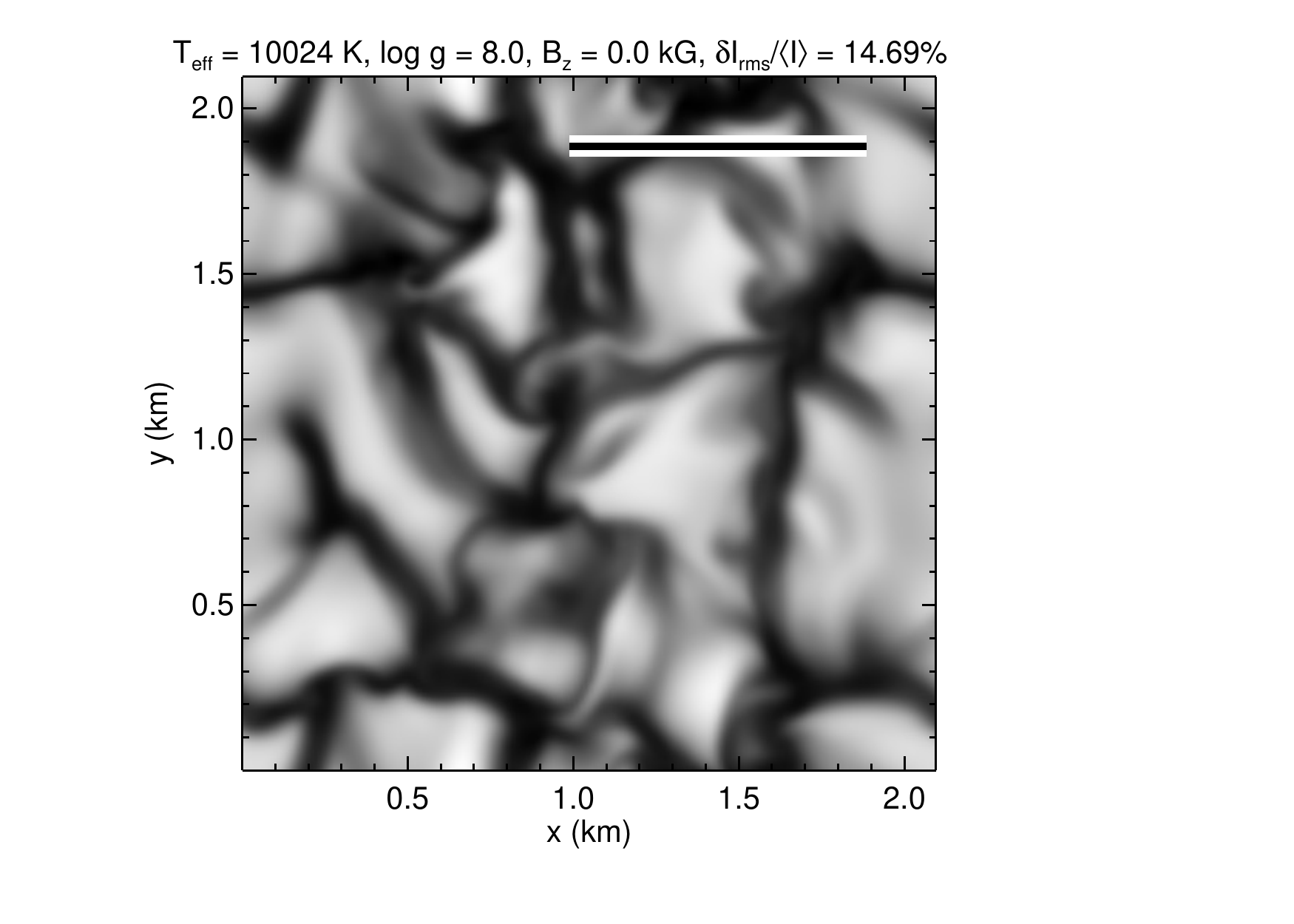}
\includegraphics[bb= 70 15 390 330,width=0.48\columnwidth]{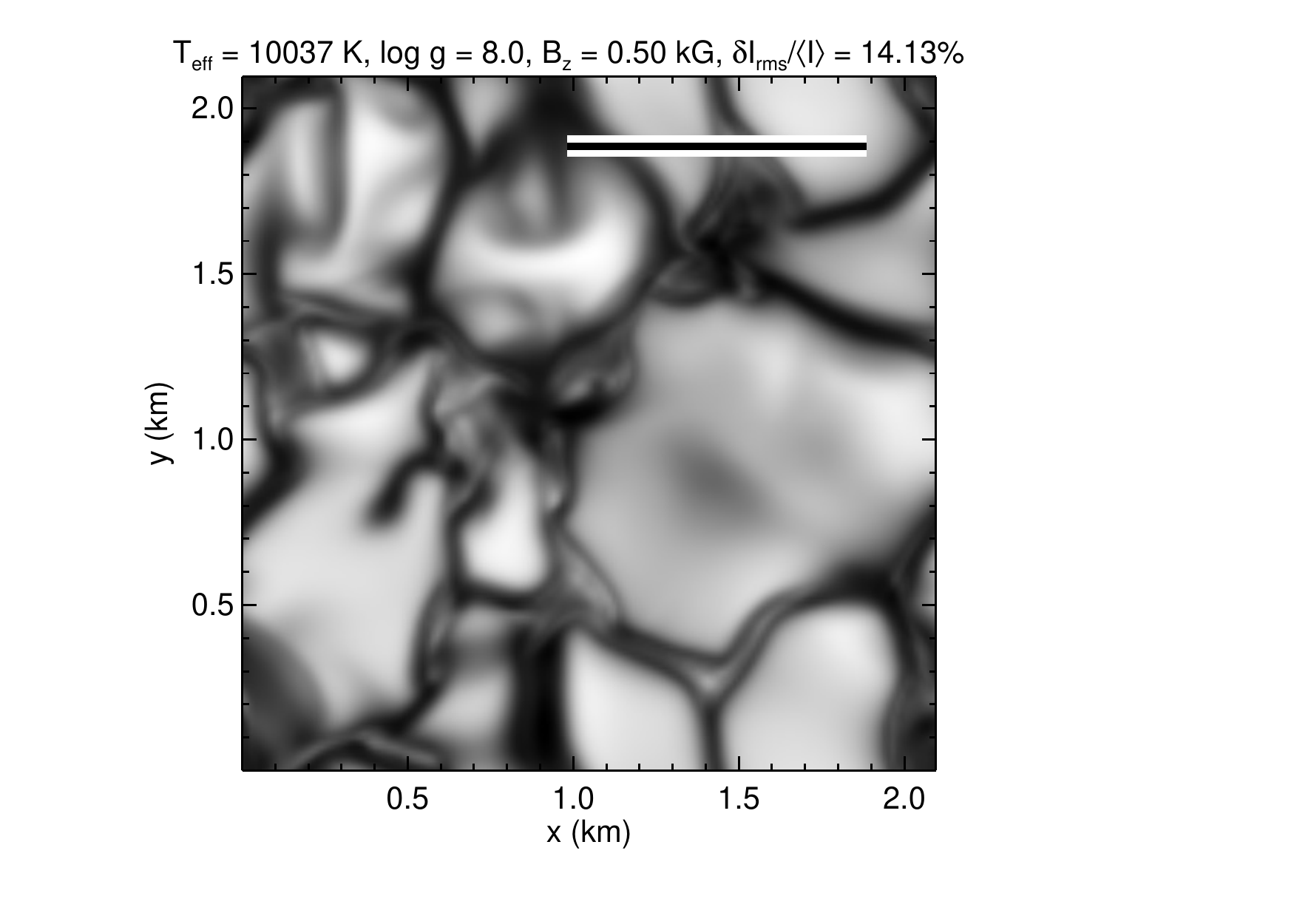}\\
\includegraphics[bb= 70 15 390 350,width=0.48\columnwidth]{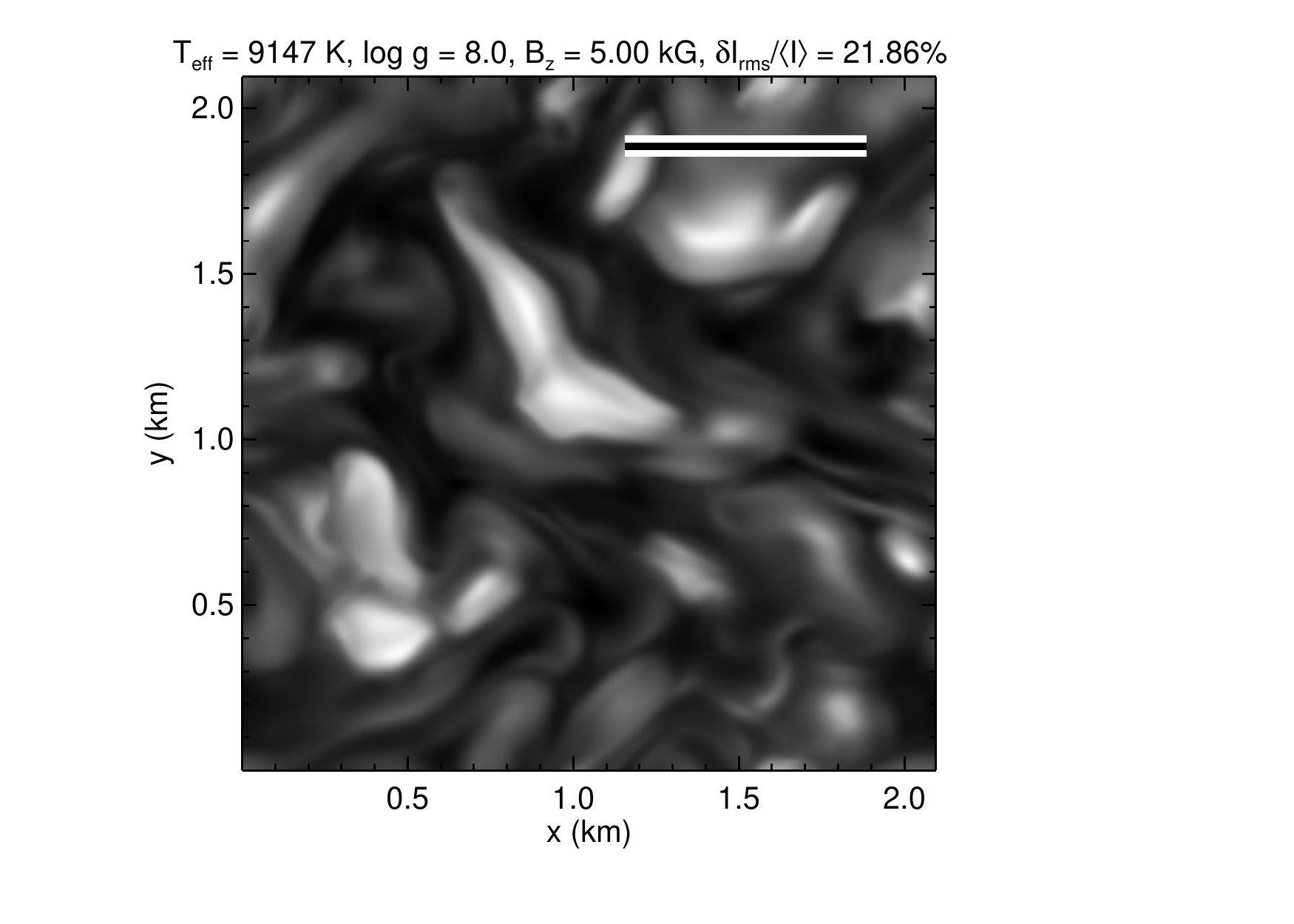}
\caption{Frequency--integrated radiative intensity emerging from the $x-y$ horizontal plane at the top of the computational domain for 3D MHD simulations. All simulations have a constant surface gravity of $\log g = 8.0$, a pure-H composition, and rely on the same entropy value for the inflowing material through the open bottom boundary, resulting in $T_{\rm eff} \approx$ 10\,000\,K. 
The panels show simulations where a constant vertical field of 0, 0.5 and 5\,kG was imposed at the lower boundary. The rms intensity contrast with respect to the mean intensities is also shown above the panels. The length of the bar in the top right is 10 times the pressure scale height at $\tau_{\rm\sss R}$ = 2/3. Reproduced from Ref. \citep{tremblay2015mag} with permission \copyright AAS. 
\label{fig:pier7}}
\end{center}
\end{figure}

It has long been predicted---and observed---that magnetic regions in the Sun and low-mass stars  locally inhibit surface convection and convective energy transport \citep{parker1974,cattaneo2003,kraus2011}. 
Based on a handful of magneto-hydrodynamics (MHD) calculations for white dwarfs at $T_{\rm eff}$ $\approx$ 10\,000\,K and pure-H composition, it has been shown that magnetic fields as low as 1\,kG can inhibit surface convective energy transfer \citep{tremblay2015mag}. This is illustrated by the disappearance of the fluctuations in surface intensity (granulation) as the field strength increases (Figure \ref{fig:pier7}).
This results in a steeper vertical temperature gradient in the atmosphere, which can be constrained using spectroscopy \citep{gentilefusillo2018}. The external magnetic fields necessary to inhibit convection at the surface varies from 1\,kG for warmer DA white dwarfs to almost 1 MG for cool He-dominated atmospheres \citep{cunningham2021}. It is important to emphasize that inhibition of bulk convective energy transport does not mean inhibition of small scale convective motions and instabilities. Three-dimensional MHD simulations show that inefficient convective flows act on smaller spatial scales in the presence of a magnetic field, but still show large local convective velocities \citep{tremblay2015mag}. This suggests that convective chemical mixing is still effective even for magnetic fields that are one or two orders of magnitude stronger than necessary to inhibit convective energy transfer. This has implications for the visibility of metals in the study of metal-polluted atmospheres.

\subsubsection{Horizontal mixing and rotation}
\label{sec:spreading}

In addition to efficient vertical mixing of chemical elements, convection can also mix material in the horizontal (or surface plane) direction. This has received little attention until recently \citep{montgomery2008,cunningham2021}, in part because this process is bound to be relatively fast given surface convective velocities reaching $\sim 1\,{\rm km\,s}^{-1}$ (both in the horizontal and vertical directions), resulting in a horizontal diffusion time scale of the order of
\begin{equation}
t_{\rm spread} \sim \frac{R_{\star}^2}{D_{\rm conv}} \sim  \frac{R_{\star}^2}{v_{\rm conv} H_{\rm\sss P}}  \sim 1~{\rm yr}~,
\end{equation}
where $D_{\rm conv}$ is the convective diffusivity and $H_{\rm\sss P}$ the pressure scale height. 
The most important parameter is the ratio between the horizontal and vertical diffusion time scales, which describes whether accreted material has time to spread horizontally before it sinks and disappears from the visible layers. It has been found that most white dwarfs are able to efficiently spread accreted material, with the exception of DA white dwarfs above 11\,000\,K. The addition of thermohaline mixing could make horizontal spreading somewhat more efficient, making that transition a few thousand degrees higher.

This prediction is possibly in tension with observations of warm ($\gtrsim 11\,000\,$K) polluted DAZ white dwarfs, which show strong evidence of a homogeneous metal-polluted surface \citep{cunningham2021}. The solutions to this problem remain open, but depend on the unknown geometry of accretion. In particular, near-spherical deposition would remove the need for an unknown source of horizontal spreading to explain the observations. Magnetic white dwarfs accreting planetary debris will be of particular interest in that context, allowing one to constrain both accretion geometry and mixing in the presence of magnetic fields. 

Rotation is slow in single white dwarfs. Ref.~\cite{hermes2017} finds an average period of 35\,h with a standard deviation of 28\,h from studies of pulsating white dwarfs. Similar results are found using periodic variability of spotted magnetic white dwarfs \cite{brinkworth2013}. Such slow rotation is not known to have observable effects on the structure, cooling age, surface composition, or magnetic field generation (except for the recent scenario of a crystallization dynamo \cite{ginzburg2022}).

\subsection{Non-ideal chemistry in cool He-dominated atmospheres}
\label{sec:He_atm}
The liquid-like density ($0.1-1\,{\rm g\,cm}^{-3}$) that characterizes cool He-rich atmospheres (Figure~\ref{fig:atmprof}) poses a significant challenge to their modeling. The equation of state $P(\rho,T)$ of He is already satisfactorily described by existing tables \citep{becker2014}. The key problem is instead characterizing the chemical equilibrium of minor ionic and molecular species that do not affect the equation of state but contribute disproportionately to the opacity under those conditions. The challenge is to calculate, for a dense interacting He fluid at density $\rho$ and temperature $T$, the ionization ratios (e.g., $n({\rm He}^+)/n({\rm He})$) and the molecular dissociation ratios (e.g., $n({\rm H})/n({\rm H}_2)$) of the different atomic and molecular species present in the atmosphere. This is a problem of the utmost importance for the study of those objects, which include some of the most intriguing white dwarfs, such as the IR-faint (or ultracool) white dwarfs and the recently discovered Li-polluted white dwarfs. The chemical equilibrium prescription used in an atmosphere model calculation determines the radiative opacities (e.g., more H$_2$ results in more CIA), and in turn the structure of the atmosphere and the resulting spectrum.


\subsubsection{The chemical picture}
All stellar atmosphere codes currently rely on the chemical picture. In this approximation, bound species (atoms and molecules) are considered as basic particles. This is to be contrasted with the physical picture, where only nuclei and electrons are the basic particles. Assuming that atoms and molecules are well-defined entities can be problematic at high densities where the boundary between bound and free states is blurred. The abundance of a molecule can depend on the very definition of a “molecule” \citep{vorberger2007}. In spite of this fundamental limitation, the chemical picture is much more practical when it comes to model atmosphere calculations. To include atomic spectral lines and molecular bands, the population of thousands of atomic and molecular levels is needed, making the physical picture impracticable.  

Under low-density conditions, the ionization and dissociation ratios are obtained with the Saha equation, which is derived assuming an ideal gas chemical potential for all species involved,
\begin{equation}
\mu_i^{\rm id} = E_{i,0} + k_{\rm\sss B} T \ln \left[ \frac{ n_i h^3}{Q_i \left(2 \pi m_i k_{\rm\sss B} T \right)^{3/2}} \right]~,
\label{eq:muideal}
\end{equation}
with $E_{i,0}$ the ground-state energy of species $i$ and $Q_i$ its internal partition function,
\begin{equation}
Q_i = \sum_j g_{i,j} \exp \left( -\frac{E_{i,j}}{k_{\rm\sss B} T} \right)~,
\label{eq:partfunc}
\end{equation}
where the sum runs over all bound states of species $i$, and where $g_{i,j}$ is the statistical weight of state $j$.\footnote{Electrons are normally classical in the atmospheres of white dwarfs but Equation \ref{eq:muideal} can be easily modified to account for degeneracy (see Figures \ref{fig:rhoT_DA}, \ref{fig:rhoT_DB}, and \ref{fig:rhoT_HDQ}).} 

The chemical equilibrium is obtained by equating the chemical potential of reacting species. For example, for the reaction H $\rightleftharpoons$ H$^+$ + $e$, we have $\mu_{\sss\rm H} = \mu_{\sss\rm H^+} + \mu_e$. The system of equation is closed by imposing overall charge neutrality and the conservation of the number of atoms of each element.
At the densities encountered in cool He-atmosphere white dwarfs, two modifications must be introduced: (1) an interaction (or non-ideal) term has to be added to Equation~\ref{eq:muideal}, and (2) the partition function should be modified to account for the perturbation of the internal degrees of freedom due to interactions with other particles. As much as possible, the latter should be consistent with the former.

\subsubsection{The occupation probability formalism}
During the last three decades, the occupation probability formalism of Hummer \& Mihalas \citep{hummer1988} has been widely used to evaluate chemical equilibrium under non-ideal conditions in white dwarf atmospheres \citep{bergeron1991,bergeron1995,koester2000,dufour2007_dz,tremblay2009,elms2022}. In this framework, an occupation probability $w_j$ is assigned to each bound state with $w_j=1$ for unperturbed levels and $w_j=0$ for levels that are completely destroyed by interactions. Those $w_j$ then multiply each term in the Boltzmann distribution,
\begin{equation}
\frac{n_{i,j}}{n_i} = w_{i,j} \frac{g_{i,j}}{Q_i} \exp \left( -\frac{E_{i,j}}{k_{\rm\sss B} T} \right)~,
\label{eq:boltzmann}
\end{equation}
and in the calculation of the partition function,
\begin{equation}
Q_i = \sum_j w_{i,j} g_{i,j} \exp \left( -\frac{E_{i,j}}{k_{\rm\sss B} T} \right)~.
\end{equation}
For cool He atmospheres, the $w_{i,j}$ are dominated by neutral interactions with neighboring particles. In this case, Hummer \& Mihalas give the prescription
\begin{equation}
w_{i,j} = \exp \left[ -\frac{4 \pi}{3} \sum_{i',j'} n_{i',j'} \left( r_{i,j} + r_{i',j'} \right)^3 \right]~,
\label{eq:wneutral}
\end{equation}
where $r_{i,j}$ is a radius assigned to species $i$ in state $j$, and the sum runs over all species and all states.\footnote{The argument of the exponential corresponds to the linear contribution in the packing fraction of the hard sphere equation of state \citep{mansoori1971}.} While very simple to implement and interpret, the excluded volume effect of Equation~\ref{eq:wneutral} is only a crude approximation of the actual interaction between two neutral particles. Moreover, there is no theoretical prescription for the radii $r_{i,j}$; they are free parameters. The $r_{i,j}$ can be arbitrarily adjusted to obtain a good agreement between the white dwarf spectra and model atmospheres \citep{bergeron1991}. However, because of the lack of predictive character of this approach, a good agreement between theory and observation does not mean that the atmosphere is being properly modeled nor that the derived atmospheric parameters are accurate.

\subsubsection{More advanced interaction models}
\label{sec:chem_recent}
Recently, chemical equilibrium calculations that are more predictive than the Hummer \& Mihalas formalism have been performed \citep{kowalski2006,kowalski2006thesis,kowalski2007,blouin2018a}. Those works have focused on accurately computing the non-ideal chemical potential $\mu_i^{\rm nid}$ that has to be added to Equation~\ref{eq:muideal} to account for interactions in a dense He fluid,
\begin{equation}
    \mu_i = \mu_i^{\rm id} + \mu_i^{\rm nid}~.
    \label{eq:mu_total}
\end{equation}
In general, $\mu_i^{\rm nid}$ is a function of $T$ and $n_i$.
Adding those $\mu_i^{\rm nid}$ to each species $i$ involved in the ionization (dissociation) process being studied effectively shifts the ionization (dissociation) potential. There are two contributions to consider in the calculation of $\mu_i^{\rm nid}$. The first is simply the interaction energy of species $i$ with the He fluid, and the second is an entropy term. The latter arises because the surrounding medium responds to the ionization or dissociation of a particle: the radial distribution function $g(r)$ of the fluid depends on the ionization or dissociation state, which in turn affects the chemical potential. Both contributions can be evaluated using the classical theory of fluids \citep{martynov1992,hansen2013}, provided that realistic  pair interaction potentials are available. Such calculations have been performed for the chemical equilibrium of H/H$_2$ \citep{kowalski2006}, H/H$^-$ \citep{kowalski2006thesis}, He$^+$/He \citep{kowalski2007}, and C$^+$/C, Na$^+$/Na, Mg$^+$/Mg, Ca$^+$/Ca, and Fe$^+$/Fe \citep{blouin2018a} in a bath of He.

Despite this recent progress, many aspects of current chemical equilibrium models should be improved. A first limitation of current calculations is that some are based on ab initio pair potentials obtained in the infinite dilution limit. This is problematic as this approach does not account for the many-body effects that become important at the liquid-like densities considered here. This problem can be solved by using potentials calibrated to high-pressure experimental data \citep{kowalski2007} or by using Density Functional Theory Molecular Dynamics (DFT-MD) to simulate the behavior of a many-body system at the relevant densities \citep{blouin2018a}.

A second limitation with the current state of the art is that the internal partition functions $Q_i$ are assumed to remain unaffected by the interactions. All calculations are performed assuming that all species remain in their ground states, and, unlike the occupation probability formalism, nothing is said about differences in how the different internal energy levels of each species are affected by non-ideal effects. This also means that there is currently no accurate prescription for how the Boltzmann distribution (Equation~\ref{eq:boltzmann}) is perturbed by non-ideal effects. Knowing the population of each internal energy level of a species is needed to predict the strength of spectral absorption lines and bands, making this problem particularly important.

\subsubsection{Pressure ionization and the phase diagram of helium}
\label{sec:He_phase_diagram}
Pure He atmosphere models have an interesting history related to the difficulty of properly modeling the pressure ionization of He at low temperatures ($T \lesssim 1$\,eV), where the ionization fraction of He is very small ($\ll 10^{-5}$) but highly consequential and very sensitive to the details of the ionization model. Atomic He has a very low opacity at optical and infrared wavelengths (a combination of He Rayleigh scattering and He--He--He collision-induced absorption, Figure~\ref{fig:kappa_atm_He}) resulting in high-density atmospheres. The presence of a small amount of free electrons, from the ionization of He or from traces of other elements with relatively low ionization potentials, introduces He$^-$ free-free electron scattering and He$_2^+$ bound-free absorption that increase the opacity considerably and decrease the overall density of the atmosphere (Figure \ref{fig:kappa_atm_He}).

Early models \citep{kapranidis1982} used an EOS and opacities 
based on the finite-temperature Thomas--Fermi--Dirac model of the isolated atom for the intra-atomic potential. In retrospect, this is a rather crude model for this $(\rho, T)$ regime, leading to a poor EOS and exotic atmosphere models where conductive energy transport is significant at the photosphere. 
A new set of pure He atmosphere models \citep{bergeron1995} introduced He--He interactions in the calculation of the chemical equilibrium using a He--He pair potential calibrated to shock data up to 56\,GPa \citep{nellis1984}. The latter ensured a realistic $P(\rho, T)$ relation for dense He that also agreed quite well with ab initio simulations of dense He \citep{kowalski2007} up to $\rho \approx 0.8\,$g\,cm$^{-3}$. This allows the evaluation of $\mu_{\sss\rm He}^{\rm nid}$ (Equation \ref{eq:mu_total}), which shifts the ionization equilibrium with respect to the ideal Saha equilibrium. The He--He interaction being strongly repulsive,  $\mu_{\sss\rm He}^{\rm nid} > 0$ and the ionization fraction of He increases. The new models also
included He Rayleigh scattering and He$^-$ free-free opacities. The resulting low-$\Teff$ models had densities as high as 4\,g\,cm$^{-3}$ and pressures of 20\,TPa at temperatures below 1\,eV.

\begin{figure}
\begin{center}
\includegraphics[width=1.0\columnwidth]{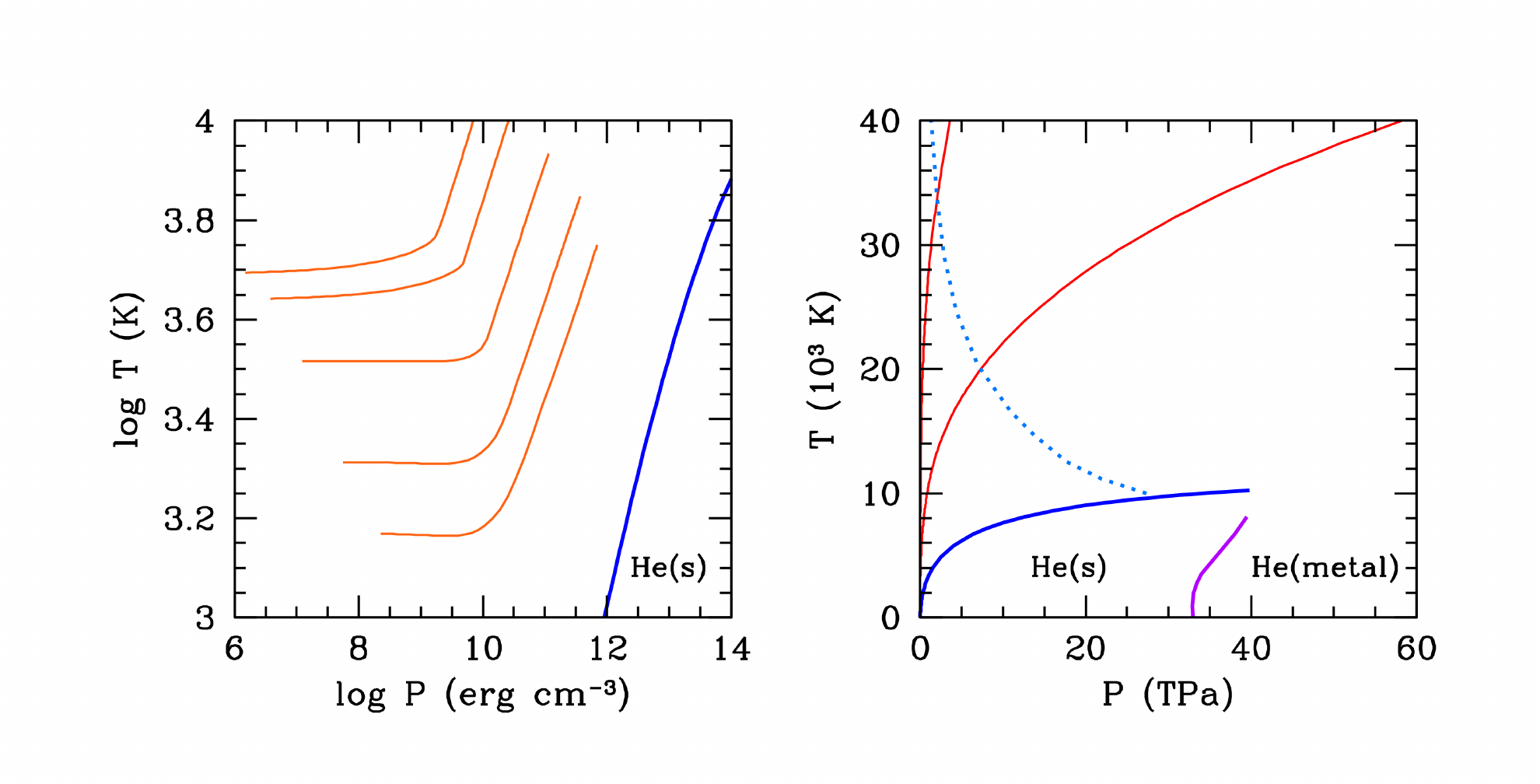}
\caption{Comparison of recent calculations of the phase diagram of He with white dwarf models. The high pressure melting curve of He \citep{preising2019} (heavy blue curve) and the insulator-metal transition in the solid phase \citep{monserrat2014} (heavy purple curve) are first order transitions. The insulator-metal transition in the fluid phase (also known as pressure ionization) is indicated by the closure of the band gap (blue dotted curve) and occurs continuously \citep{preising2020}. These transitions in dense He are compared to pure He white dwarf atmosphere models of $\Teff=4000$\,K to 8000\,K (left panel, orange lines) and He-rich white dwarf interior models with $\Teff=4000$\,K and 5000\,K (right panel, red curves). All models shown have a typical white dwarf surface gravity of $10^8\,$cm\,s$^{-2}$.  The models stay clear of the first order phase transitions but those with $\Teff \lesssim 6000\,$K cross the regime of pressure ionization (right panel) at depths well below the atmosphere. 1\,TPa = 10\,Mbar = $10^{13}$\,erg\,cm$^{-3}$.}
\label{fig:He_phase}
\end{center}
\end{figure}

Based on those high densities and pressures, the intriguing possibility that phase transitions in He at high pressures occur in white dwarfs was explored. Two recent ab initio calculations pertain to this question. One is the insulator-metal transition of solid He \citep{monserrat2014}, which occurs around 35\,TPa, the other is the melting curve of atomic He \cite{preising2019}. However, the improved ionization model of Ref.~\cite{kowalski2007}, which also includes He--$e$ and He--He$^+$ interactions --- thus providing a more complete description of the ionization equilibrium of He by introducing additional contributions to $\mu_{\sss\rm He}^{nid}$ and additional charged species --- predicts an increase of the ionization fraction and of the He$^-$ free-free opacity of about 3 orders of magnitude. Consequently, current pure He atmosphere models have maximum densities of $\sim 1$\,g\,cm$^{-3}$ and barely reach 100\,GPa. As a result, the models do not approach the melting and metallization coexistence curves and remain well into the fluid/plasma regime; no phase transition is expected (Figure \ref{fig:He_phase}).

The sensitivity of the atmospheric profile to the very small ionization fraction illustrates the importance of accurately modeling the onset of pressure ionization of He in cool atmospheres. We stress that the ionization model of Ref.~\cite{kowalski2007}, which represents the state of the art in the context of white dwarf atmosphere models, remains provisional. An indirect validation of the ionization model of He might be possible by measuring the opacity of He compressed dynamically or in a laser-heated diamond anvil cell (Section~\ref{sec:exp_opacities}).

A third set of simulations focused on the transition of He from an insulating atomic fluid to a conducting, fully ionized plasma \citep{preising2020}. This recent work found that the EOS, the conductivity and the ionization fraction all vary smoothly as density is increased at fixed temperature. A useful definition of pressure ionization that is borrowed from condensed matter physics is the density at which the band gap between the valence and conduction electron bands closes, which corresponds to a vanishing ionization energy. Figure \ref{fig:He_phase} shows the calculated locus of $(\rho,T)$ points of the band gap closure of He. As we saw in Figure \ref{fig:rhoT_DB}, the coolest He-rich white dwarf models cross the regime of pressure ionization but at a depth well below that of the atmosphere. This careful study of pressure ionization provides a benchmark for tables of the EOS and transport properties of He in the regime of 1--20\,g\,cm$^{-3}$ and 1--5\,eV. Note that the method used is not suited for quantifying the very low ionization fraction in cool He-rich atmospheres.

\subsection{Theoretical atomic line profiles}
\label{sec:line_prof}
\begin{figure}
\begin{center}
\includegraphics[width=0.6\columnwidth]{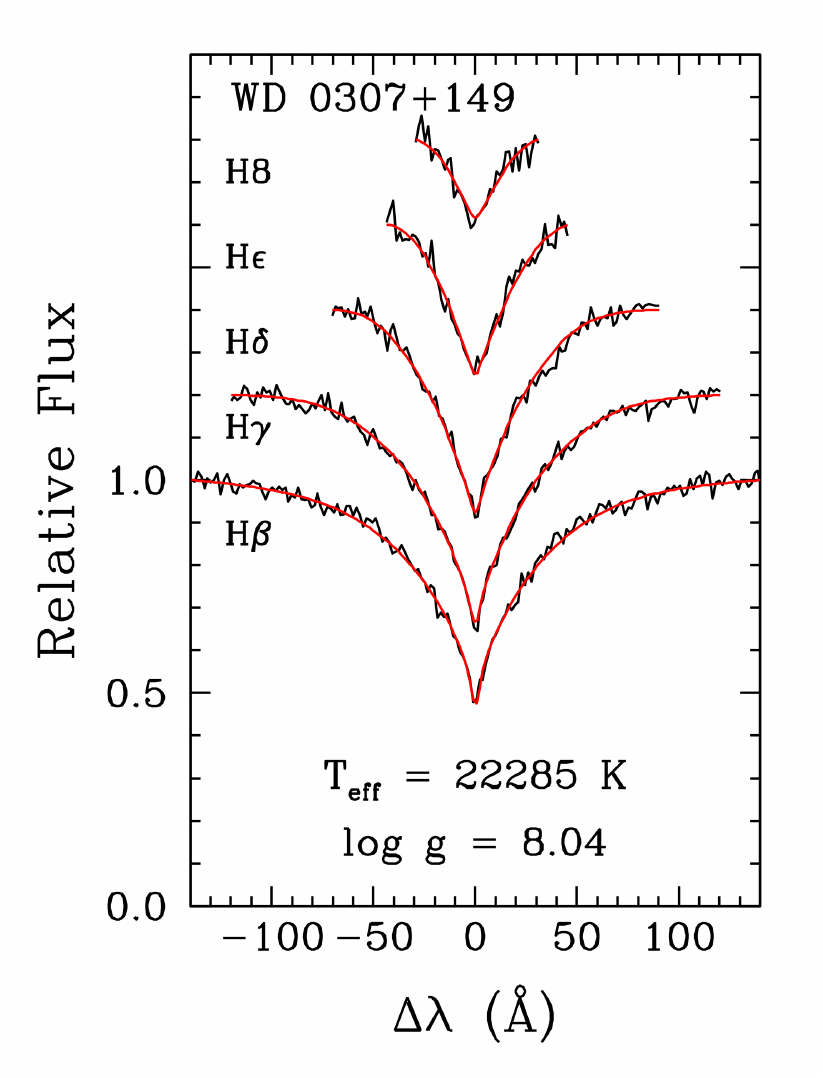}
\caption{Example of the spectroscopic method applied to fitting the Balmer lines of the DA white dwarf WD 0307+149. Data are shown in black and the best fitting model (with $\Teff=22\,285$\,K and $\log g ({\rm cm}\,{\rm s}^{-2})=8.04$) in red. The five individual absorption lines shown have been extracted from a continuous optical spectrum. Figure courtesy of P. Bergeron.}
\label{fig:sp_method}
\end{center}
\end{figure}

The determination of the basic physical characteristics of white dwarfs, such as their surface temperature, gravity, and composition, from which their mass,
radius and age are derived, is usually based on a detailed analysis of their spectra. The method of fitting the absorption lines developed three decades ago \citep{bergeron1992} has become a work horse
in the field and has been applied to over 18\,000 white dwarfs (Figure \ref{fig:sp_method}). The success of this spectroscopic  method depends on the reliability of the modeled line profiles, primarily 
in the UV and optical. This puts stringent demands on the modeling of pressure broadening, particularly of H and He lines, the two most common 
elements seen in white dwarf spectra, and on the atomic physics of dilute and weakly non-ideal plasmas. In the fully or partially ionized plasmas typical of
white dwarf atmospheres, Stark broadening dominates but in the coolest white dwarfs broadening by collisions with atoms (H or He) and even molecules (H$_2$) takes precedence.
Because of the relatively simple atomic structure of H and He, they have been studied extensively both theoretically and experimentally. 

Note that natural and thermal (Doppler) broadening are generally negligible in white dwarf atmospheres and pressure broadening dominates. Natural line broadening is due to the uncertainty principle, $\Delta E \tau \sim \hbar$, where $\tau$ is the atomic state lifetime. Typically, we have $\tau \sim 10^{-8}\,{\rm s}$ and
\begin{equation}
\frac{\Delta E}{E} = \frac{\Delta \nu}{\nu} = \frac{\hbar}{E \tau} \simeq 10^{-8}~.
\end{equation}
for a transition at $5000\,$\AA. Thermal broadening is given by
\begin{equation}
    \frac{\Delta \nu}{\nu} = \frac{v_{\rm thermal}}{c} = \frac{\sqrt{2 k_B T /m}}{c} \simeq 10^{-4}
\end{equation}
for a H gas at $20{\,}000$\,K. From those estimates, we see that natural and thermal broadening do not broaden spectral lines by more than $1\,$\AA, a much smaller value than the line widths observed in white dwarfs (Figure~\ref{fig:sp_method}).

\subsubsection{Broadening by collisions with charged particles}
\label{sec:line_prof_stark}
Interactions between a radiating atom or ion with ions and electrons in the surrounding plasma lead to Stark broadening of spectral lines. Even after decades of studies of atomic line broadening in plasmas, the theory, as implemented in several line shape codes, remains unsatisfactory and a work in progress \cite{stambulchik2019}. The theory is complex, which motivates several approximations of unknown accuracy that are variously implemented in codes \citep{alexiou2018, gomez2021}. This is of concern as Stark broadening is used as a diagnostic of conditions in laboratory plasmas and various codes/models lead to different inferred values of the electron density. The problem is compounded by the very limited amount of experimental data that can benchmark the theory and even fewer that have independent diagnostics of the temperature and electron density (Section \ref{sec:line_prof_exp}).

Applications of the unified Stark broadening theory \cite{vcs71} gives good fits of the spectra of DA white dwarfs (Figure \ref{fig:sp_method}) but fits of individual Balmer lines 
leads to systematic differences in the derived stellar parameters. Closer attention to the perturbed population of the H atom levels in a slightly 
non-ideal gas regime partly resolved this internal inconsistency \citep{tremblay2009}, yet resulted in a white dwarf mass distribution that is dependent on $T_{\rm eff}$, 
which is unreasonable on astrophysical grounds \citep{giammichele2012}. Modeling the line formation through an inhomogeneous convective atmosphere (Section \ref{sec:rhd}), rather 
than under the standard assumption of a one-dimensional, vertically stratified atmosphere model, appears to have largely resolved this issue \citep{tremblay2013}. 
Nonetheless, this sparked a renewed interest in the experimental measurement and the theory of line profiles of H and He relevant to white dwarfs that revealed some surprises.

Recent developments in the theory of Stark broadening with a view toward applications to denser plasmas and white dwarfs started with an investigation of the common dipole interaction approximation \citep{gomez2016}. The Coulomb potential energy between a perturber particle (ion or electron) and the radiator's nucleus and electrons can be expanded as a series of multipole interaction terms, which is usually truncated after the second (dipole) term. By evaluating higher order terms and treating the ions and electrons on the same footing, it was found that the dipole approximation is inaccurate and that the quadrupole term is important to match the experimental line profile of H$\beta$ at $n_e \sim 10^{18}$\,cm$^{-3}$ and $T=1$\,eV \cite{gomez2016}. The line profile is converged to $< 1$\% when including the octupole term. When using H$\beta$ as a diagnostic of the electron density under those conditions, the quadrupole moment brings a 12\% correction to $n_e$. 

A series of improvements followed. The application of a different numerical method to solve the Hartree--Fock equations improved the accuracy of the wave functions of the bound and the perturber electrons used to calculate the transition probability \citep{gomez2018b}. The inclusion of correlations between the radiator and the plasma in the relaxation theory of electron broadening \citep{gomez2018a} resulted in a considerable improvement in the agreement with the line shift for $n_e = 2$--$10 \times 10^{16}$\,cm$^{-3}$. This effort has culminated in a systematic study \citep{gomez2021} of the three main approximations in Stark line shape modeling: 1) The second order expansion of the broadening operator (T-matrix), 2) The dipole approximation of the radiator-plasma interaction (see above), and 3) The classical treatment of the perturber electron. Various combinations of these approximations have been used in previous work but this is the first to have lifted all three. Besides allowing a systematic assessment of the accuracy of each approximation, the application of this new model to the Lyman $\alpha$ and Lyman $\beta$ lines in a hydrogen plasma ($T= 1$--7\,eV, $n_e = 10^{18}$\,cm$^{-3}$) shows significant differences with the model widely used in white dwarf atmospheres \citep{vcs71} and the more recent work removing the second order approximation \citep{gomez2016}. These recent improvements in Stark line shape theory have proved very valuable in the analysis of the He$\beta$ line in a  highly ionized, solid density Ti plasma ($T_{\rm ion} \approx 600$\,eV, $n_e \approx 1.1 \times 10^{24}$\,cm$^{-3}$) \citep{kraus2021}. This systematic progress in the theory and calculations of Stark line profiles in weakly coupled plasmas is an important development in plasma physics\footnote{Ref. \cite{gomez2022} gives a comprehensive introduction to the current state of the theory.}. A preliminary implementation of the latest line profiles for H$\alpha$ through H$\gamma$ shows very little change in the synthetic spectra of DA white dwarfs but they are found to be sensitive to the treatment of the occupation probability and its effect on the pseudo-continuum opacity \citep{cho2022}. The impact on the analysis of white dwarf spectra has not yet been assessed.

The more complex case of He lines in the optical was recently revisited \citep{tremblay2020}. For over two decades, the Stark-broadened He line profiles used in white dwarf atmosphere codes have been those of Ref. \citep{beauchamp1997}, who applied the semi-classical (and semi-analytical) theory of Stark broadening to 21 lines in the optical. This standard approach to Stark broadening relies, among other approximations, on the quasi-static approximation for the ions and the impact approximation for the electrons. The latter was complemented with the so-called "one electron" approximation to improve the transition between the impact approximation for the electrons, which is valid in the line wings, and the quasi-static approximation for the electrons in the line core. By applying a simulation method to model the collisions between the radiating atom and the perturbers (ions and electrons, treated as classical particles), it becomes possible to do away with the quasi-static approximation for the ions and include the dynamics of ion-radiator collisions on the same footing as electron-radiator collisions. Weak and strong collisions are also treated without distinction, resulting in a naturally self-consistent description of the line core and line wings \citep{tremblay2020}.  This simulation approach is similar to that used for modeling Stark broadening of H lines \citep{cho2022, gomez2016} although Ref. \citep{tremblay2020} retained many common approximations in the theory of Stark line broadening, such as the dipole interaction between radiator and perturber. Six optical lines of neutral He were modeled under a range of conditions found in white dwarf atmospheres. As was found for the lines of H, the ion dynamics lead to variations from the profiles of Ref. \citep{beauchamp1997}, particularly near the line core and the forbidden components, but the effect is very small. Inclusion of the new profiles for the 4471\,{\rm \AA} and 4922\,{\rm \AA}\ He lines in atmosphere models shows minimal deviations from models spectra  based on the standard profiles. It will be interesting to see whether increasing the level of physics fidelity by lifting other important approximations --- as was done for hydrogen --- will lead to significant effects in He lines. As is the case for the new  H Stark line profiles, additional calculations are needed to quantify the impact of this work on the analysis of white dwarf spectra.

\subsubsection{Broadening by collisions with neutral particles: Cool white dwarfs}
\label{sec:line_prof_cool}

In cool stellar atmospheres, the broadening of atomic absorption lines is dominated by collisions between the radiating atom and perturbing atoms or molecules (predominantly H, He, and H$_2$). This broadening mechanism can usually be described using a simple Lorentzian profile. This assumption is well justified in the low-density limit ($\rho \lesssim 10^{-4}\,{\rm g}\,{\rm cm}^{-3}$), where the impact approximation can be made, i.e., when the duration of a collisions is small compared to the time elapsed between collisions. However, cool white dwarf atmospheres (especially those dominated by He) can reach much higher densities (Figure~\ref{fig:atmprof}). Under those conditions, the impact approximation breaks down, and a more sophisticated approach is required to describe the collisionally broadened lines. In this regime, white dwarfs exhibit severely distorted and shifted absorption lines, and Lorentzian profiles are a very poor description of the observations. This is particularly problematic for cool DZ white dwarfs, where an accurate modeling of the spectral lines of the heavier elements is needed to reliably extract the composition of the accreted exoplanetary material.

The so-called unified line broadening theory has been developed to describe collisionally broadened line profiles under those challenging conditions. It allows one to compute the complete line profile (from the core to the far wings) of an atom interacting with other atoms or molecules in its environment, and it remains valid up to the densities reached in the line-forming regions of most cool He-atmosphere white dwarfs. This theory is described at length in Refs.~\citep{allard1982,allard1999}. It requires two important inputs: a potential energy curve and a transition dipole moment curve of the perturbing atom or molecule interacting with the radiating atom in its relevant states (i.e., the lower and upper states involved in the atomic transition being modeled).

\begin{figure}
\begin{center}
\includegraphics[width=0.8\columnwidth]{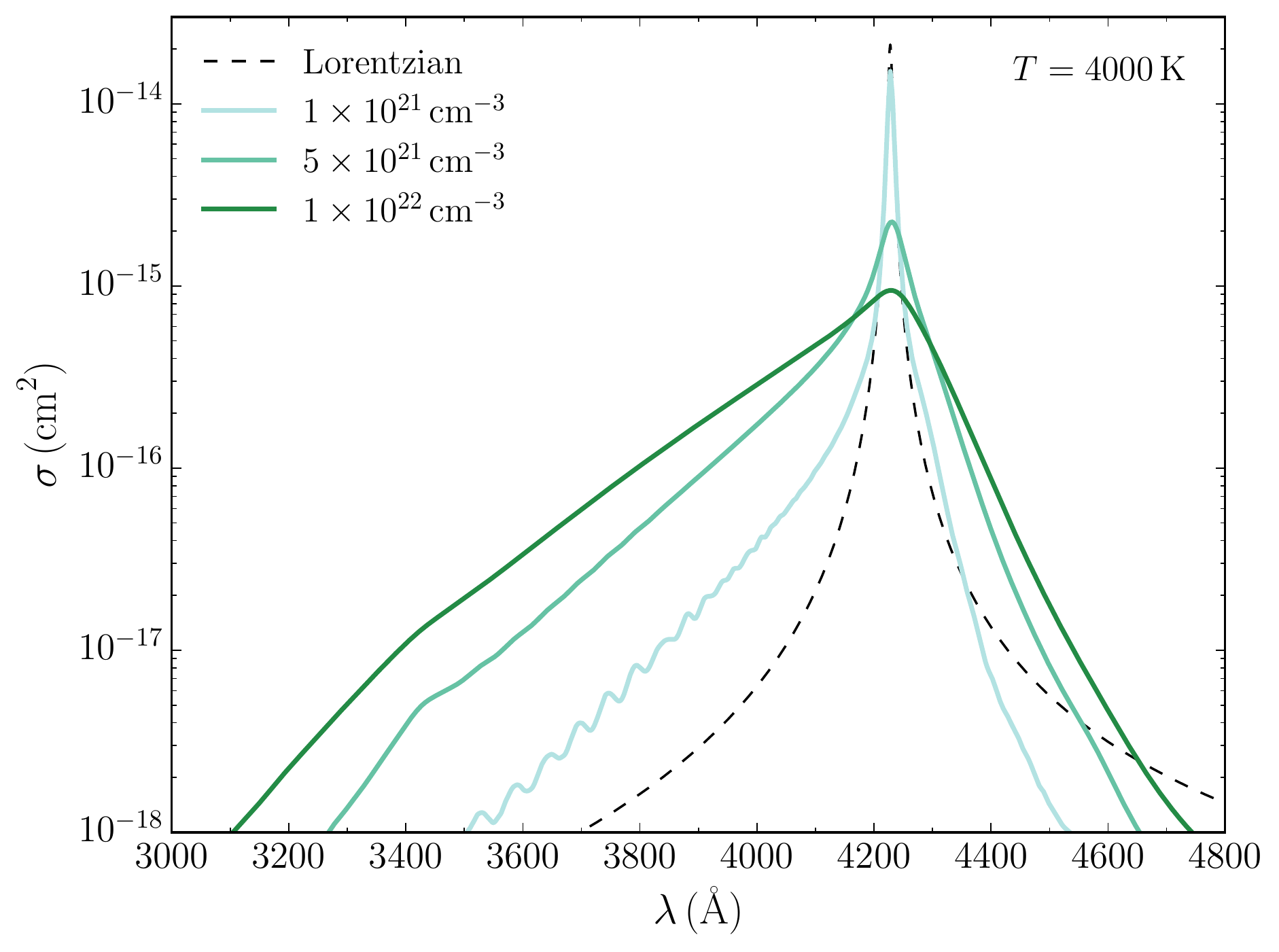}
\caption{Absorption cross section of the Ca I 4226\,{\AA} line perturbed by neutral He atoms. The Lorentzian profile is compared to profiles obtained with the unified line broadening theory for different He densities relevant to the modeling of cool DZ white dwarfs \citep{blouin2019ca}. \label{fig:ca4226}}
\end{center}
\end{figure}

The unified line broadening theory has been very successful at modeling absorption lines in cool DZ white dwarfs \citep{allard2016,hollands2017,allard2018,blouin2019na,blouin2019ca,coutu2019,kawka2021}, and its use is crucial to reach sound conclusions on the composition of rocky bodies accreted by those stars \citep{turner2020,blouin2020mg}. Despite its success and importance, this theory has so far only been applied to a small fraction of all the atomic transitions observed in cool white dwarfs. Most transitions are still modeled using Lorentzian profiles even in a regime where this is a very poor approximation (Figure~\ref{fig:ca4226}). The chief reason for this state of affairs is the lack of appropriate potential energy and transition dipole moment calculations, which are obtained using high-accuracy quantum chemistry methods. Potential energy and transition moment curves for more transitions observed in cool white dwarfs would solve this problem and allow for a more robust determination of the composition of the very old planetary debris accreted by those stars. All C, Fe, Ni, Ti, Cr, K, and Li transitions observed in cool white dwarfs at visible and UV wavelengths are currently modeled with Lorentzian profiles and are therefore all good candidates for future work.

One caveat of the unified line broadening theory is that it is based on potential energy and transition moment curves obtained in the infinite dilution limit (i.e., just one perturber interacting with the radiating atom). At the high densities characterizing the coolest DZ white dwarfs, this approximation could become problematic as on average more than one perturber may be inside the interaction volume of the radiating atom and the interactions may not be linearly additive. This issue becomes particularly important when considering spectral lines produced by excited atomic states that are characterized by a larger spatial extent.

\subsubsection{Broadening by collisions with neutral particles: DB white dwarfs}
\label{sec:line_prof_DB}
In the He-rich atmospheres of DB white dwarfs, broadening of neutral He lines by collisions with neutral He atoms becomes significant below $T_{\rm eff}=16\,000\,$K. Even though the density remains low near the photosphere ($n_{\sss\rm He} < 10^{21}\,{\rm cm}^{-3}$) and Lorentz line profiles are a good approximation to the line profile, the line width from collisions with neutral He atoms is poorly described in atmosphere models. This results in large uncertainties, increased scatter, and systematic deviations in the masses inferred from fitting DB white dwarf spectra for stars with $T_{\rm eff} \sim 11\,000$ -- $16\,000\,$K  \citep{gbb2019a,cukanovaite2021}. The models use a combination of comparatively simple line broadening models \citep{unsold1955, deridder1974, deridder1976} and ad hoc prescriptions. The widths based on the more advanced model of Ref.~\cite{leo1995} are limited to $T \le 300\,$K and not applicable to white dwarf atmospheres. Finally, the unified line broadening theory has been applied to broadening collisions with neutral He atoms but only for transitions in the UV and far red part of the spectrum \citep{allard2013} that are difficult to observe with ground based telescopes, if at all possible. There is a strong need for reliable calculations of the line widths of He broadened by collisions with He (and H) atoms for the transitions seen in the visible part of the spectrum.

\subsection{Molecular opacities}
\label{sec:mol_opac}
\subsubsection{C$_2$ absorption bands}
\label{sec:swan}
\begin{figure}
\begin{center}
\includegraphics[width=0.8\columnwidth]{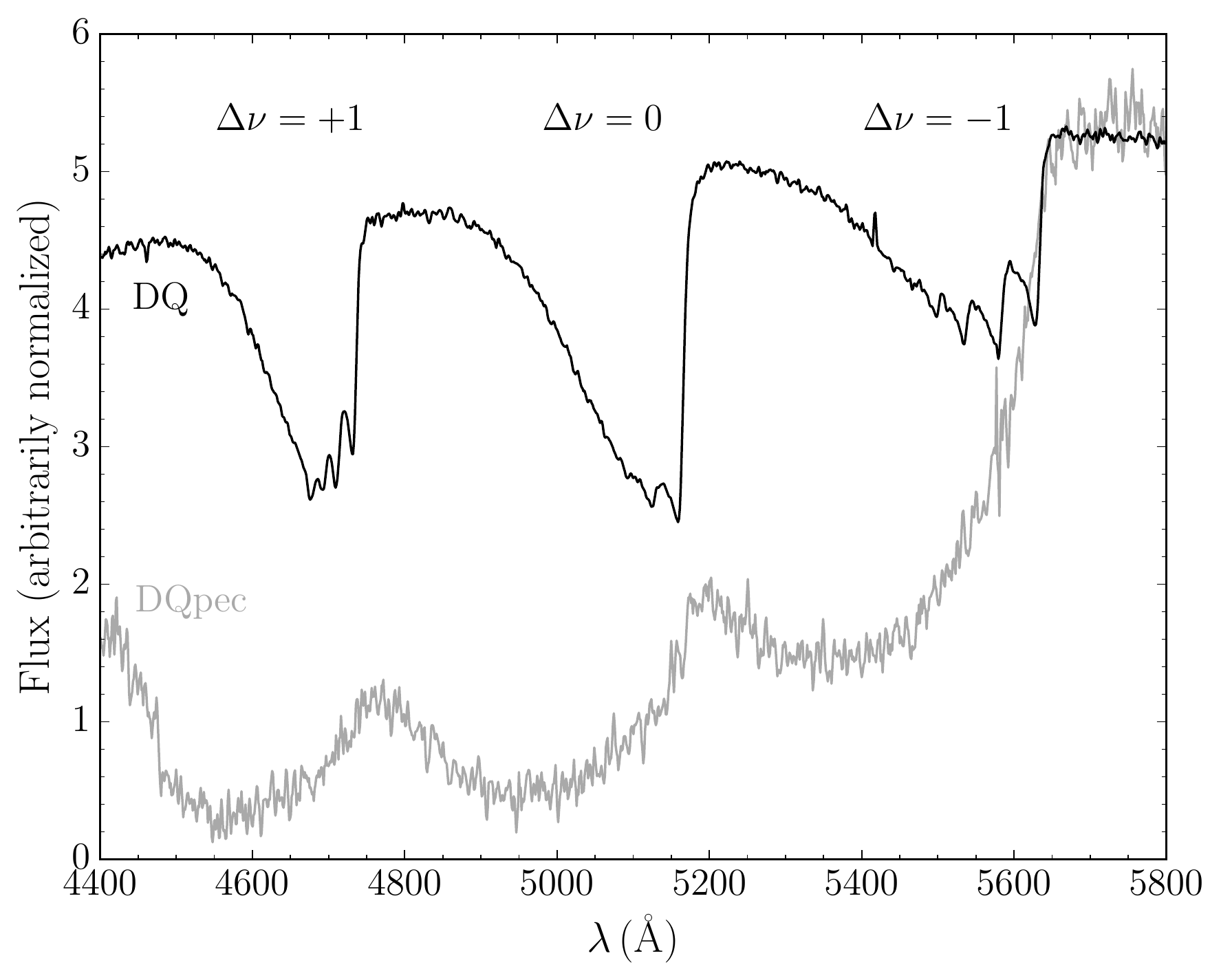}
\caption{Observed spectra of the DQ white dwarf BPM~27606 (black) showing the C$_2$ Swan bands and of the DQpec white dwarf LP~315--42 (gray) where those bands are severely distorted. The spectra were taken from \cite{bergeron2001} and the Sloan Digital Sky Survey (SDSS) archive, respectively.} \label{fig:swan}
\end{center}
\end{figure}

In cool white dwarf atmospheres, the formation of diatomic molecules can lead to observable molecular absorption bands in the optical spectrum. In general, those molecular bands are the result of a combination of electronic, vibrational, and rotational transitions. Bands from a few molecular species (e.g., CH, MgH) have been identified in white dwarfs but the most prominent features are by far the C$_2$ Swan bands detected in cool DQ white dwarfs, which have atmospheres of He with traces of C. The C$_2$ Swan bands are caused by the $a^3 \Pi_u \rightarrow d^3 \Pi_g$ electronic transition and hundreds of different vibrational and rotational transitions. Figure~\ref{fig:swan} shows a typical cool DQ spectrum (in black), where the different vibrational bands are identified; the rotational transitions correspond to the finer structure within each of those vibrational bands.

Current C$_2$ line lists allow model atmospheres to precisely reproduce the Swan bands observed in most DQ white dwarfs \citep{blouin2019,blouin2019dq}. However, at very low temperatures ($T_{\rm eff} \lesssim 6000\,{\rm K}$), the C$_2$ Swan bands are shifted and distorted compared to the predictions based on standard line lists \citep{bergeron1994,schmidt1995,hall2008}. Such objects are known as peculiar DQ, or DQpec, white dwarfs (an example is shown in gray in Figure~\ref{fig:swan}). The cause of the DQ$\rightarrow$DQpec transition appears to be a density-driven shift of the electronic levels of the C$_2$ molecule under the liquid-like densities that characterize DQpec white dwarfs. This interpretation is supported by DFT calculations that reveal a monotonic increase of the separation between the $a^3 \Pi_u$ and $d^3 \Pi_g$ electronic levels with respect to the He density \citep{kowalski2010}. This results in a blue shift of the absorption bands. Once integrated over the densities of the atmospheric layers along our line of sight, this effect not only shifts the Swan bands, but also distorts them as indicated by the observations (Figure~\ref{fig:swan}).

While existing DFT calculations can qualitatively explain the DQpec phenomenon, the electronic energy transition shift is one order of magnitude too strong compared to what is needed to reproduce DQpec spectra \citep{kowalski2010,blouin2019dq}. There are two possible explanations for this tension: (1) model atmospheres overestimate the atmospheric density of DQpec white dwarfs by an order of magnitude or (2) current DFT-based calculations of the C$_2$ energy spectrum in dense He are inadequate. The first explanation appears unlikely since atmosphere models reproduce precisely the density-sensitive features of cool DZ white dwarfs in the same physical regime \citep{blouin2018a,blouin2018b,blouin2019,blouin2019ca,blouin2019na} and since the presence of H (which reduces the atmospheric density) is well constrained by the absence or presence of a CH molecular band \cite{blouin2019}. In practice, the distortion of Swan bands in DQpec models is included by simply linearly shifting the reference low-density C$_2$ absorption spectrum. This approximation is most likely too crude.  For instance, the distortion of the electronic energy spectrum should in turn affect the vibrational energies. More work on the absorption of C$_2$ in dense He is needed to solve this problem and to enable the reliable modeling of those peculiar objects.

Another related question in need of further theoretical work is that of the effect of strong magnetic fields on C$_2$ Swan bands. Fields of the order of $100\,{\rm MG}$ could provide another mechanism to distort Swan bands in white dwarfs \citep{bues1999}, possibly explaining why some DQpec white dwarfs exist even in a regime where the photosphere is not particularly dense \citep{blouin2019dq}. This effect remains poorly quantified.

\subsubsection{Collision-induced absorption}
\label{sec:CIA}
In a dilute gas, the H$_2$ molecule has no dipole moment and it produces only very weak quadrupole absorption features that are not detected in white dwarfs. At higher densities however, the H$_2$ molecules undergo frequent collisions with other particles (H$_2$, H, and He in white dwarf atmospheres) that induce temporary dipole moments and lead to ``collision-induced absorption'' (CIA) \citep{frommhold1993}. In cool white dwarf atmospheres, this can give rise to broad and strong absorption features in the infrared \citep{lenzuni1991,bergeron1995,saumon1999}. With CIA being driven by collisions in the gas, it becomes stronger at increasing densities, and therefore it increases with decreasing effective temperature (Figure~\ref{fig:atmprof}). In fact, in very cool white dwarfs, CIA is so strong at near infrared wavelengths that it starts shifting the peak of the spectral energy distribution into the visible with decreasing temperature (at odds with the expectation from a simple blackbody spectrum) \cite{hansen1998,kilic2020}. White dwarfs that are past this turn-off point (i.e., when they start becoming bluer instead of redder as they age) are known as IR-faint (or ultracool) white dwarfs. These may be the oldest white dwarfs of our Galaxy, but their properties remain very uncertain \citep{bergeron2002,kilic2009,gianninas2015,kilic2020,lam2020,bergeron2022,elms2022}.

\begin{figure}
\begin{center}
\includegraphics[width=0.9\columnwidth]{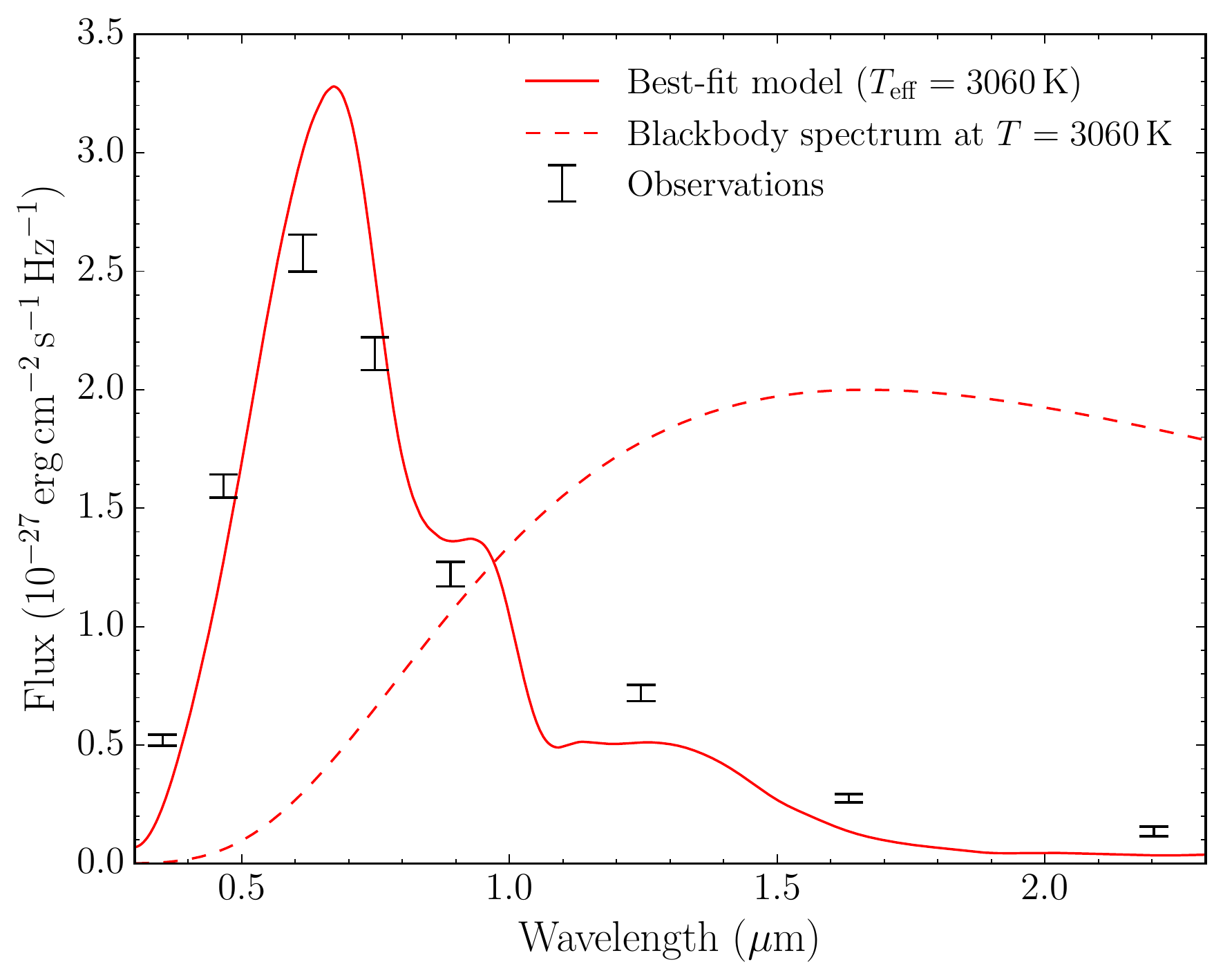}
\caption{Best-fit model of the IR-faint white dwarf LHS~3250 \cite{gianninas2015}. The temperature, mass, and atmospheric composition of the model atmosphere were adjusted to obtain the best agreement possible with the observations (black error bars), but severe discrepancies remain. In most bandpasses, the discrepancies between the model and the observations are several times larger than the $1\sigma$ uncertainties. For non-IR-faint white dwarfs, a reduced chi-square $\chi_r^2$ for such fits of $\sim 1$ is the norm; here $\chi_r^2=75$. Note that a blackbody spectrum at the best-fit temperature ($T_{\rm eff}=3060\,{\rm K}$) would peak at $1.7\,\mu$m. Here, the model spectrum peaks at much shorter wavelengths due to strong H$_2$ CIA in the near-infrared. Note that Ref.~\cite{bergeron2022} recently obtained better agreement with the observations, and found a much warmer $T_{\rm eff}$ of 4990\,K for this star. \label{fig:ultracool}}
\end{center}
\end{figure}

Collision-induced absorption being a collisional process, its absorption coefficient can be expressed as a density expansion
\begin{equation}
    \alpha(\nu) = a_1(\nu) n + a_2(\nu) n^2 + a_3(\nu) n^3 + ...
\end{equation}
where $n$ is the number density of the CIA absorber (we assume a single species here for simplicity) and the $a_i(\nu)$ are the frequency-dependent absorption coefficients involving $i$ particles. By definition, $a_1(\nu)=0$ for a CIA process.
Throughout the years, the accuracy of the theoretical CIA opacities used in model atmospheres has steadily increased \citep{borysow1997,jorgensen2000,abel2012}. However, those opacity calculations are normally based on a molecular scattering approach where only binary collisions (e.g., between H$_2$ and He) and only $a_2(\nu)$ is considered. This is a valid approximation for cool H-dominated atmospheres (Figure~\ref{fig:atmprof}), but it breaks down at the higher densities that characterize He-dominated atmospheres, where many-body collisions become important.\footnote{Due to their high densities, even a trace of H$_2$ in cool He-dominated atmospheres is sufficient to generate strong CIA features.} The higher order coefficients are extremely difficult to evaluate with the techniques used to compute $a_2(\nu)$. This problem has been addressed using DFT-MD simulations, and many-body collisions were indeed found to significantly alter the H$_2$--He CIA opacities for conditions relevant to cool He-atmosphere white dwarfs \citep{blouin2017} (at least at $\lambda \gtrsim 1\,\mu$m, CIA at shorter wavelengths has yet to be revisited). Similarly, it was shown that CIA can occur in cool He atmospheres even in the absence of H$_2$ through three-body He--He--He collisions ($a_2(\nu)=0$ due to the symmetry of the He--He collision) \citep{kowalski2014}. Despite those advances, model atmospheres generally struggle to reproduce the observed spectral energy distributions of IR-faint white dwarfs (Figure~\ref{fig:ultracool}). Recently, Ref.~\cite{bergeron2022} presented model atmosphere calculations that markedly improved the agreement with observations. However, to achieve this result, they found that they had to resort to older CIA opacity tables \cite{jorgensen2000}. Whether this is due to inadequacies in the current CIA opacities or to problems with another microphysical ingredient remains an open question.

\subsection{Phase diagrams of ultra-dense plasmas}
\label{sec:phase_diag}
Modeling the fractionation (or phase separation) that takes place in crystallizing white dwarfs (Section~\ref{sec3:crystallization}) requires accurate phase diagrams of multi-component plasmas. Those phase diagrams yield the crystallization temperature of a given mixture and quantify the composition change between the coexisting solid and liquid phases upon crystallization. The problem of calculating those phase diagrams was first addressed in the 1980s and 1990s \citep{stevenson1980,mochkovitch1983,segretain1993,segretain1994}, but here we will focus on three more modern techniques that have been used during the last decade.

\subsubsection{Phase diagram calculation techniques}
\label{sec:dphase_methods}
A first phase diagram calculation method is the semi-analytic approach described in Ref.~\citep{medin2010}, which is based on analytic fits to the free energies of multi-ionic mixtures obtained with Monte Carlo simulations \citep{ogata1993}. Analytic fits allow a rapid determination of the phase diagram and makes this method applicable to mixtures with a large number of ionic species. Recent applications of this approach to white dwarf interiors include a calculation of the C-O-Ne phase diagram \citep{caplan2020} and the study of the phase separation and settling of traces of actinides early in the crystallization phase \citep{horowitz2021}. Its main drawback is that it is only as accurate as the free energy fits on which it relies, and, unfortunately, those fits are unsatisfactory in several aspects. For instance, they do not include the effect of electron--ion screening (they are all based on the Monte Carlo or Molecular Dynamics simulations where the ions interact through bare Coulomb potentials); they implicitly assume that the vibration modes of the multi-component zero-temperature solid are identical to that of the one-component zero-temperature solid; and it is not always clear which functional form should be adopted for the free energy \citep{ogata1993,dewitt1996}. Another issue with this approach is that those analytic fits are based on linear mixing rules that often require knowledge of the free energy of a strongly super-cooled liquid or super-heated solid (this problem is particularly important for ionic mixtures with high charge ratios). Not only do current fits not extend into those regimes, but it is conceptually problematic to refer to a liquid one-component plasma at an extreme coupling regime that is well within the solid regime (e.g. $\Gamma=1000$). This issue was recently discussed in Ref.~\cite{jermyn2021}.

A more direct approach that circumvents the need for analytic fits to the free energy is two-phase Molecular Dynamics simulations. Typically, those simulations are initialized with a solid and a liquid phase in contact with each other. The ions are allowed to diffuse through the solid--liquid interface, and eventually a state of equilibrium is reached. The compositions of the two coexisting phases then directly yield a point on the liquid--solid coexistence line. This approach has been successfully applied to mixtures relevant to white dwarf crystallization \citep{horowitz2010,schneider2012,hughto2012}. Unfortunately, the 
``first-principles'' aspect of this method comes with a very steep computational cost. Molecular Dynamics simulations that include a solid--liquid interface must be performed on very large systems to mitigate finite-size effects. As a consequence, only a very partial sampling of a given phase diagram can be achieved, which makes this approach more suitable to verify the results produced by other, more efficient methods than to calculate the detailed phase diagrams implemented in white dwarf codes. Applying this technique to study the fractionation of minor trace species such as $^{22}$Ne in a C-O white dwarf, is also challenging since a very large total number of particles is needed to include a reasonable number of ions of the trace species.

Finally, a third approach is to directly integrate the Clapeyron relation that describes the shape of the liquid--solid phase boundary. When applying the Gibbs--Duhem relation 
\begin{equation}
   s dT - v dP + \sum_i x_i d\mu_i = 0~,
\end{equation}
where $s$ and $v$ are the entropy and volume per particle, respectively, and $x_i$ is the number fraction of species $i$, to the coexisting solid and liquid phases of a pure substance, we obtain the well-known Clausius--Clapeyron relation for the coexistence curve
\begin{equation}
    \Bigg( \frac{dP}{dT}\Bigg)_{\rm coex} = \frac{\Delta h}{T \Delta v}~,
\end{equation}
where the derivative is along the coexistence curve and $\Delta h$ and $\Delta v$ are the changes in specific enthalpy and specific volume across the transition. For mixtures, the Gibbs--Duhem relation gives rise to several forms of Clausius--Clapeyron equations. For a two-component system, we obtain a differential equation giving the slope $dT/d\mu_2$ along the coexistence curve in terms of $\mu_2$,  $\Delta x_2$ and $\Delta h$ \citep{hitchcock1999}. Given a starting point on the coexistence curve (usually $x_2=0$ or 1), the coexistence curve $T(x_2)$ can be obtained by integration at a fixed pressure. The $(P,T,x_2)$ phase diagram of a binary mixture can be generated by repeating the integration over a range of values for the pressure.

This method has been used for some time in other domains \citep{kofke1993,hitchcock1999}, but was only recently adapted to electron--ion plasmas \citep{blouin21b}. Distinct Monte Carlo simulations for the two coexisting phases are used to evaluate the quantities needed to perform this integration. In the implementation of Ref.~\cite{blouin21b}, all calculations are performed at constant pressure (the appropriate choice for phase transitions in white dwarfs) and explicitly include the electron background (which is usually neglected with other methods where the constant-volume approximation is ordinarily made). Just like the Molecular Dynamics approach, the Clapeyron method avoids the need for thorny analytic fits of the free energy. But it is also much less computationally costly than the two-phase Molecular Dynamics technique. This is mainly due to the fact that the two phases are simulated separately. This means that there is no solid--liquid interface in the Monte Carlo simulations, thereby greatly reducing the number of particles needed to mitigate finite-size effects (from 10{,}000s of ions to just under 1000). This approach has therefore allowed the calculation of highly accurate and well-sampled phase diagrams for a variety of two- and three-component mixtures relevant to white dwarfs interiors \citep{blouin20,blouin21,blouin21c}.

\begin{figure}
\begin{center}
\includegraphics[width=0.9\columnwidth]{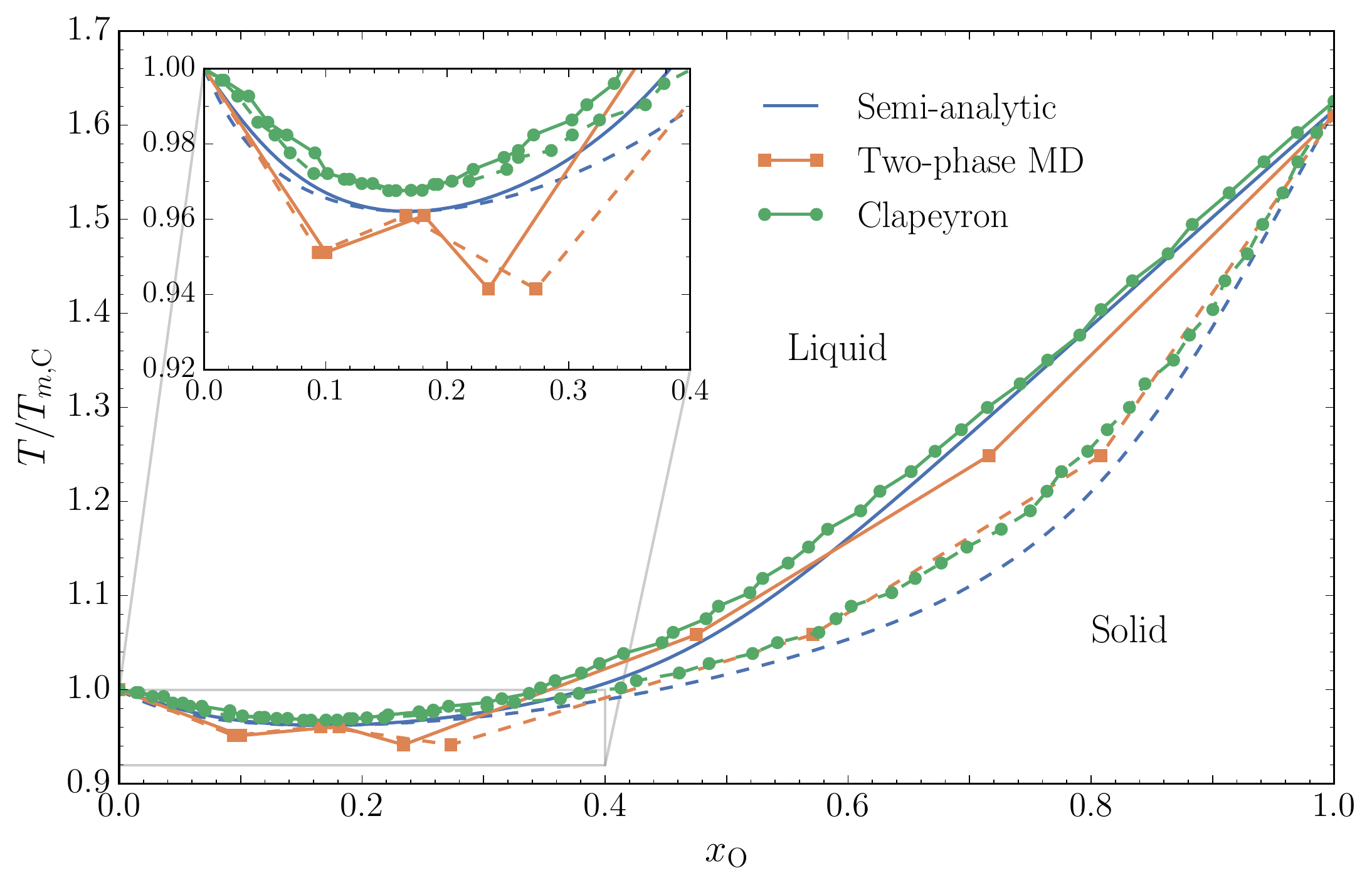}
\caption{Carbon-Oxygen phase diagram. The solid lines are the liquidus (above which the plasma is liquid), and the dashed lines are the solidus (below which the plasma forms a stable solid). The temperature is given in units of the melting temperature of a pure C plasma, and the horizontal axis gives the O number concentration (pure C at $x_{\rm\sss O}=0$ and pure O at $x_{\rm\sss O}=1$). The separation between the solidus and the liquidus at constant temperature gives the composition change upon solidification of the plasma. Note that this phase diagram is virtually independent of the pressure under white dwarf conditions because the pressure is entirely dominated by the degenerate electron gas and therefore decoupled from the ions. The results showed here are taken from Refs.~\cite{horowitz2010,medin2010,blouin20}.} \label{fig:CO}
\end{center}
\end{figure}

Figure~\ref{fig:CO} shows three versions of the C-O phase diagram, obtained using the three techniques just described. It is reassuring to see that these three totally independent approaches lead to very similar results. In particular, all three methods predict an azeotropic phase diagram with the azeotropic point at $x_{\rm\sss O} \approx 0.2$ (inset). Note that for this mixture, the semi-analytic approach is not expected to be affected by the super-cooled liquid (and super-heated solid) problem, due to the small charge ratio ($Z_{\rm O}/Z_{\rm C} = 4/3$) of the mixture. The coarseness of the Molecular Dynamics-based phase diagram makes it inappropriate for direct use in white dwarf codes, but is nevertheless useful to verify the results of the other methods.

\subsubsection{Some recent results: Crystallization of trace species}
\label{sec:trace_crystal}
The excellent agreement shown above for the two-component C-O phase diagram suggests that this problem is mostly solved. Accordingly, the most recent works on phase diagrams of white dwarf interiors have instead focused on the role played by trace species. Even if C and O largely dominate most white dwarf cores, the fractionation of traces species can play a disproportionate role on the evolution of the star. Three recent works demonstrate this point and show that there are still important physical processes to uncover by calculating phase diagrams for mixtures found in white dwarfs interiors.

{\bf $^{22}$Ne distillation} --- Ref.~\cite{blouin21} investigated the fractionation of $^{22}$Ne in C-O white dwarfs \footnote{\ $^{22}$Ne makes up $1-2\%$ of the mass of C-O cores.} and showed that the solid phase can be significantly depleted in $^{22}$Ne compared to the coexisting liquid. Because $^{22}$Ne has a higher mass per electron than $^{16}$O and $^{12}$C, the $^{22}$Ne-poor crystals are lighter than the liquid in which they form. This is expected to lead to a ``distillation" process \citep{isern1991} that can transport large quantities of $^{22}$Ne toward the central layers. This in turn liberates the gravitational energy stored in $^{22}$Ne and can lead to multi-Gyr cooling delays that appear to be needed to explain recent astronomical observations \citep{tremblay2019_crystal,cheng2019,kilic2020}.

{\bf Fe-rich crystallite formation} --- $^{56}$Fe is another neutron-rich trace species present in white dwarf cores (with a mass fraction of the order of 0.1\%). Ref.~\cite{caplan2021} found that $^{56}$Fe should separate into Fe-rich crystals before the rest of the ionic mixture freezes. Those heavier crystals should sink rapidly to the center of the white dwarf, thereby releasing gravitational energy and slowing down the cooling of the white dwarf by a few hundred million years \cite{salaris2021}.

{\bf Actinide fractionation} --- Ref.~\cite{horowitz2021} looked at the case of even lesser species, actinides. Because of their much higher ion charge, resulting in strong plasma coupling (Equation \ref{eq:Gamma}), actinides crystallize at temperatures well above that of C and O. They argue that when the white dwarf begins to crystallize, the solid phase could be so enriched in actinides that it could support a fission chain reaction, thereby potentially igniting C burning and leading to a supernova explosion. While this scenario needs to be supported by more calculations, the prospect of a completely new supernova mechanism is tantalizing.

\subsubsection{Remaining challenges}
\label{sec:dphase_challenges}
The three methods described above treat the ions as classical particles. This approximation is valid in the limit where the mean interparticle distance is larger than the thermal de Broglie wavelength of the ions, but it starts to break down in white dwarf cores (Figures \ref{fig:DA_panel3}a and \ref{fig:DB_panel1}b) \citep{baiko2019,baiko2021}. Future work should therefore investigate how quantum effects impact phase diagrams under white dwarf conditions. Very recently, Ref.~\citep{jermyn2021} included the quantum behavior of ions in their equation of state and find a very weak effect on the crystallization temperature, with $\Gamma$ decreasing from 175 to 168 for carbon. However, they could not study the impact of quantum effects on fractionation due to limitations in their EOS implementation. 

Another important approximation that is made by all three methods listed in Section~\ref{sec:dphase_methods} concerns the lattice structure of the solid phase. For a one-component plasma, it is well established that the most stable solid structure is a bcc lattice \citep{hamaguchi1997}. However, in multi-component plasmas, other more stable configurations may be possible \citep{ogata1993,igarashi2001,caplan2020b,kozhberov2021}. It remains unclear to what extent the bcc lattice is a good approximation of the real crystalline structure, especially for highly asymmetric mixtures. Moreover, the solid may well not even form a regular lattice structure but could instead form a more complex polycrystalline mixture.

Finally, reliably modeling the fractionation of high-$Z$ trace species remains challenging. Molecular Dynamics and Monte Carlo-based methods demand simulations with a very high total number of ions in order to include a reasonable number of ions of the high-$Z$ trace species, and the semi-analytic approach is precarious due to the limitations of the analytic free energy fits.

\subsection{Diffusion coefficients for trace species}
\label{sec:diffusion}

In most instances of diffusion in white dwarfs, we are concerned with the inter-diffusion in a binary mixture of ions. For a mixture with number concentrations $x_1$ and $x_2=1-x_1$, arbitrary electron degeneracy and plasma coupling, the equation for the relative velocity of the ion species $w_{12}$ due to diffusion is 
\begin{equation}
\begin{aligned}
	w_{12} = D_{12} (1+\gamma) \bigg[
	    & -\frac{d \ln x_2}{d r} + 
		\Big( \frac{A_1 Z_2- A_2 Z_1}{Z_1 + \gamma Z_2} \Big) \frac{m_0 g}{\kb T} +
		\Big( \frac{Z_2-Z_1}{Z_1+\gamma Z_2} \Big) \frac{d \ln P_i}{d r} \\
		&+ \alpha_{\sss T} \frac{d\ln T}{dr} \bigg]~,
	\label{eq:w12}
	\end{aligned}
\end{equation}
where $D_{12}$ is the coefficient of inter-diffusion, $A_i$ and $Z_i$ the atomic mass (in a.m.u.) 
and charge of ions of species $i$, $P_i$ the total {\it ionic} pressure, $m_0$ the atomic mass unit, $\gamma=x_2/x_1$, and
$\alpha_{\sss T}$ is the (dimensionless) thermal diffusion factor \citep{pelletier1986, bauer2019}. The first term on the right-hand side of Equation \ref{eq:w12} corresponds to ``ordinary'' chemical diffusion. For trace elements ($x_2 \ll x_1$), this term is negligible in white dwarfs \citep{fontaine1979}. The second and third terms describe barodiffusion, known as ``gravitational settling'' in stars, caused by the pressure gradient associated with the star's gravitational field and the induced electric field that maintains local charge neutrality for a mixture of charged species of different masses in a gravity field. This is generally the dominant term in white dwarf envelopes \citep{paquette86a}.  The last term is the contribution of thermal diffusion. The coefficient of thermal diffusion $D_{\sss T} = D_{12} \alpha_{\sss T}$ is determined by ion-ion and electron-ion collisions and can be written as $\alpha_{\sss\rm T}=\alpha_{12} + \alpha_{1e} + \alpha_{2e}$.  The contributions of $\alpha_{1e}$ and $\alpha_{2e}$ can be comparable or even much larger than the ion-ion term $\alpha_{12}$ \citep{paquette86a}.

When Equation \ref{eq:w12} is applied to a fully ionized plasma, the value of $Z_i$ is unambiguous. However, it is common for one of the elements to be partially ionized \citep{bauer2019}, in which case a separate equation is necessary for each charge state, treated as a separate species. An average diffusion velocity for a given element can be defined for such a multi-component diffusion problem but it is a common approximation to use a single diffusion equation for an element with an average ion charge instead \citep{dupuis1992, koester2009, bauer2019}. 

\subsubsection{Diffusion in the envelope}
\label{sec:diffusion_env}

In stellar astrophysics, the evaluation of diffusion coefficients is usually based on the expression derived from the solution of the Boltzmann kinetic equation \citep{burgers1969, chapman1970}. The resulting transport coefficients are expressed in terms of collision integrals involving scattering cross sections for isolated, binary collisions. In systems where particles interact with a pure Coulomb potential, these integrals diverge. In a real plasma however, the pair interactions are screened by correlations between particles, leading to a short-range effective pair potential and finite collision integrals. An
effective pair potential is appropriate if the dynamics of charged particles can be described as a succession of binary
collisions in this effective potential, a condition that corresponds to a weakly coupled plasma ($\Gamma < 1$). The simplest form of effective interaction is a pure Coulomb potential with an upper radius cutoff, usually chosen to be the Debye screening length $\lambda_{\rm\sss D}$. A more physically correct description of the pair potential uses the statically screened Coulomb (or Yukawa) potential \citep{eliezer2002}
\begin{equation}
V_{\rm sc}(r) = \frac{Z_1 Z_2}{r} e^{-r/\lambda}~,
 \label{eq:yukawa}
\end{equation}
which is found to be applicable for $\Gamma < 0.3$.
The most common source of collision integrals used in white dwarf models are the fits of Ref.~\cite{paquette86b} that use the Yukawa effective potential with a screening length given by
\begin{equation}
  \lambda={\rm max}[ \lambda_{\rm\sss D}, a ]~,
  \label{eq:paquette_length}
\end{equation}
where $\lambda_{\sss D}$ is the Debye screening length and $a$ is the ion sphere radius. This choice provides a heuristic correction to the screening length that extends the applicability of the approach into the strong coupling regime. Ref.~\cite{stanton2016} improved on this calculation by accounting for electron degeneracy in $\lambda_{\rm\sss D}$ and substituting a smooth, physically motivated interpolation form for Equation \ref{eq:paquette_length}. Note that the models of Refs.~\cite{paquette86b} and \cite{stanton2016} require a separate input of the ion charges $Z_1$ and $Z_2$. In white dwarf envelopes, the dominant ions  (H and He) are usually fully ionized but trace species are typically low-Z ions ($Z \sim 6$ -- 26) that are partially ionized. The calculation of the charge state(s) of such ions in the warm dense matter regime is an ongoing challenge \citep{murillo2013}.

More recently, a rigorous extension of the Boltzmann kinetic equations into the moderate coupling regime, called the effective potential theory has been developed. It can be shown that the proper effective potential is the ion-ion potential of mean force $V_{\rm\sss MF}(r)$, defined as
\begin{equation}
     g(r) = \exp( {-V_{\rm\sss MF}(r)/\kb T} )~,
    \label{eq:Vmf}
\end{equation}
where $g(r)$ is the pair distribution function of the ions \citep{baalrud2019}. The potential of mean force includes the direct binary interaction between a pair of particles $V(r)$ and corrections that account for the presence of the surrounding correlated particles. This formalism is general and can be applied to the calculation of collision integrals for a system with any direct pair potential (e.g. Equation \ref{eq:yukawa}), as long as $g(r)$ can be calculated or experimentally measured. The effective potential theory has been extended to plasma mixtures \citep{beznogov2014, shaffer2017} and applied to thermal diffusion \citep{kagan2017}. It accounts for ion screening through $V_{\rm\sss MF}(r)$ while the electron screening is implicit in the direct pair potential $V(r)$, as the electron contribution to the screening length in Equation \ref{eq:yukawa} for example. The principal remaining difficulty is to develop a realistic model for electron screening in regimes where the Yukawa potential is not appropriate.

\begin{figure}
\begin{center}
\includegraphics[width=0.75\columnwidth]{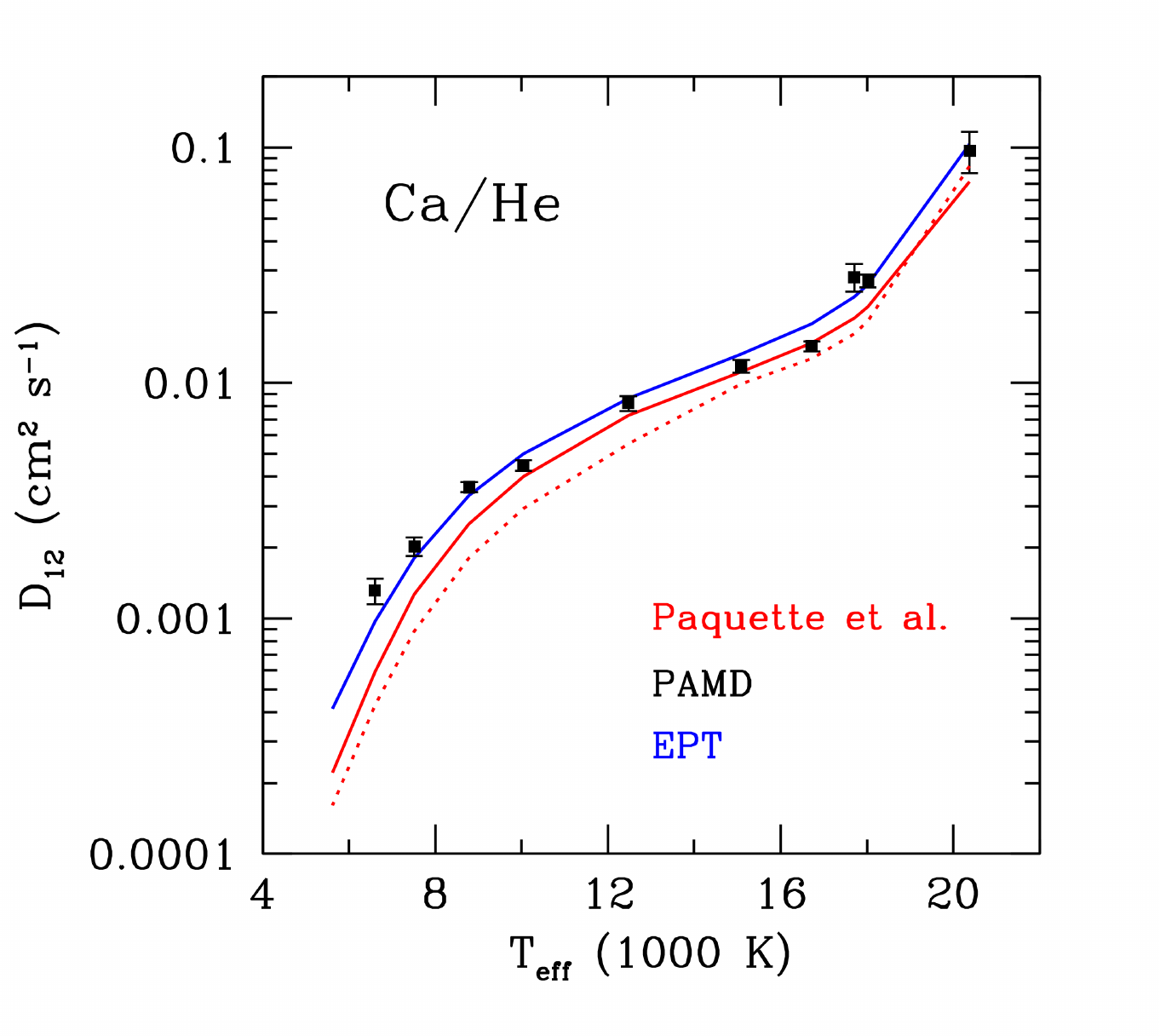}
\caption{Interdiffusion coefficient of a trace of Ca in He for the conditions at the bottom of the convection zone of He-rich white dwarf models as a function of the model $\Teff$. The $(\rho,T)$ conditions vary as a function of $\Teff$. The values from the Effective Potential Theory combined with the average atom model of the plasma are shown in blue (``EPT''). The black symbols (``PAMD'') show the values obtained with classical Molecular Dynamics simulations using the direct potential from the same average atom model. Ideally, these would agree perfectly with the EPT result. The differences and scatter in these data show the difficulty of obtaining converged molecular dynamics results for the diffusion of a trace species. The simpler model of Paquette et al. \cite{paquette86b} {\it with the same ion charges as the EPT and PAMD calculations} is shown with the dotted red line. Changing the ion charges to those from a simpler, somewhat heuristic ionization model \citep{fontaine2015} results in the solid red line. In this case, it fortuitously improves the agreement with the EPT calculation. Adapted from Figure 5 and Table 3 of Ref.~\cite{heinonen2020}.} 
\label{fig:D12}
\end{center}
\end{figure}

 Average atom models explicitly solve for the quantum mechanical states (bound and free) of the electrons surrounding a nucleus, including the effects of the surrounding plasma. While approximate, average atom models naturally account for pressure ionization and strongly non-linear screening (see \cite{callow2021} for a recent overview). They provide self-consistent ion charge, electron cloud spatial distribution and the ion-ion interaction potential.  They have been very useful for the description of states of warm dense matter, including the EOS and transport properties.  Recently, Ref.~\cite{heinonen2020} combined the potential of mean force extracted from an average atom model \citep{starrett2014} with the effective potential theory to compute interdiffusion coefficients and ionic thermal diffusion coefficients ($\alpha_{12}$) of Si and Ca under the conditions found in white dwarfs. In the weakly coupled regime, which is relevant to all H-rich white dwarfs  and those He-rich white dwarfs hotter than $\Teff \sim 14\,000$\,K, the results agree with those of Ref.~\cite{paquette86b} to within 25\%. In the cooler He-rich stars, however, the background He plasma becomes strongly coupled and the differences grow to a factor of 2.4, which is significant (Figure \ref{fig:D12}). These values are obtained when the ion charges calculated with the average atom model are used as input to the Ref.~\cite{paquette86b} formalism. With the charge state obtained with a more heuristic ionization model \citep{fontaine2015}, the coefficients of interdiffusion of Ref. \cite{paquette86a} can change by as much as 30\% across all coupling regimes. This shows that a realistic ionization model to compute diffusion coefficients is as important as using an advanced theory of transport coefficients \citep{bauer2019, grabowski2020}. This effect is compounded when the {\it ratio} of interdiffusion coefficients of different elements is the more relevant quantity, such as in the estimation of the elemental composition of the accreted material onto polluted DZ white dwarfs. It is also important when computing the upward diffusion tail of C at the C-O/He interface when modeling the dredge up of C observed in DQ and DQpec white dwarfs.

The state of the art in the calculation of transport properties are ab initio computer simulations of plasmas that consider only nuclei and quantum mechanical electrons. These methods involve the fewest number of approximations but are computationally very expensive. It is difficult to obtain converged values for interdiffusion coefficients when considering a trace species or if the mass ratio or charge ratio between the two species is large. At present, ab initio simulations are impractical to generate the large tables of diffusion coefficients for the many different elements of interest to model white dwarfs. 

To summarize, the combination of an average atom model with the Effective Potential Theory \citep{heinonen2020} is a significant advance in the calculation of interdiffusion coefficients compared to the approach of Ref.~\cite{paquette86b} and variations thereof (e.g. Ref.~\cite{stanton2016}).  These improvements matter primarily for the cooler He-rich white dwarfs polluted with low-Z elements (type DZ) and  those with traces of dredged-up C (spectral types DQ and DQpec). The contribution of ion-electron interactions to the coefficient of thermal diffusion ($\alpha_{1e}$ and $\alpha_{2e}$) remains to be evaluated with a comparable level of theory for quantum mechanical electrons (e.g. Ref.~\cite{rightley2021}). Benchmark ab initio calculations (e.g. with Density Functional Theory - Molecular Dynamics simulations) for a few representative cases would be very valuable to 
validate the more efficient but approximate models that can be applied realistically to the extensive computations of diffusion coefficients.

\subsubsection{Diffusion in the core}
Diffusion (or gravitational settling) of trace species in the core is also an important process. Neutron-rich species (with $A>2Z$) experience a net downward force in white dwarf cores and therefore tend to diffuse toward the centre ($w_{12}<0$ in Equation \ref{eq:w12}). In C-O white dwarfs, the most important species to undergo this process is $^{22}$Ne, which makes up $\sim 1-2$\% of the core mass. This transport releases a sizeable amount of gravitational energy, which slows down white dwarf cooling \citep{bildsten2001, garciaberro2008,althaus2010b,camisassa2016}. This effect must be considered in the building of white dwarf cooling sequences used for age-dating applications, and its efficiency depends on the relevant diffusion coefficients.

Gravitational settling of $^{22}$Ne in the core takes place in a fully ionized plasma, eliminating some of the complications discussed above for the diffusion of low-Z elements in the envelope. Many works have focused on determining self-diffusion coefficients in a one-component Yukawa plasma \citep{daligault2005,hughto2010,hughto2011,daligault2012,khrapak2013,caplan2021b} and this quantity is now well known. However, the inter-diffusion coefficients needed to model diffusion in the multi-component plasma that makes up the core are less well constrained. Linear mixing rules are often used, but their accuracy should be further validated.

\subsection{Thermal conductivity}
\label{sec:thermal_cond}

\begin{figure}
\begin{center}
\includegraphics[height=5.5cm]{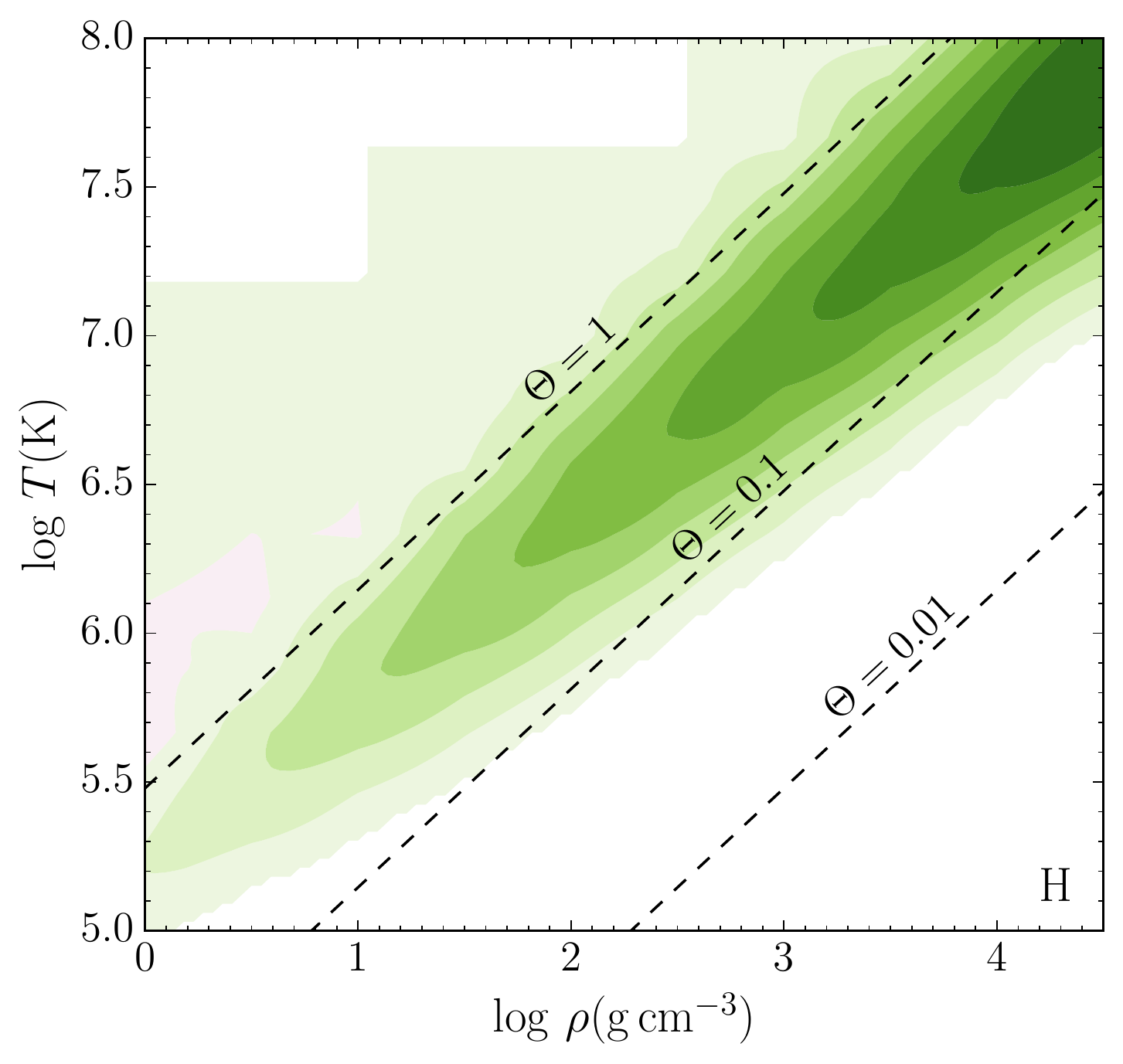}
\includegraphics[height=5.5cm]{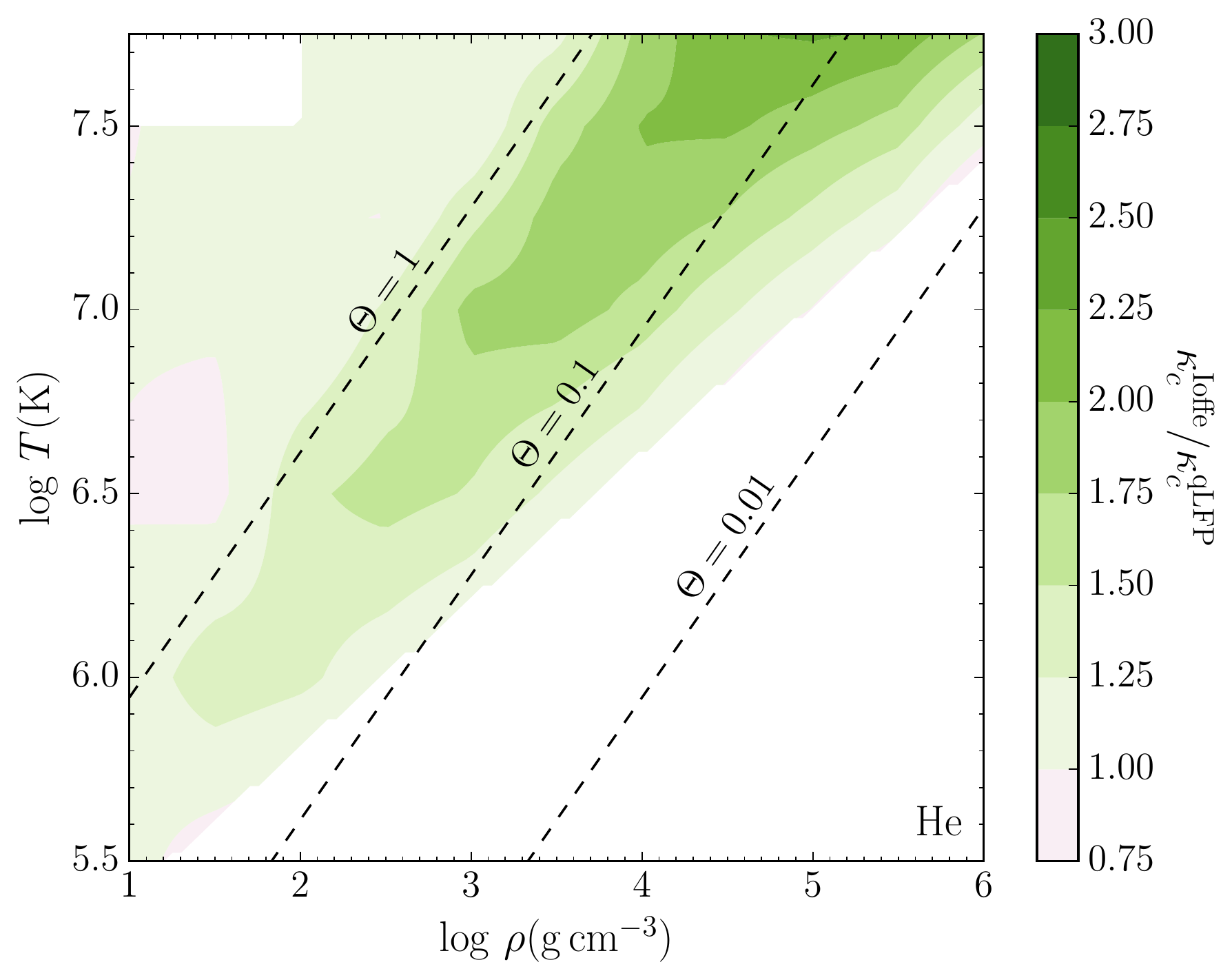}
\caption{Contours of the ratio of the standard conductive opacity $\kappa_{\rm c}^{\rm Ioffe}$ \citep{cassisi2007} to the more recent opacities $\kappa_{\rm c}^{\rm qLFP}$ \cite{blouin2020con}. The long axis of the green band corresponds to $\Theta \sim 0.3$ for both H (left panel) and He (right panel). Adapted from Ref.~\cite{blouin2020con}.} 
\label{fig:blouin_cond}
\end{center}
\end{figure}

The calculation of the electron thermal conductivity in stellar astrophysics and in white dwarfs in particular has a long history, most recently summarized in Ref.~\cite{cassisi2021}. The current standard source of conductive opacity used in white dwarf models is Ref. \cite{cassisi2007}. In the C-O core of white dwarfs, conductive opacities accounting for the strong electron degeneracy, strong plasma coupling and partial relativity (Figures \ref{fig:DA_panel2}, \ref{fig:DA_panel3}, \ref{fig:hDQ_panel2}) are well understood. The regime of partial degeneracy and moderate coupling found in the He and H layers of white dwarfs is more challenging. 
The simple Spitzer--H\"arm expression \cite{spitzer1953} for the electron conductivity in a weakly coupled, non-degenerate and non-relativistic plasma can be extended to stronger coupling with the Chapman--Enskog solution of the kinetic equation for transport using the Landau--Fokker--Planck collision operator.
The recent development of a quantum mechanical (accounting for electron degeneracy) form of the Landau--Fokker--Planck kinetic equation for plasmas (qLFP; \cite{daligault2018}), combined with the potential of mean force from an average atom model \cite{starrett2017} to describe electron-ion and electron-electron collisions, has allowed the extension of calculations of the electron thermal conductivity into the partially degenerate regime where both electron-electron and electron-ion collisions contribute to a comparable degree \cite{shaffer2020}. Previously, the electron-electron contribution was interpolated in this regime \citep{cassisi2007}. The conductive opacities for H and He in white dwarfs computed with this new model \citep{blouin2020con} depart from those of Ref.~\cite{cassisi2007} in a  narrow region of intermediate degeneracy with $\Theta \sim  0.1$ -- 1 (Figure \ref{fig:blouin_cond}). The lower conductive opacities result in faster cooling of white dwarf models. A typical 0.6\,$M_\odot$ H-rich white dwarf reaches the end of the cooling sequence ($\Teff \approx 4000\,$K) about 1\,Gyr earlier, while a massive 0.9\,$M_\odot$ H-rich  white dwarf will reach the same temperature about 2\,Gyr sooner. The correction in cooling time is about 3 times smaller for He-rich  white dwarfs, owing to the smaller change in the conductive opacity of He. Given that the cooling age of such stars is 8--10\,Gyr, these corrections are substantial. The correction to the thermal conductivity due to electron-electron collisions in higher $Z$ plasmas such as carbon will be smaller than for He and are likely negligible.

While Ref.~\cite{blouin2020con} limits their calculation to a range of validity estimated to be $\Theta \gtrsim 0.1$, Ref.~\cite{cassisi2021} points out that the bulk of the change in the conductive opacity calculated by Ref.~\cite{blouin2020con} occurs for $\Theta \lesssim 1$ where the qLFP model is less accurate and likely departs too strongly from the standard conductive opacities of Ref.~\cite{cassisi2007}.  Exploring a range of
possible behaviors of the conductive opacity for $\Theta < 1$ while keeping the opacities of Ref.~\cite{blouin2020con} for $\Theta > 1$, they find that the effect on the cooling time of old white dwarfs can range from slightly shortened to close to the large values reported by Ref.~\cite{blouin2020con}. This recent work points to the need for a better model for the calculation of the thermal conductivity of H and He for $\Theta \sim 0.1 - 1$.

\subsection{Laboratory experiments}
\label{sec: experiments}

The extreme physical conditions encountered in white dwarfs are very challenging---when not impossible---to reproduce in the laboratory. As a result, 
there is an extreme paucity of experimental data that is directly relevant to white dwarfs. Much of 
the constitutive physics that is input to the models is based on physical theories and  models that are of unknown 
accuracy\footnote{Interestingly, the problem of the lack of experiments under white dwarf conditions can sometimes be inverted by using white dwarfs
as laboratories to test physical theory in otherwise inaccessible regimes such as the crystallization of the core (Section \ref{sec:phase_diag}) \citep{tremblay19, blouin20, blouin21}.}.
Contemporary white dwarf research is increasingly concerned with their detailed properties and the stellar models are confronted with a very rich trove of observations that calls for increased physical realism and detail. Much of white dwarf astrophysics relies on our confidence in the validity of the underlying physical models.

Unique, world-class experimental facilities built to create and study matter at extreme conditions, with a focus on nuclear fusion research, have become 
available for research of a more academic nature.\footnote{Two programs that provide access to such facilities to the broader community are the Z Fundamental Science Program at 
Sandia National Laboratories and the National Ignition Facility Discovery Science program at the Lawrence Livermore National Laboratory.} Ongoing experiments at these facilities study various properties of white dwarf plasmas. This new, exciting development in white dwarf science heralds a fresh evaluation of some of the constitutive physics and provides opportunities for original experimental concepts.

\subsubsection{Line profiles}
\label{sec:line_prof_exp}

Experiments that study the Stark effect in H plasmas can achieve temperatures and electron densities found in white dwarf atmospheres
($T \sim 1$ -- 3$\,$eV, $n_e \sim 5 \times 10^{16}$ -- $10^{20}\,$cm$^{-3}$) with a variety of techniques such as Laser Induced Breakdown 
Spectroscopy, gas-liner Z-pinch, and arc discharge. The vast majority of these studies are limited to
measurements of a single emission line (typically H$\alpha$ or H$\beta$) and often focus on the Stark line width and line shift
\cite{buscher2002, djurovic2005, kielkopf2014, parigger2018}. On the other hand, accurate modeling of the absorption 
lines in white dwarf spectra requires detailed theoretical line profiles for several lines in a series up to the merging of the pressure-broadened 
lines near the limit of the series. For five decades, the main experimental reference for Stark broadening of the Balmer lines of hydrogen under 
conditions similar to those in white dwarf atmospheres has been Ref. \cite{wiese1972}, who measured the line profiles of H$\alpha$ to H$\epsilon$ in emission as well as the merging of lines near the Balmer limit. 

\begin{figure}
\begin{center}
\includegraphics[width=0.75\columnwidth]{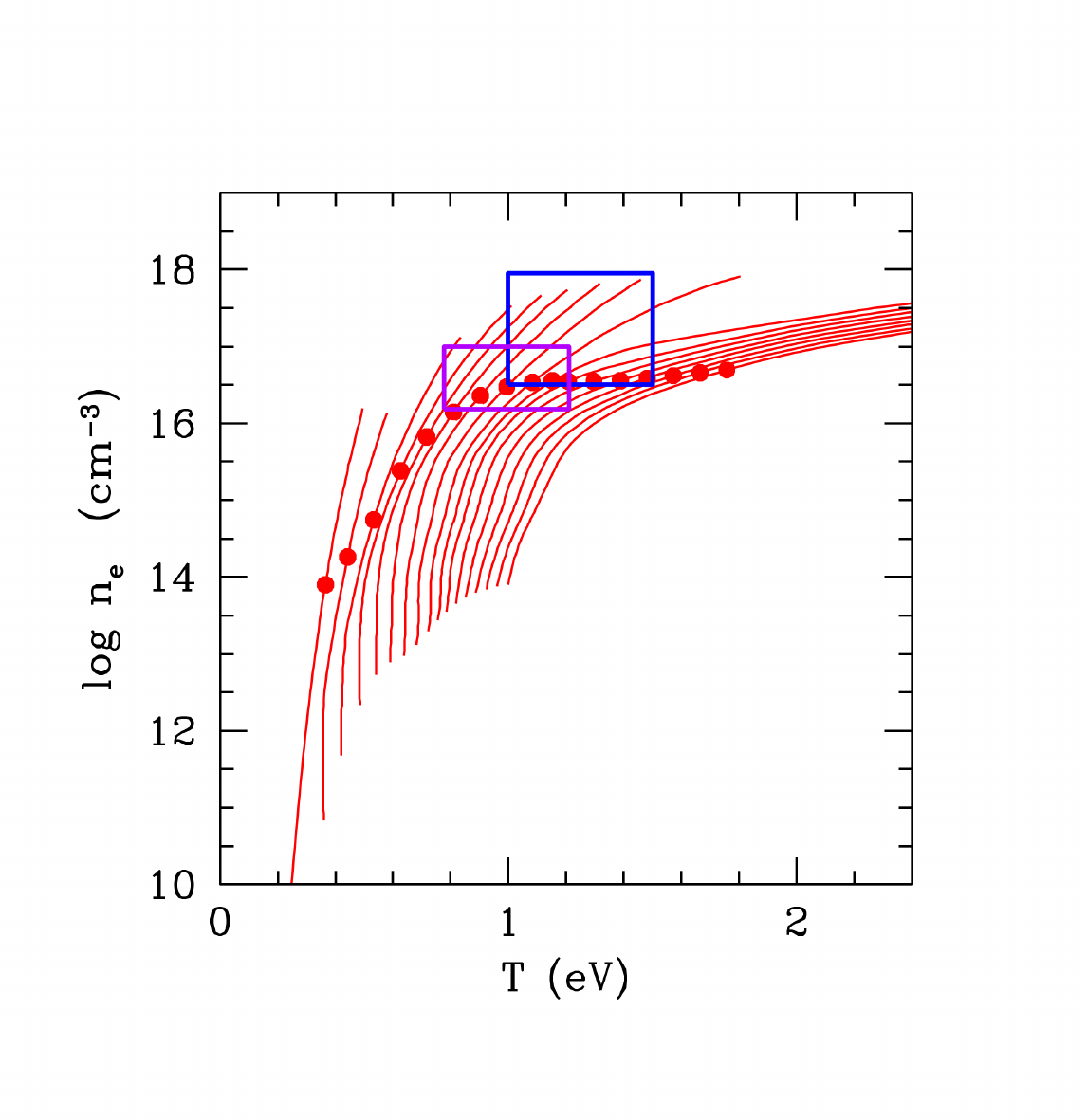}
\caption{Electron density along pure hydrogen white dwarf atmosphere models of $T_{\rm eff}$ = 4000$\,$K to 20\,000$\,$K and $\log g=8$ (from left to right). 
The depth in the atmosphere is represented by the temperature, which increases downward in the star. The solid dot on each curve indicates the location 
of the photosphere, the nominal level where the spectrum is formed. The purple and blue rectangles represent the experimental conditions achieved by Ref.~\cite{wiese1972} and at Z \citep{falcon2015, montgomery2015}.} 
\label{fig:DA_atmos}
\end{center}
\end{figure}

New line profile experiments are underway at the Z-pinch pulsed-power facility at Sandia National Laboratories. The Z machine is a unique and versatile 
experimental facility that can create extreme physical conditions for a wide variety of high energy density experiments including equation of state, 
opacities, and magnetic confinement fusion \cite{sinars2020}. This facility consists of a bank of capacitors that can deliver an electrical 
pulse of up to 30\,MA on time scales of $\sim 100\,$ns onto cm-scale targets. It can be used to generate a hot plasma emitting a 330\,TW flash of soft X-rays. A new experimental platform was developed to use this X-ray source to heat and ionize hydrogen in a large ($\sim 10\,$cm) gas cell. Conditions achieved in the H plasma are $n_e \sim 0.3 $ -- 9 $\times 10^{17}\,$cm$^{-3}$ and 
$T \sim 1$ -- 1.5$\,$eV which are typical of DA white dwarf atmospheres (Figure \ref{fig:DA_atmos}). Simultaneous spectroscopic measurements of the H$\beta$, H$\gamma$ 
and H$\delta$ lines in both emission and absorption are taken with a streak camera which allows tracking of the time evolution of the plasma 
conditions \cite{falcon2015, schaeuble2019}. The experiment is designed to minimize spatial inhomogeneities in the plasma. Separate fits 
of the H$\beta$ and the H$\gamma$ lines observed in absorption give inconsistent values of $n_e$, however
\cite{schaeuble2019}, which echoes the results of fitting different Balmer lines in white dwarf spectra. These results have prompted refinements in the experimental 
design and a new effort in modeling line profiles in 
low-$Z$ elements (Section \ref{sec:line_prof_stark}) \cite{gomez2016, gomez2018a, gomez2018b, gomez2020}. This experimental approach has now been extended to the neutral He lines 
5015$\,$\AA\ and 5875$\,$\AA\ \cite{schaeuble2021} that are among those used in characterizing DB white dwarfs. Because of the large difference in the ionization potentials 
of H and He, it is possible to probe both the broadening of He lines by collisions with neutral atoms and the Stark broadening caused by the surrounding plasma 
by mixing H (a source of electrons and protons) and He (a source of neutral atoms) by controlling the experimental conditions and the H/He mixing ratio. This is a particularly valuable approach in view of the 
uncertainties in the neutral broadening of He lines (Section \ref{sec:line_prof_DB}).  Measurements of C and O lines relevant to hot DQ white dwarfs are also under way \citep{winget2020}.

\subsubsection{Equation of state and K-shell ionization}
A very recent development is the capability to produce plasmas at conditions found well below the atmosphere of white dwarfs. The National Ignition Facility (NIF) is the 
most energetic laser facility in the world and was designed to achieve nuclear fusion by inertial confinement of a deuterium-tritium mixture. The 192 laser beams 
deliver up to 1.8\,MJ of energy on 
a millimeter-size target on a time scale of a few nanoseconds \citep{moses09}. Typically, the UV lasers shine on the inside surface of a small cylindrical 
hohlraum that produces thermal X-ray radiation that fills the cavity. The target is a sphere of $\sim 1\,$mm diameter suspended at the center of the 
hohlraum. Its outer layers are ablated by the X-ray radiation, 
sending a strong converging shock wave towards its center which rapidly compresses to a few 100\,Gbar of pressure and heats up to several keV. Inertial 
confinement fusion (ICF) targets  contain a mixture of deuterium and tritium, and fusion reactions occur in the compressed target \citep{zylstra2022}. 

\begin{figure}
\begin{center}
\includegraphics[width=0.75\columnwidth]{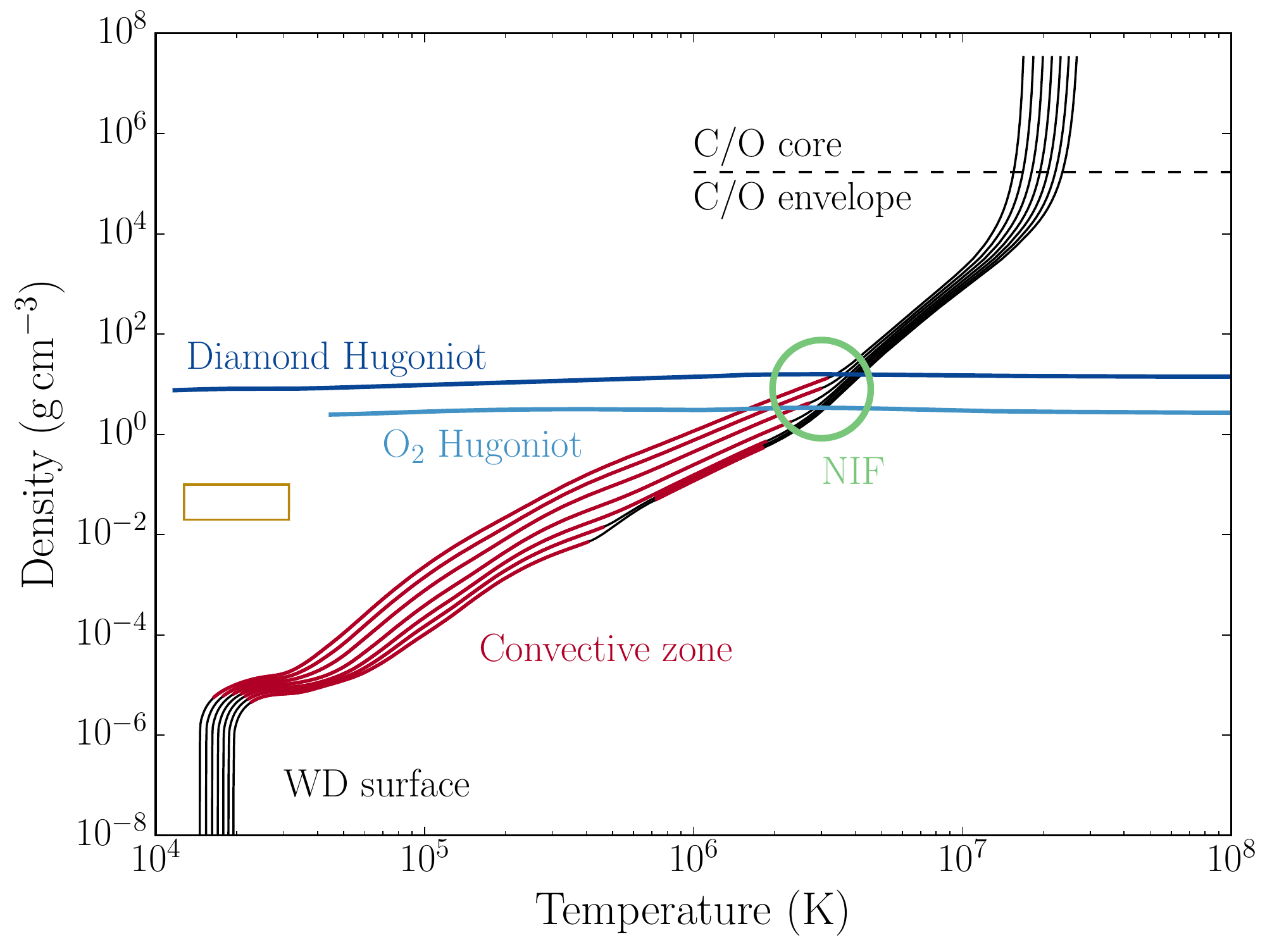}
\caption{Sequence of temperature--density structures of a cooling hot DQ white dwarf of $1\,M_{\odot}$ from $T_{\rm eff}=24\,000$ (bottom profile) to $18\,000\,{\rm K}$ (top profile).
Upon cooling, the structure evolves from right to left. Regions of convective mixing are shown in red. Calculated Hugoniots of diamond (dark blue) and oxygen 
(light blue) are shown \citep{militzer_fpeos}. The conditions to be achieved in the NIF Gbar experiments are highlighted by the green circle, and the conditions reached by Ref.~\cite{roycroft2021} are represented by the box on the left.}
\label{fig:hugoniots}
\end{center}
\end{figure}

This basic concept of spherical shock compression within a hohlraum has been adapted to conduct equation of state measurements at pressures approaching a billion atmospheres (Gbar). The Gbar platform on NIF uses X-ray streak radiography to monitor the time evolution of the spherically imploding capsule, the convergence of the shock front and the density contrast between the unshocked and shocked material. Combined with the Rankine--Hugoniot shock conditions, these measurements 
provide a time-resolved record of the pressure and density along the principal Hugoniot well into the warm dense matter regime. The Gbar platform has produced 
precise Hugoniot data that can discriminate between different equation of state models at unprecedented pressures up to 400\,Mbar for C$_9$H$_{10}$ 
\citep{Kritcher20} and 800$\,$Mbar for diamond. Pressures of $\sim 1\,$Gbar on the principal Hugoniot correspond to the ionization of the $K$ electronic shells 
of C and O. This is a regime that is critical to many white dwarf processes (convection and mixing, the driving of pulsations, diffusion) and most 
challenging to model. Yet, the pressure in EOS tables used in white dwarf models can differ
by 40\% in this regime.  State-of-the-art equations of state \citep{militzer_fpeos} remain untested under those conditions. In a ongoing experimental 
campaign, the Gbar platform on NIF is used to measure the shock Hugoniot of C and O well into the 1$\,$Gbar regime to constrain 
models of the partial ionization of the $K$ shell electrons.
Typical conditions will be $T \sim 100-200\,$eV, $P \sim 0.1 - 1.5\,$Gbar and $\rho$ up to 5\,g\,cm$^{-3}$ (oxygen) and 15$\,$g\,cm$^{-3}$ (carbon). These conditions correspond to the base of the convection zone of hot DQ white dwarfs (Figure \ref{fig:hugoniots}). The prospect of experimentally probing the conditions higher up in the envelopes of hot DQs (at lower pressures and temperatures) was recently explored by Ref.~\cite{roycroft2021} using isochorically heated carbon foams.

This new development opens interesting possibilities for other experiments at those extreme conditions. The charge of a low-$Z$ trace element that is 
gravitationally settling in a fully ionized background of H or He is a key factor in determining its diffusivity \citep{heinonen2020}.  
Under the conditions of interest, elements such as Mg, Si, Ca, Fe are partially ionized and the equation of state models used to obtain their average charge 
make widely different predictions. An experiment that could measure the charge state of ions under white dwarf or similar conditions to challenge these models would be 
very valuable. This could be achieved by measuring the free electron density $n_e$ if the background species (H or He) is fully ionized. While the average ion charge in a warm dense plasma is an ill-defined parameter \citep{murillo2013}, the relevant quantity for 
ionic inter-diffusion is the effective (screened) ion-ion interaction, which can be probed experimentally by measuring the pair distribution function or 
the structure factor.  A direct measurement of diffusivity would be even more valuable but is likely far more challenging.

\subsubsection{Opacities}
\label{sec:exp_opacities}
We have seen that the opacity in cool, pure He atmospheres of spectral type DC presents difficulties from the point of view of the underlying chemical equilibrium, particularly in estimating the number density of free electrons (Section \ref{sec:He_phase_diagram}). Similarly, the cross-sections of all four important absorption processes in He atmospheres (He Rayleigh scattering, He$^-$ free-free, He$_2^+$ bound-free, and triple-He collision-induced opacity, Figure~\ref{fig:kappa_atm_He}) are affected by the correlations in fluid-density atmospheres \citep{iglesias2002, kowalski2014}. Opacity measurements are very much necessary to validate the models for both the chemical equilibrium and the frequency dependence of these continuum absorption cross-sections.  This is particularly important in DC white dwarfs because of the absence of spectral lines that provide powerful diagnostics of conditions in the atmosphere. 
Of particular interest are laser-heated diamond anvil cell experiments that measured the transmittance of heated and compressed He \citep{mcwilliams2015}. The data were obtained along two isobars (22 and 52\,GPa) at temperatures ranging from $\sim 3200$ to 16\,000\,K,\footnote{The corresponding densities are $\sim 0.5$ to 1.2\,g\,cm$^{-3}$.} which match the conditions in atmospheres of $\Teff =4000$ and 5000\,K (Figure \ref{fig:atmprof}). The data show a rapid transition from a transparent insulating fluid to an opaque poor conductor around 10\,000\,K --- a lower temperature than expected from models. In addition, the absorption coefficient increases toward shorter wavelengths, which is opposite to the trend of the dilute gas He$^-$ free-free cross-section (Figure \ref{fig:kappa_atm_He}). These are important results that would affect the predicted spectral energy distribution of He-atmosphere DC white dwarfs. Unfortunately, the range of data is too narrow in wavelength and pressure to confidently develop and calibrate a model that can be implemented in white dwarf atmosphere models. Additional He data with this technique would be very helpful.

A well-known problem in stellar astrophysics is the tension between the increasingly reliable determination of the composition of the Sun that leads to 
interior models that do not agree with the exquisite data of helioseismology \citep{serenelli2009}. An increase in the opacity of the solar plasma could resolve this discrepancy.  A considerable experimental effort has been invested in
measuring the opacity of iron (the principal suspect) at the conditions found at the radiative/convective boundary in the Sun  \citep{bailey15}. The experiments conducted at the Z machine give surprising results that have motivated a very detailed series of experiments and theoretical work in atomic physics \citep{nagayama19}.  They have also prompted the development of the new Opacity platform at the NIF to investigate the opacity of Fe with a completely different technique \citep{perry20}. This Opacity platform is also being applied to white dwarf physics to measure the opacity of O in the partial ionization regime, reaching temperatures of 100--200$\,$eV and densities of 0.01--1$\,$g\,cm$^{-3}$ (Figure \ref{fig:rhoT_HDQ}). The opacity is difficult to model in this dense, partially ionized regime. These X-ray opacity measurements, spanning photon energies of 0.6 to 1.6\,keV, will provide strong constrains on the theory.

The Z pulsed-power machine and the NIF are unique, world-class facilities that complement powerful ``table top'' techniques like laser-heated diamond anvil cells. Together, they  allow for the first time direct measurements of equation of state, opacities and line profiles of key materials (H, He, C and O) under the conditions found in white dwarf stars and bring strong constraints on untested theoretical calculations. The challenging experiments described above are the first efforts in this new area of laboratory astrophysics. We anticipate that during the next decade, more advanced techniques and creative approaches will be developed at these and other facilities with powerful lasers \cite{miquel2018, nicolaizeau2019, jiang2019}, X-ray free electron lasers \citep{bostedt2016, ishikawa2012,tschentscher2017}, and the next generation of pulsed-power machines \citep{sinars2020} to probe white dwarf matter in new ways.

\section{Conclusion}
\label{sec:conclusion}
Astrophysics is a vast playground for the application of physical theory in exotic systems and challenging regimes. Stars of all types have been studied for well over a century and we understand them well enough to ask very detailed questions whose answers require close attention to the constitutive physics of the models. This is particularly true of white dwarfs where the physical conditions are quite different from those in other types of stars.  

As we have seen, there has been steady progress in the modeling of the physics of white dwarfs---particularly over the past decade and a half---that has resulted in tangible improvements in white dwarfs models, better agreement with the observations, the clarification of astrophysical puzzles and, in a few instances, their solution. Yet, many problems remain untouched or unsolved, or the advances, while tangible, are not sufficient to match the observations. In other instances, the physical theory has been developed but not yet applied to white dwarf modeling. It is important to keep in mind that state-of-the-art stellar models are the product of complex multi-physics codes and that the adoption of a better physics model requires that a practical implementation be possible. 

As a summary and for convenience, we list the various current problems that we have discussed with references to the relevant sections. 
\begin{itemize}
    \item Application and experimental validation of Zeeman--Stark line broadening and line shifts in the non-perturbative regime to magnetic white dwarfs (Section~\ref{sec3:magnetic})
    \item Ionization and opacity of He at $\rho \sim 0.1-1\,{\rm g\,cm}^{-3}$ and $T \sim 0.1-1\,{\rm eV}$ (Section~\ref{sec:He_phase_diagram})
    \item Dissociation equilibrium of diatomic molecules (H$_2$, C$_2$, He$_2^+$, HeH$^+$) in dense He (Section~\ref{sec:chem_recent})
    \item Line broadening of all the most important transitions in cool metal-polluted white dwarfs (Section~\ref{sec:line_prof_cool})
    \item Broadening of optical lines of He by collisions with He atoms (Section \ref{sec:line_prof_DB})
    \item Modeling and measurement of the distortion of C$_2$ molecular bands in dense He and in the presence of magnetic fields (Section~\ref{sec:swan})
    \item Ion quantum effects on phase diagram calculations (Section~\ref{sec:dphase_challenges})
    \item Crystallization dynamics and lattice structure in crystallized white dwarfs for different mixtures of species including impurities (Section~\ref{sec:dphase_challenges})
    \item Robust calculations of high-$Z$ (such as uranium) impurity fractionation in crystallizing white dwarfs (Section~\ref{sec:dphase_challenges})
    \item Inter-diffusion coefficients: Benchmark ab initio calculations and new tables of $D_{12}$ and $\alpha_{\rm\sss T}$ based on modern theory, experimental measurement of the charge state of low-Z impurities in a H or He plasma (Section~\ref{sec:diffusion_env})
    \item Calculations of the thermal conductivity of H and He in the $\Theta \sim 0.1$ -- 1 regime (Section~\ref{sec:thermal_cond})
    \item Experimental measurements of the absorption coefficient of dense, heated He under the conditions in cool He-rich DC atmospheres (Section \ref{sec:exp_opacities})
    \item Experimental validation of the latest Stark line shapes for H and He (Section~\ref{sec:line_prof_stark})
    \item Experimental measurement of the H$_2$-He collision-induced absorption in cool white dwarf atmosphere conditions (Section~\ref{sec:CIA})
    \item Line formation in the context of an inhomogeneous convective atmosphere (Section~\ref{sec:rhd})
    \item 3D hydrodynamics simulations of convective overshoot and thermohaline instabilities in the context of chemical mixing (Section~\ref{sec:overshoot})
    \item Magnetic effects on convection and envelope structure, as well as vertical diffusion and horizontal spreading of accreted material (Sections~\ref{sec:conv-magnetic} and \ref{sec:spreading})
\end{itemize}

This long enumeration is by no means exhaustive. Many other problems arise in other types of white dwarfs that are not discussed herein. For instance, non-local thermodynamic equilibrium (NLTE) prevails in the atmospheres of hot white dwarfs with $\Teff \gtrsim 40\,000$\,K. The characterization of those stars, particularly their composition, is hampered by our incomplete knowledge of the relevant atomic physics. 
In young white dwarfs, stellar winds driven by radiation gently blow out the surface layers, compete with diffusion and alter the surface composition. Modeling of these winds also depends on atomic physics data, some of it poorly known. White dwarfs are commonly found in binary star systems where matter from the other star is transferred onto the white dwarf. This leads to many phenomena not seen in single white dwarfs. The transferred material may trigger nuclear fusion under a wide variety of scenarios. If the accreting white dwarf has a strong magnetic field, the infalling matter is funneled to the magnetic poles in two accreting columns. Because of their very high density, white dwarfs have been proposed as potential ``detectors'' of various dark matter particles whose accumulation in their core could affect the cooling rate. The breadth of interesting physics associated with white dwarfs is quite remarkable.

The challenges that we have explored in some detail cover a very wide range of sub-disciplines of physics, from atomic physics to equations of state, non-ideal chemistry, warm dense matter, particle transport, radiative opacities, computer simulations of dense matter,  computational fluid dynamics, and high energy density experiments; all of which are necessary to advance the astrophysics of white dwarfs and address the new questions that will surely flow from new powerful observatories such as the James Webb Space Telescope, the CASTOR space telescope, the Vera Rubin Observatory, the SDSS-V, 4MOST, DESI and WEAVE multi-object spectroscopic sky surveys and the ongoing mining of the data from the {\it Gaia} mission. Advances in solving the problems we have discussed will have a direct impact in several areas of white dwarf astrophysics. Line profiles and line broadening, continuum and molecular opacities, ionization and dissociation equilibria are all essential ingredients in modeling atmospheres and for the calculation of model spectra that are directly compared to observations to determine $\Teff$, surface gravity, composition and magnetic field strength. These basic stellar parameters inform the tool of cosmochronology to determine the age of stellar populations and contribute to the timeline of the formation of structures in our galaxy. Diffusion, convective mixing, overshoot and thermohaline convection are all critical to interpret the abundances of the white dwarfs whose atmospheres are polluted by infalling planetary solids and to draw conclusions about the nature of old exoplanetary systems. Crystallization coupled with diffusion in the envelope and thermal conductivity affect the cooling time scale of old white dwarfs and the determination of white dwarf ages. Accurate collision-induced opacities are crucial to understand the sub-class of IR-faint white dwarfs about which we know so little. Better models of the pressure and magnetic distortion of the Swan bands of C$_2$ and of the C$_2$ dissociation in dense He will allow an accurate characterization of the C dredge up mechanism that involves both diffusion at the C/He interface and convective mixing. Equations of state, in the warm dense matter regime of partial ionization of H, He and C affect the extent of convection zones, the driving of pulsation modes, and the onset of convective coupling that is a critical event in white dwarf cooling. The ionization state of low-Z elements is an important factor in inter-diffusion of trace elements, which arcs back to the composition of exoplanetary material orbiting white dwarfs.

It remains a source of fascination that the application of fundamental physics principles at the atomic scale, combined with clever models, powerful computing resources and experiments on small, laboratory-scale samples can lead directly to an understanding of systems as unreachable as stars. In this endeavor, we look forward to the contributions of scientists with new ideas from related fields of research to further our knowledge of these very small, dim, yet numerous stars imbued with cosmic significance.

We dedicate this review to the memory of our teacher, colleague and friend Gilles Fontaine who, animated by a great passion for stellar astrophysics and training young scientists, left an indelible legacy in this field.

\section*{Acknowledgments}

We are grateful to J\'er\^ome Daligault, Christopher Fontes and Paul Bradley for carefully reading the manuscript and for their constructive comments. We thank the participants in the workshop on ``Current Challenges in the Physics of White Dwarf Stars'', held in Santa Fe in 2017, who planted the seed of this review by sharing their expertise and enthusiasm with colleagues from unfamiliar fields of research.
Part of this work was performed under the auspices of the U.S Department of Energy under contract No. 89233218CNA000001 and by the
Laboratory Directed Research and Development program of Los Alamos National Laboratory under project number 20190624PRD2. SB is a Banting Postdoctoral Fellow and a CITA National Fellow, supported by the Natural Sciences and Engineering Research Council of Canada (NSERC). PET received funding from the European
Research Council under the European Union’s Horizon 2020 research and
innovation programme number 101002408 (MOS100PC), the Leverhulme Trust Grant (ID RPG-2020-366) and the UK STFC consolidated grant ST/T000406/1.

\bibliographystyle{elsarticle-num} 
\bibliography{references}

\end{document}